\documentclass[12pt,epsf]{article}
\usepackage{psfrag,graphicx,subfigure}

\makeatletter
  
  \@addtoreset{equation}{section}
\makeatother

\begin{document}

%
%
\begin{titlepage}

\begin{flushright}
OU-HET 384 \\
hep-th/0105272 \\
May 2001 \\
Revised version
\end{flushright}

\bigskip

\begin{center}
{\Large \bf CLASSICAL OPEN-STRING FIELD THEORY :}\\
{\large \bf  A$_{\infty}$-Algebra, Renormalization Group} 
{\large \bf and Boundary States}

\bigskip
\bigskip
Toshio NAKATSU \\
\bigskip
{\small \it
Department of Physics,\\
Graduate School of Science, Osaka University,\\
Toyonaka, Osaka 560, JAPAN
}
\end{center}

\bigskip
\bigskip

\begin{abstract}
We investigate classical bosonic open-string field theory from 
the perspective of the Wilson renormalization group of world-sheet theory. 
The microscopic action is identified with Witten's covariant cubic action 
and the short-distance cut-off scale is introduced by length of 
open-string strip which appears in the Schwinger representation of 
open-string propagator. {\it Classical open-string field theory} 
in the title means open-string field theory governed by a classical part 
of the low energy action. It is obtained by integrating out suitable 
tree interactions of open-strings and is of non-polynomial type. 
We study this theory by using the BV formalism. It turns out to be deeply 
related with deformation theory of $A_{\infty}$-algebra. 
We introduce renormalization group equation of this theory and 
discuss it from several aspects.  
It is also discussed that this theory is interpreted as a boundary 
open-string field theory. Closed-string BRST charge and boundary states 
of closed-string field theory in the presence of open-string field 
play important roles.

\end{abstract}

\end{titlepage}



%
%
\newtheorem{definition}{Definition}[section]
\newtheorem{proposition}{Proposition}[section]
\newtheorem{theorem}{Theorem}[section]
\newtheorem{lemma}{Lemma}[section]
\newtheorem{remark}{Remark}[section]
\newcommand{\Bll}{\Bigl\langle}
\newcommand{\Brr}{\Bigr\rangle}
\newcommand{\bll}{\bigl\langle}
\newcommand{\brr}{\bigr\rangle}
\newcommand{\Bl}{\Bigl |}
\newcommand{\Br}{\Bigr |}
\newcommand{\bl}{\bigl |}
\newcommand{\br}{\bigr |}


\section{Introduction}

In this paper we investigate classical bosonic 
open-string field theory from the perspective of 
the Wilson renormalization group of world-sheet theory. 
The microscopic action is identified with 
Witten's covariant cubic action \cite{Witten NCG} of open-string field 
and the short-distance cut-off scale parameter is introduced 
by length of open-string strip which appears 
in the Schwinger representation of open-string propagator.

Low energy action at the cut-off scale $\zeta$ 
is obtained by integrating out all the contributions 
of open-string interactions at length scales less than $\zeta$. 
It becomes the cubic action as $\zeta$ goes to zero, and provides 
a macroscopic description of open-string field theory as 
$\zeta$ goes to $\infty$. The low energy action is of non-polynomial 
type similarly to the case of closed-string field theory 
\cite{Zwiebach BV, HIKKO} and has an expansion by 
open-string loops. {\it Classical open-string field theory} in the 
title means open-string field theory governed by the classical 
part of the low energy action. 
It is understood to be obtained 
by integrating out the tree contributions. 
$n$-valent tree interactions of open-strings beyond the cubic 
become elementary at the cut-off scale $\zeta$ ($>0$) and 
contribute to the action. 
Geometric structure of these interactions can be handled 
by homotopy associative algebra ($A_{\infty}$-algebra) \cite{Stasheff}. 
Study of {\it classical open-string field theory} turns out to be 
deeply related with deformation theory of $A_{\infty}$-algebra which 
is recently developed by K. Fukaya et al \cite{Fukaya,FOOO}.

In Sections $4$ and $5$ 
we describe {\it classical open-string field theory} 
from the viewpoint of the deformation theory. 
Differential graded algebra in the microscopic description 
is open-string gauge algebra  \cite{Witten NCG}. 
It is non-commutative associative algebra. 
The gauge algebra is deformed in the macroscopic description 
or flows in the sense of renormalization group,  
to $A_{\infty}$-algebra which is non-commutative and non-associative. 
Our description is based on 
the Batalin-Vilkovisky formalism \cite{BV formalism}. 
It was elegantly used in quantization of open-string field theory 
\cite{Thorn,Bochicchio}. It also played an important role 
in the construction of closed-string field theory \cite{Zwiebach BV}.

{\it Off-shell} open-string fields at different length scales 
are related with one-another by renormalization group flows. 
In Section $6$ we introduce renormalization group equation of 
{\it classical open-string field theory}. 
This is a rational concept since the action is the classical part 
of low energy effective action.  
Our treatment of renormalization group is based on that 
given in \cite{Wilson-Kogut,Polchinski}.

~

Nevertheless it may be important to explain the above idea 
in the context of field theories and to make a comparison between two. 
For this purpose let us recall briefly the renormalization 
group treatment in field theories. 
We consider a real scalar field theory in $d$-dimensions 
$(2< d \leq 4)$. We use the momentum representation. 
Let $\varphi_p$ be a momentum mode of 
$\varphi(x)=\frac{1}{(2\pi)^{d/2}}\int dp 
e^{ipx}\varphi_p$. 
We put $\varphi \equiv \left\{ \varphi_p \right\}$. 
Let $\Lambda$ be a UV cut-off. 
We put $\varphi_{\leq \Lambda}$ 
$\equiv \left\{ \varphi_p \right\}_{|p|\leq \Lambda}$. 
Let $S[\varphi_{\leq \Lambda_0}:\Lambda_0]$ be an Euclidean action 
at $\Lambda_0$. For $\Lambda < \Lambda_0$ the Wilsonian 
action $S[\varphi_{\leq \Lambda}:\Lambda]$ is defined 
in principle by the following integration on 
$\varphi_p(\Lambda<|p|\leq \Lambda_0)$ in the original action. 
\begin{eqnarray}
\int \prod_{\Lambda < |p| \leq \Lambda_0} d\varphi_p
e^{-S[\varphi_{\leq \Lambda_0}:\Lambda_0]}= 
const. \times 
e^{-S[\varphi_{\leq \Lambda}:\Lambda]}.
\label{Wilsonian for field theory}
\end{eqnarray}
The integration on the high energy modes 
changes the values of the coupling constants or generates 
several interactions at $\Lambda$.

It is convenient to describe the cut-off theory by using 
a regularized propagator $\Delta(p:\Lambda)$. 
We may put $\Delta(p:\Lambda)=\frac{K(p^2/\Lambda^2)}{p^2+m^2}$, 
where $K$ is a cut-off function and $m$ is a bare mass. 
We put ${\cal S}(\Lambda)$ be a set of (running) coupling 
constants, 
\begin{eqnarray}
{\cal S}(\Lambda)= 
\Bigl \{ S_n(p_1,\cdots,p_n:\Lambda) \Bigr \}_{n \geq 0},  
\label{set of coupling constants for field theory}
\end{eqnarray}
where $S_n(p_1,\cdots,p_n:\Lambda)$ are symmetric functions with 
respect to $p_a$. The relevant action has the following form. 
\begin{eqnarray}
S[\varphi:{\cal S}(\Lambda):\Lambda]&=& 
\frac{1}{2}\int dp \Delta(p:\Lambda)^{-1}\varphi_p\varphi_{-p}
+\sum_{n=0}^{\infty}\frac{1}{n!} 
\int \prod_{a=1}^n dp_a 
S_n(p_1,\cdots,p_n:\Lambda) 
\prod_{a=1}^n \varphi_{p_a} 
\nonumber \\
&\equiv& 
\frac{1}{2}\int dp \Delta(p:\Lambda)^{-1}\varphi_p\varphi_{-p}
+S_{int}[\varphi:\Lambda]. 
\label{regularized action for field theory}
\end{eqnarray}
Let $Z({\cal S}(\Lambda):\Lambda)$ be the partition function 
\begin{eqnarray}
Z({\cal S}(\Lambda):\Lambda)= 
\frac{\int D\varphi e^{-S[\varphi:{\cal S}(\Lambda):\Lambda]}}
{\int D\varphi e^{-S[\varphi:\emptyset:\Lambda]}}
~~~~~~~~~~(D\varphi \equiv \prod_p d\varphi_p), 
\label{partition funtion for field theory}
\end{eqnarray}
where the denominator is a simple Gaussian integral. 
The above construction of the Wilsonian action implies  
\begin{eqnarray}
\Lambda \frac{d}{d \Lambda} 
Z({\cal S}(\Lambda):\Lambda)=0, 
\label{RG principle}
\end{eqnarray}
which states that the coupling constants flow so that the partition 
function is actually independent of $\Lambda$.  
This equation can be brought to a perturbatively tractable form  
by using the expression 
$Z=
\left.
e^{-S_{int}[\frac{\delta}{\delta j}]}
e^{\frac{1}{2}\int dp \Delta(p:\Lambda)j_pj_{-p}}
\right|_{j=0}$.  
After some manipulation we can rewrite equation (\ref{RG principle}) 
in the following form. 
\begin{eqnarray}
0=
\int D\varphi
\left\{ 
\Lambda \frac{dS_{int}}{d\Lambda}
-\frac{1}{2}\int dq 
\Lambda \frac{\partial \Delta(q:\Lambda)}{\partial \Lambda}
\left( 
\frac{\partial S_{int}}{\partial \varphi_q}
\frac{\partial S_{int}}{\partial \varphi_{-q}}-
\frac{\partial^2S_{int}}{\partial \varphi_q \partial \varphi_{-q}}
\right)
\right\}e^{-S[\varphi]}. 
\end{eqnarray}
This shows that the flow of ${\cal S}(\Lambda)$ needs to satisfy 
the Wilson-Polchinski's renormalization group equation 
\cite{Wilson-Kogut,Polchinski}
\begin{eqnarray}
\Lambda \frac{dS_{int}}{d\Lambda}[\varphi:{\cal S}(\Lambda)]
=\frac{1}{2}\int dq 
\Lambda \frac{\partial \Delta(q:\Lambda)}{\partial \Lambda}
\left( 
\frac{\partial S_{int}}{\partial \varphi_q}
\frac{\partial S_{int}}{\partial \varphi_{-q}}-
\frac{\partial^2S_{int}}{\partial \varphi_q \partial \varphi_{-q}}
\right)
[\varphi:{\cal S}(\Lambda)].
\label{Wilson-Plochinski's RG equation for field theory}
\end{eqnarray}

In the text we regulate open-string propagator utilizing 
the Schwinger representation. 
The counterpart in the field theory is 
\begin{eqnarray}
\Delta(p:\Lambda)= 
\int_{1/\Lambda^2}^{\infty}ds e^{-s(p^2+m^2)}
~~~~~
\left(
= \frac{e^{-\frac{p^2+m^2}{\Lambda^2}}}{p^2+m^2}
\right). 
\label{Schwinger rep of propagator for field theory}
\end{eqnarray}
Advantage in taking this regularization is that the action 
(\ref{regularized action for field theory}) becomes equivalent 
to the following one after the similarity transformation, 
$\varphi_p \rightarrow e^{\frac{p^2+m^2}{2\Lambda^2}}\varphi_p$ : 
\begin{eqnarray}
&&
\hat{S}[\varphi:{\cal S}(\Lambda):\Lambda]
\nonumber \\
&&= 
\frac{1}{2}\int dp (p^2+m^2)\varphi_p\varphi_{-p}
+\sum_{n=0}^{\infty}\frac{1}{n!}\int \prod_{a=1}^n dp_a 
\left( 
\prod_{a=1}^n e^{-\frac{p^2_a+m^2}{2\Lambda^2}}
\right)
S_n(p_1,\cdots,p_n:\Lambda)
\prod_{a=1}^n\varphi_{p_a} 
\nonumber \\
&&\equiv 
\frac{1}{2}\int dp (p^2+m^2)\varphi_p\varphi_{-p}
+\hat{S}_{int}[\varphi:{\cal S}(\Lambda):\Lambda]. 
\label{regularized action 2 for field theory}
\end{eqnarray}
Instead of the propagator the interaction vertices acquire 
the regularization factors (external legs in the Feynman diagrams).  
The effective action of open-string field in the text is expressed   
in a way analogous to this form. 
The renormalization group equation 
(\ref{Wilson-Plochinski's RG equation for field theory})
can be written down in terms of $\hat{S}_{int}$ as follows. 
\begin{eqnarray}
&&
\Lambda \frac{d\hat{S}_{int}}{d\Lambda}
[\varphi:{\cal S}(\Lambda):\Lambda]
\nonumber \\
&&
=
\frac{1}{2}\int dq 
\frac{1}{\Lambda^2}
\left( 
\frac{\partial \hat{S}_{int}}{\partial \varphi_q}
\frac{\partial \hat{S}_{int}}{\partial \varphi_{-q}}-
\frac{\partial^2\hat{S}_{int}}{\partial \varphi_q \partial \varphi_{-q}}
\right)
[\varphi:{\cal S}(\Lambda):\Lambda]
+\Lambda \frac{\partial \hat{S}_{int}}{\partial \Lambda}
[\varphi:{\cal S}(\Lambda):\Lambda].
\label{Wilson-Polchinski's RG equation 2 for field theory}
\end{eqnarray} 
The evolution of the coupling constants 
$S_n(p_1,\cdots,p_n:\Lambda)$ can be read from this equation. 
\begin{eqnarray}
&&
\Lambda \frac{\partial S_n}{\partial \Lambda}(p_1,\cdots,p_n:\Lambda)
\nonumber \\
&&
=\int dq \frac{e^{-\frac{q^2+m^2}{\Lambda^2}}}{\Lambda^2}
\left\{ 
\sum_{I,J}
S_{|I|+1}(q,P_I:\Lambda)S_{|J|+1}(-q,P_J:\Lambda)
-S_{n+2}(q,-q,p_1,\cdots,p_n:\Lambda)
\right\},
\nonumber \\
\label{RG equation for field theory}
\end{eqnarray}
for $n=0,1,2,\cdots$.  Here $I,J$ are the multi-indices 
satisfying $I \cup J=\left\{ 1,\cdots,n \right\}$.

\begin{figure}
\psfrag{zeta}{$\zeta$}
\psfrag{s_1}{$s_1$}
\psfrag{s_2}{$s_2$}
\psfrag{s_3}{$s_3$}
\psfrag{sum}{$\sum$}
\psfrag{A}{$s_1$}
\psfrag{B}{$s_1,s_2$}
\psfrag{C}{$s_1,s_2,s_3$}
\psfrag{1}{$1$}
\psfrag{2}{$2$}
\psfrag{3}{$3$}
\psfrag{n}{$n$}
\psfrag{cdots}{$\cdots$}
\psfrag{vdots}{$\vdots$}
\psfrag{D}
{$n$-valent vertex at $\Lambda=\Lambda_0e^{-\zeta}$}
\psfrag{E}
{$4$-valent vertex at $\Lambda_0$}
\begin{center}
\includegraphics[height=16cm]{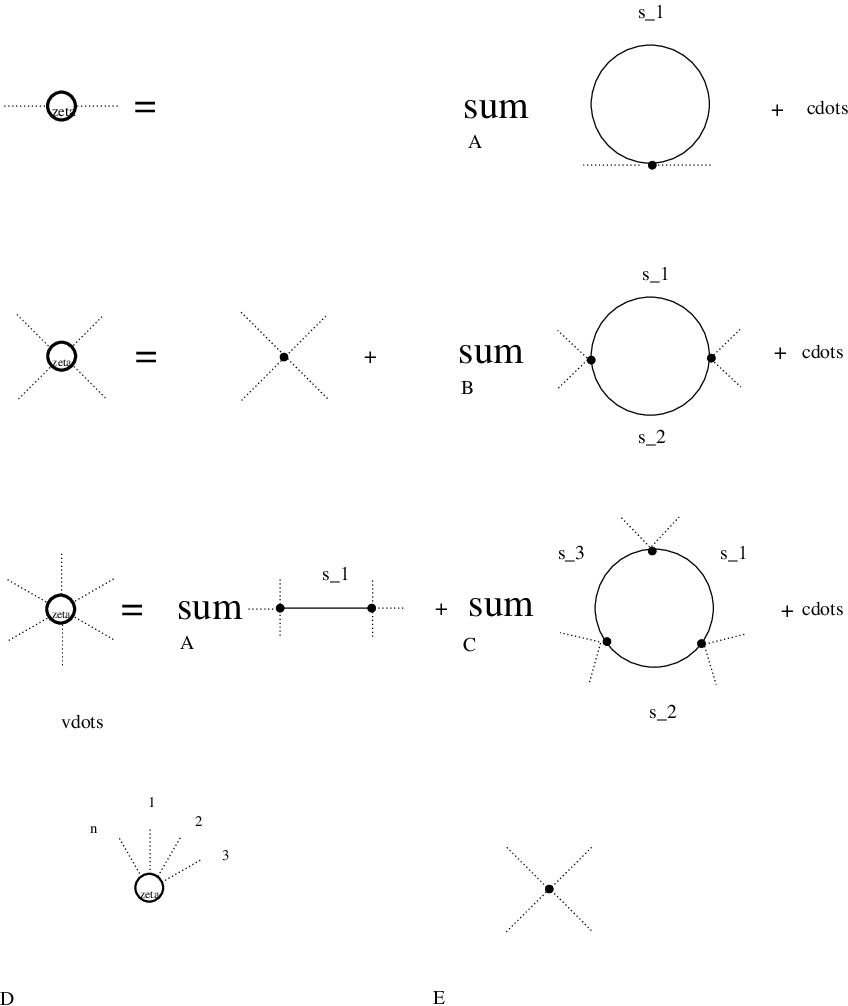}
\caption{
{\small 
Perturbative expansion of the solution 
: The Schwinger parameters $s_i (i=1,2,3)$ are integrated over 
the region 
$(\frac{1}{\Lambda_0})^2<s_i \leq (\frac{e^{\zeta}}{\Lambda_0})^2$. 
The dotted lines are deleted external legs.} }
\label{solutionofRG}
\end{center}
\end{figure}

Let us take a closer look on the evolution of the coupling 
constants. To be definite, we start at $\Lambda_0$ with the 
condition, 
\begin{eqnarray}
S_4(p_1,p_2,p_3,p_4:\Lambda_0)=v_0 \times \delta(\sum_{a=1}^4p_a) 
~~~~~
\mbox{and all other $S_k$ vanish.}
\end{eqnarray}
[Note : 
To study the renormalizability of $\phi^4$-theory in $d$-dimensions 
($^{i.e.}$ to search an existence of the limit 
$\Lambda_0 \rightarrow \infty$), other coupling constants 
$S_2(p,q:\Lambda_0)=(\eta_0p^2+u_0)\delta^d(p+q)$ are required. 
We take the above condition just to explain the underlying idea 
in the text.] 
Solutions of the evolution equation 
(\ref{RG equation for field theory}) admit to have 
perturbative expansions. 
At $\Lambda=\Lambda_0e^{-\zeta}$($\zeta >0$),  
solution in this particular example has the expansions 
as depicted in Fig. \ref{solutionofRG}.
The Feynman graphs are interpreted as world-lines in the figure  and  
the Schwinger parameter $s$ of $e^{-s(p^2+m^2)}$ is regarded as length 
of the world-line. 
The higher vertices are generated at $\Lambda$ 
by the contribution of graphs 
all the internal propagators in which have length 
$(\frac{1}{\Lambda_0})^2<s \leq (\frac{e^{\zeta}}{\Lambda_0})^2$.  
The appearance of loops is originated in the term 
$\frac{\partial^2\hat{S}_{int}}
{\partial \varphi_q \partial \varphi_{-q}}$ of the renormalization 
group equation. The proliferation of trees is due to the term 
$\frac{\partial \hat{S}_{int}}{\partial \varphi_q}
\frac{\partial \hat{S}_{int}}{\partial \varphi_{-q}}$.

At this present stage we introduce a concept of {\it classical} 
renormalization group flow. 
It is defined by dropping out the loop effect 
in the original equation. 
\begin{eqnarray}
\Lambda \frac{d\hat{S}_{int}}{d\Lambda}
[\varphi:{\cal S}(\Lambda):\Lambda]
=
\frac{1}{2}\int dq 
\frac{1}{\Lambda^2}
\frac{\partial \hat{S}_{int}}{\partial \varphi_q}
\frac{\partial \hat{S}_{int}}{\partial \varphi_{-q}}
[\varphi:{\cal S}(\Lambda):\Lambda]
+\Lambda \frac{\partial \hat{S}_{int}}{\partial \Lambda}
[\varphi:{\cal S}(\Lambda):\Lambda].
\label{Classical Wilson-Polchinski RG equation  for field theory}
\end{eqnarray} 
The evolution of the coupling constants can be read as    
\begin{eqnarray}
\Lambda \frac{\partial S_n}{\partial \Lambda}(p_1,\cdots,p_n:\Lambda)
=\int dq \frac{e^{-\frac{q^2+m^2}{\Lambda^2}}}{\Lambda^2}
\sum_{I,J}
S_{|I|+1}(q,P_I:\Lambda)S_{|J|+1}(-q,P_J:\Lambda),
\label{Classical RG equation for field theory}
\end{eqnarray}
for $n=0,1,2,\cdots$. 
With the same initial condition at $\Lambda_0$ 
the evolution of the coupling constants becomes 
that obtained by neglecting the loops in the previous solution.

In the case of open-string field theory 
we take a route which is slightly different 
from the above field theory argument. 
This is partially because in open-string field theory 
there are no coupling constants except the string field itself.
It indicates that wave function renormalizations suffice 
for the renormalization group argument. 
In fact the classical effective action $S[\Phi:\zeta]$, 
where $\Phi$ is open-string field and $\zeta$ is the cut-off scale, 
is given in Section 5 and shown to satisfy 
the master equation, $\left\{ S,S \right\}=0$. 
This confirms the above observation. Since, as one can find 
in the standard argument on renormalizability of 
gauge field theories, 
the master equation ensures that the theories maintain 
gauge symmetries in each step of renormalization group 
transformation.  
It is useful to point out that equation (\ref{RG principle}) 
can be expressed as 
\begin{eqnarray}
\Lambda 
\frac{d \Gamma}{d \Lambda}[\varphi:{\cal S}(\Lambda):\Lambda] 
=0, 
\end{eqnarray}
where $\Gamma$ is the effective action. 
Guided by this expression we then define 
the renormalization group equation of 
{\it classical open-string field theory} by 
\begin{eqnarray}
\frac{dS}{d\zeta}[\Phi(\zeta):\zeta]=0.  
\end{eqnarray}
Description of renormalization group equation  
of string-field theory along this line was first given 
for closed-string in \cite{Alwis}. 
The renormalization group equation turns out to give 
an infinitesimal variation of the $A_{\infty}$-algebra but 
its precise relation with the deformation theory  
is still left unclear. 
Open-string field theory analogue of equation 
(\ref{Classical Wilson-Polchinski RG equation  for field theory}) 
is derived in equation (\ref{RG equation 2}) as  
\begin{eqnarray}
\omega \left(\frac{d\Phi}{d\zeta},
~\left|\frac{\partial S[\Phi\!:\!\zeta]}{\partial \Phi}
\right \rangle \right)=
\omega \left(  
b_0
\left|
\frac{\partial S_{int}[\Phi\!:\!\zeta]}{\partial \Phi}\right \rangle 
+L_0\Phi,~
\left |
\frac{\partial S_{int}[\Phi\!:\!\zeta]}{\partial \Phi}\right \rangle 
\right). 
\end{eqnarray}
Here $\omega$ is the odd symplectic structure of open-string field 
theory and the ket vectors are the hamiltonian vectors of 
the classical effective action and its interaction part.
They are explained in Section 5 and used to develop 
the Batalin-Vilkovisky formalism of open-string field theory.

World-lines in field theories are replaced 
by world-sheets of strings in string field theories.  
The open-string world-sheets are identified with ribbon graphs while 
the world-lines are with graphs. The tree ribbon graphs are mapped 
to $2$-disk by the Mandelstam mapping. This cause a great difference 
with field theory. By this mapping the effective action 
$S[\Phi:\zeta]$ acquires an interpretation 
in terms of two-dimensional $\sigma$-model.
And the classical renormalization group flow can be identified with 
renormalization group flow of the underlying two-dimensional 
$\sigma$-model. This aspect is studied in the next section.

~

We start Section 7 with a further investigation on the 
renormalization group flow by imposing the Siegel gauge condition 
on open-string field $\Phi$.  
We express the renormalization group equation 
in a form $dT^i/d\zeta=\beta^i(T,\zeta)$, 
where $T^i$ are the coefficients in a suitable expansion 
of open-string field $\Phi$ (or the space-time variables ). 
The beta functions are shown in Proposition 
\ref{def beta function} to have the form, 
\begin{eqnarray}
\beta^i=\sum_jg^{ij}\frac{\partial S}{\partial T^j},
\end{eqnarray}
where $g_{ij}$ is roughly the renormalized BPZ metric 
of two-dimensional field theories. 
Since $T^i$ are open-string field $\Phi$ itself 
this expression implies that 
zeros of the beta functions are nothing but classical 
solutions of open-string field theory. 
Contrary to our naive expectation, the beta functions depend on 
the cut-off scale parameter $\zeta$ {\it explicitly}. 
This originates in our regularization scheme of open-string 
field theory. Open-string field theory is formulated using 
{\it two-dimensional} conformal field theory \cite{LPP}. 
But the regularization we choose is simply to put a restriction 
on length of open-string evolution, which is a regularization 
of {\it one-dimension}. Actually we have two length scales.  
The missing scale is length of open-string itself. 
$\zeta$ is the ratio of these two length scales 
and is a dimensionless parameter. 
In the end of Section $7$ 
the regularization employed so far  
is examined from the perspective of world-sheet boundary theories. 
It corresponds to a point-splitting regularization of 
short-distance on the boundary when $\zeta$ is sufficiently large. 
The point-splitting is prescribed by the boundary length scale 
$\sim e^{-\zeta}$. 
The action $S[\Phi\!:\!\zeta]$ is interpreted as an analogue of 
a generating function of all correlation function of a world-sheet 
boundary theory regularized by the point-splitting method.

Our discussions in Sections $4-7$ are based on a conjecture given 
in the end of Section $3$. 
The conjecture is related with construction of 
open-string $n$-valent vertices at the 
cut-off scale $\zeta$ for $n \geq 3$. 
Section $3$ besides Appendix are devoted to test the conjecture.

One of the motivations of this paper is Sen's conjecture 
\cite{Sen conjecture} on open-string tachyon condensation. 
The conjecture has been studied by using open-string field theory 
in two different formulations. 
One \cite{tachyon cubic} is based on Witten's covariant 
cubic open-string field theory and 
the other \cite{tachyon BOSFT} \cite{Kutasov-Marino-Moore}
is based on the so-called boundary open-string field theory 
\cite{Witten BOSFT}. 
Relation between these two treatments has not been clarified yet. 
In Section $8$ we investigate their relation. It is discussed 
that the macroscopic open-string field theory studied in this paper  
is interpreted as a boundary open-string field theory. 
Boundary states of closed-string field theory in the presence 
of open-string field and the closed-string BRST charge play 
important roles in the discussion. 
In particular the following equality is found :
\begin{eqnarray}
Q_c 
\Bl 
B[\Phi \!:\! \zeta]
\Brr
=
\delta_{{\bf BRS}}
\Bl 
B[\Phi \!:\!\zeta]
\Brr.
\end{eqnarray}
The ket vector is a boundary state of closed-string 
in the presence of open-string field $\Phi$ and 
is given in equation (\ref{boundary state bphi}). 
$Q_c$ is the (free) closed-string BRST charge. 
$\delta_{{\bf BRS}}$ is the BRST transformation of open-string 
field $\Phi$ which is determined by the action 
$S[\Phi:\zeta]$ in the framework of the Batalin-Vilkovisky 
formalism.

We provide a brief review on the cubic open-string field theory 
in Section $2$ to make the paper self-contained as far as possible.


\section{Microscopic Open-String Field Theory}
In this section we give a brief review on 
Witten's covariant bosonic open-string field theory \cite{Witten NCG}. 
It is described along the work of LeClair, Peskin and Preitschopf 
\cite{LPP}. Our goal is Definition \ref{def micro action}, 
where the covariant action is given. 
This section is also devoted to a preparation for later discussions. 
Basic machinery and concepts which will be used in the subsequent sections 
are explained here.

Bosonic string field theory is formulated using a two-dimensional 
conformal field theory ($2D$ CFT). 
This conformal field theory consists of 
the matter sector described by $X^{\mu}$ 
$(0\leq \mu \leq 25)$ and the ghost sector 
described by world-sheet reparametrization ghosts $(b,c)$. 
$X^{\mu}$ and $(b,c)$ are respectively grassmann-even and -odd 
variables of string coordinates $(X^{\mu},b,c)$. 
The matter and ghost sectors are $2D$ CFTs of central charge 
respectively equal to $26$ and $-26$. 
The total conformal field theory therefore has central charge 
equal to zero. 
Consider the conformal field theory formulated in the $z$-plane 
with $z\!=\!e^{\tau+i\sigma}$. 
Mode expansion of the string coordinates on a unit disk $|z|\leq 1$ is 
\begin{eqnarray}
&&
X^{\mu}(z)=x^{\mu}-ip^{\mu}\ln z
-i\sum_{n\neq 0}\frac{\alpha_{-n}^{\mu}}{n}z^n, 
\nonumber \\
&&
b(z)=\sum_{n}b_{-n}z^{n-2}, 
~~~~~~~~~
c(z)=\sum_nc_{-n}z^{n+1}, 
\label{mode expansion of coordinates}
\end{eqnarray} 
where we set string length scale $l_s$ equal to one. 
Their first quantization gives the following 
(anti-)commutation relations.
\begin{eqnarray}
&&
\mbox{[} x^{\mu},p^{\nu} \mbox{ ]}=i\eta^{\mu \nu}, ~~~~~~~
[\alpha_n^{\mu},\alpha_{m}^{\nu}]=n \eta^{\mu \nu} \delta_{n+m,0},
\nonumber \\
&&
\left\{b_n,c_m\right\}=\delta_{n+m,0},
\label{CCR}
\end{eqnarray}
where $\eta^{\mu \nu}\!=\!\mbox{diag}(-1,1,\cdots,1)$ 
is the Minkowski metric of ${\bf R}^{1,25}$. 
For open-string, 
the $\sigma$ of $z=e^{\tau+i\sigma}$ originally 
runs only over $0 \leq \sigma \leq \pi$. 
The conformal field theory may be formulated on the upper half-plane 
$\mbox{Im}~z \geq 0$. But we can extend the $\sigma$ in the equations 
to run over $0 \leq \sigma \leq 2\pi$ just as for closed-string.  
\subsubsection*{Open-string Hilbert space}
Open-string Hilbert space ${\cal H}$ is the tensor 
product ${\cal H}_{matter}\otimes {\cal H}_{ghost}$, 
where ${\cal H}_{matter}$ and ${\cal H}_{ghost}$ are respectively 
the Fock spaces of the matter and ghost CFTs. It consists of 
the following vectors. 
\begin{eqnarray}
{\cal H}=
\Bigl\{
b_{-n_1}\cdots b_{-n_p}
c_{-m_1}\cdots c_{-m_q}
\alpha_{-l_1}^{\mu_1}\cdots \alpha_{-l_r}^{\mu_r}
|k \rangle \Bigr\},
\end{eqnarray}
where $|k \rangle \equiv e^{ik^{\mu}x_{\mu}}|0 \rangle$. 
We introduce the $SL_2$-invariant vacuum $|0 \rangle$ 
by the conditions,  
\begin{eqnarray}
&&
\alpha_{n}^{\mu}|0\rangle =0~~~~\mbox{for}~~ n \geq 0,~~~~
(\alpha_0^{\mu}\equiv p^{\mu}), 
\nonumber \\
&&
c_n|0 \rangle =0~~~~\mbox{for}~~  n \geq 2,~~~~
b_n|0 \rangle =0~~~~\mbox{for}~~  n \geq -1. 
\label{SL2-vacuum}
\end{eqnarray}
The open-string Hilbert space is ${\bf Z}$-graded 
by ghost number $G$.  The string coordinates have the ghost numbers, 
$G(X^{\mu})\!=\!0,G(b)\!=\!-1$ and $G(c)\!=\!1$. 
The vacuum state $\bl 0 \brr$ is set to have no ghost number 
and to be a grassmann-even vector. 
The BRST charge $Q$ acts on this Hilbert space. 
It is a grassmann-odd operator and has the ghost number equal to one.
The vacuum state becomes a BRST-invariant vector. 
The BRST charge obeys the usual relations, 
\begin{eqnarray}
Q^2=0,~~~~~\Bigl\{Q,b(z)\Bigr\}=T(z), 
\end{eqnarray}
where $T$ is the total energy-momentum tensor.

The dual Hilbert space 
${\cal H}^* \equiv {\cal H}_{mattter}^* \otimes {\cal H}_{ghost}^*$ 
consists of the following vectors. 
\begin{eqnarray}
{\cal H}^*=
\Bigl\{ 
\langle k|
\alpha_{l_1}^{\mu_1}\cdots \alpha_{l_r}^{\mu_r}
c_{m_1}\cdots c_{m_q}
b_{n_1}\cdots b_{n_p} 
\Bigr\},
\end{eqnarray} 
where $\langle k | \equiv  \langle 0|e^{ik^{\mu}x_{\mu}}$. 
We introduce the state $\langle 0|$ as the $SL_2$-invariant vacuum of 
${\cal H}^*$ by imposing the conditions,
\begin{eqnarray}
&&
\langle 0| \alpha_{n}^{\mu}=0~~~~\mbox{for}~~ n \leq 0,
\nonumber \\
&&
\langle 0|c_n =0~~~~\mbox{for}~~  n \leq -2,~~~~
\langle 0|b_n =0~~~~\mbox{for}~~  n \leq 1. 
\label{dual SL2-vacuum}
\end{eqnarray} 
These conditions also make $\bll 0 \br$ 
a BRST-invariant state. 
Dual pairing between ${\cal H}$ and ${\cal H}^*$ 
is prescribed based on 
$\langle k' |c_{-1}c_0c_1| k \rangle =\delta_{k'+k,0}$,  
where $c_{\pm 1,0}$ is the ghost zero modes on ${\bf CP}_1$.  
The pairing between any two vectors can be computed by using 
the (anti-)commutation relations (\ref{CCR}) and taking account 
of the conditions (\ref{SL2-vacuum}) and (\ref{dual SL2-vacuum}). 
For the consistency the $SL_2$-invariant vacuum $\bll 0 \br$ must 
be grassmann-odd. We set $\bll 0 \br$ to have the ghost number 
equal to $-1$.

\subsubsection*{BPZ conjugation}
The Belavin-Polyakov-Zamolodchikov (BPZ) conjugation 
is a general operation of two-dimensional conformal field theories.   
First we explain this operation 
in a generic CFT and then 
apply it to the case of open-string field theory.

Let $U_{0,\infty}$ be two-disks on ${\bf CP}_1$ 
respectively given by  
$|z| \!\leq\! 1 $ and 
$|z| \!\geq\! 1 $. 
For the convenience of later application to 
open-string field theory, 
local coordinates around 
$0$ and $\infty$ 
are chosen to be 
$z_0\! \equiv\! z$ 
and 
$z_{\infty}\!\equiv\! -1/z$. 
The coordinate transform is given by a map 
$I$ : $z_{\infty} \mapsto z_{0}\!=\!-1/z_{\infty}$. 
We have a set of local field operators 
on each coordinate patch. 
Let ${\cal O}\Bigl[(U_0, z_0)\Bigr]$ 
and 
${\cal O}\Bigl[(U_{\infty},z_{\infty})\Bigr]$ 
be sets of local field operators 
associated with the patches 
$(U_0,z_0)$ and $(U_{\infty},z_{\infty})$. 
Quantum field $\varphi$ of the theory on 
${\bf CP}_1$ is considered as a collection of  
$\varphi^{(0)}\!\in\!
{\cal O}\Bigl[(U_0, z_0)\Bigr]$
and 
$\varphi^{(\infty)}\!\in\!
{\cal O}\Bigl[(U_{\infty},z_{\infty})\Bigr]$. 
These two field operators are not independent. 
In particular, 
when $\varphi$ is primary, 
their relation becomes simple. 
It is given by  
\begin{eqnarray}   
\varphi^{(0)}(z_0)~(dz_0)^{\Delta}=
\varphi^{(\infty)}(z_{\infty})~(dz_{\infty})^{\Delta},
\label{primary varphi}
\end{eqnarray} 
where $\Delta$ is the conformal dimension of $\varphi$.

We attach the Hilbert space ${\cal H}$ 
(representation of the Virasoro algebra) to each coordinate patch 
in the operator formalism. 
Let ${\cal H}^{(0)}$ and ${\cal H}^{(\infty)}$ 
be the Hilbert spaces attached  to 
$(U_0,z_0)$ and $(U_{\infty},z_{\infty})$. 
Consider expansions of the primary field $\varphi$ 
at $0$ and $\infty$. 
They are given by  
$\varphi^{(0)}(z_{0})
\!=\!
\sum_{n}\varphi_n^{(0)}z_{0}^{-n-\Delta}$ 
and 
$\varphi^{(\infty)}(z_{\infty})
\!=\!
\sum_{n}\varphi_n^{(\infty)}z_{\infty}^{-n-\Delta}$.
The coefficients 
$\varphi_n^{(0)}$ and $\varphi_n^{(\infty)}$ 
are operators which generate the building blocks of 
${\cal H}^{(0)}$ 
and 
${\cal H}^{(\infty)}$. 
The transformation (\ref{primary varphi}) determines  
a map between 
$\varphi_n^{(0)}$ and $\varphi_n^{(\infty)}$, 
\begin{eqnarray} 
\varphi_n^{(\infty)} 
\longmapsto 
(\varphi_n^{(\infty)})^T 
\equiv 
(-)^{n+\Delta}\varphi_{-n}^{(0)}.
\label{T of modes of varphi}
\end{eqnarray} 
It can be generalized to any product of operators by letting 
$(\varphi_{n_1}^{(\infty)}\cdots \varphi_{n_l}^{(\infty)})^T$
$\!=\!(\varphi_{n_1}^{(\infty)})^T\cdots(\varphi_{n_l}^{(\infty)})^T$. 
This generalization induces a linear map $T$~:
${\cal H}^{(\infty)}$$\rightarrow$$({\cal H}^{(0)})^*$ by 
\begin{eqnarray} 
|A \rangle_{\infty}
\!=\!
O_A^{(\infty)}|0\rangle_{\infty} 
\longmapsto 
{}_0\langle A^T|
\!\equiv\!
{}_0 \langle 0|(O_A^{(\infty)})^T, 
\label{T of CFT}
\end{eqnarray} 
where $|A\rangle$ is arbitrary state of ${\cal H}$, 
and $O_A$ is an operator which gives $|A\rangle$ 
when it acts on the vacuum.   
The map (\ref{T of CFT}) is the BPZ conjugation 
in the operator formalism. 
In the particular case of open-string field theory  
the BPZ conjugation becomes as follows. 
\begin{eqnarray}
{\cal H} &\ni&
|A \rangle = 
b_{-n_1}\cdots b_{-n_p}
c_{-m_1}\cdots c_{-m_q}
\alpha_{-l_1}^{\mu_1}\cdots \alpha_{-l_r}^{\mu_r}
|k \rangle
\nonumber \\
&&~~
\longmapsto 
\langle A^T | 
=\langle k |
b_{-n_1}^T\cdots b_{-n_p}^T
c_{-m_1}^T\cdots c_{-m_q}^T
(\alpha_{-l_1}^{\mu_1})^T\cdots 
(\alpha_{-l_r}^{\mu_r})^T 
\in {\cal H}^* ,
\label{T for OSFT 1}
\end{eqnarray}
where the conjugations of the oscillator modes 
are given by   
\begin{eqnarray}
\left(\alpha_n^{\mu}\right)^T
=(-)^{n+1}\alpha_{-n}^{\mu},~~~
b_n^{~T}= (-)^{n}b_{-n},~~~
c_n^{~T}= (-)^{n-1}c_{-n}. 
\label{T for OSFT 2}
\end{eqnarray}
Since the BRST current and total energy-momentum tensor 
are primary fields of $\Delta\!=\!1$ and $2$, 
the BPZ conjugates of the BRST charge and Virasoro generators become 
\begin{eqnarray}
Q^{~T}=-Q,~~~~~~~
L_n^{~T}=(-)^nL_{-n} .
\label{T for OSFT 3}
\end{eqnarray}

The linear map $T$ gives a map at the level of local field operators. 
One-to-one correspondence between states and local field operators 
is known in two-dimensional conformal field theories. 
We have a local field operator 
$\varphi_{A}^{(\infty)}(z_{\infty})$
for arbitrary state 
$|A \rangle_{\infty}$$\in\!{\cal H}^{(\infty)}$.  
The correspondence may be seen by 
$\lim_{z_{\infty}\rightarrow 0}
\varphi_A^{(\infty)}(z_{\infty})|0 \rangle_{\infty}
=|A \rangle_{\infty}$. 
For the BPZ conjugate 
${}_{0}\langle A^T|$$\in({\cal H}^{(0)})^*$,  
we have another local field operator, 
which we call $I[\varphi_A]^{(0)}(z_0)$. 
The correspondence may be seen by 
$\lim_{z_0 \rightarrow \infty}{}_0\langle 0|
I[\varphi_A]^{(0)}(z_0)$ $={}_0\langle A^T|$. 
Thus we obtain \cite{LPP} a linear map $I$ : 
${\cal O}\Bigl[(U_{\infty},z_{\infty})\Bigr]\rightarrow$
${\cal O}\Bigl[(U_{0},z_{0})\Bigr]$. 
(We use the same name as the coordinate transform.)
We also obtain the following commutative diagram.  
\begin{eqnarray}
\begin{array}{ccc}
{\cal H}^{(\infty)} & \stackrel{T}{\longrightarrow} & 
({\cal H}^{(0)})^* \\
\downarrow &  & \downarrow \\
{\cal O}\Bigl[(U_{\infty},z_{\infty})\Bigr] & 
\stackrel{I}{\longrightarrow} &
{\cal O}\Bigl[(U_0,z_0)\Bigr]. \\
\end{array}
\end{eqnarray}

The BPZ conjugation gives a non-degenerate pairing 
between ${\cal H}^{(\infty)}$ and ${\cal H}^{(0)}$, 
which we call the BPZ pairing. 
It is defined by ${}_0\langle A^T|B \rangle_0$ 
for any two states $|A \rangle_{\infty}$ and $|B \rangle_0$.
The BPZ pairing is equal to the following 
two-point function on the $z$-plane. 
\begin{eqnarray}
\langle A^T | B \rangle =
\Bll I[\varphi_A](\infty)\varphi_B(0)\Brr.  
\end{eqnarray}
In the particular case of open-string field theory,   
existence of the ghost zero-modes $c_{\pm 1,0}$ indicates 
the following selection rule. 
\begin{eqnarray} 
\langle A^T|B \rangle \neq 0 
\Rightarrow 
G(A)+G(B)=3.
\label{selection rule of BPZ}
\end{eqnarray}

\subsubsection*{Reflector $\Bll \omega_{12}\Br$}  
Reflector $\Bll \omega_{12}\Br$ is 
a vector of $({\cal H}^{ \otimes 2})^*$ 
and is determined by the condition,  
\begin{eqnarray}
\Bll \omega_{12} \Br
\bigl. A \brr_1 
\bl B \brr_2 
=\bll A^T \br \bigl. B \brr.  
\label{def of reflector}
\end{eqnarray}
The subscripts in the LHS of the equation label 
the open-string Hilbert spaces appearing in the tensor product,
$^{i.e.}$
$\Bll \omega_{12}\Br \in$
$({\cal H}^{(1)} \otimes {\cal H}^{(2)})^*$ 
and   
$|A\rangle_1|B\rangle_2 \in $
${\cal H}^{(1)}\otimes {\cal H}^{(2)}$. 
The Hilbert spaces 
${\cal H}^{(1)}$ and ${\cal H}^{(2)}$ 
are understood to be attached respectively 
to the coordinate patches 
$(U_{\infty},z_{\infty})$ and $(U_0,z_0)$. 
By the condition (\ref{def of reflector}) 
the reflector enjoys the following properties. 
\begin{eqnarray}
&& 
\Bll \omega_{12}\Br O^{(1)}
=\Bll \omega_{12}\Br (O^{T})^{(2)}
~~~~(\mbox{$O$ is arbitrary operator} ),
\label{T and reflector}
\\
&&
\Bll \omega_{12}\Br \bigl. k \brr_1
= {}_{2}\bll  k \br. 
\label{T and ket}
\end{eqnarray}
The superscripts of $O$ in eq.(\ref{T and reflector}) 
indicate the Hilbert spaces on which the operator acts. 
If we take the BRST charge $Q$ as $O$ 
in eq.(\ref{T and reflector}),  
we obtain the BRST invariance of the reflector. 
\begin{eqnarray}
\Bll \omega_{12}\Br \Bigl(Q^{(1)}\!+\!Q^{(2)}\Bigr)=0.
\label{BRST invariance of reflector}
\end{eqnarray}
If we regard the reflector as an element 
of {\it Hom}$({\cal H}^{(1)},({\cal H}^*)^{(2)})$, 
it gives the BPZ conjugation.
\begin{eqnarray}
\Bll \omega_{12}\Br \bigl. A\brr_1=
{}_2\bll A^T \br.
\label{conjugate and reflector}
\end{eqnarray}
This is shown by using the properties 
(\ref{T and reflector}) and (\ref{T and ket}) 
as follows. 
\begin{eqnarray}
\Bll \omega_{12}\Br \bigl. A\brr_1
&=&
\Bll \omega_{12}\Br O^{(1)}_A \bl 0\brr_1
\nonumber \\
&=&
\Bll \omega_{12}\Br 
\left(O_A^T\right)^{(2)}
\bl 0\brr_1
\nonumber \\
&=&
\Bll \omega_{12} \Br 
\bigl. 0\brr_1
\left(O_A^T\right)^{(2)}
\nonumber \\
&=&
{}_2\bll 0\br \left(O_A^T\right)^{(2)}
\nonumber \\
&=&
{}_2\bll A^T \br.
\end{eqnarray}

The reflector has the following 
oscillator representation \cite{Thorn}. 
\begin{eqnarray}
\Bll \omega_{12} \Br &=&
\sum_k {}_{1}\langle k |c_{-1}^{(1)}
\otimes {}_{2}\langle -k |c_{-1}^{(2)}
\left(
c_0^{(1)}\!+\!c_0^{(2)}
\right)
\nonumber \\
&& ~~~~~~
\times \prod_{n=1}^{\infty}
\exp 
\left\{
\frac{(-)^{n+1}}{n}
\sum_{\mu}
\alpha_n^{\mu (1)}\alpha_{n \mu}^{(2)}
\!+\!(-)^{n+1}
\left(
c_n^{(1)}b_n^{(2)}\!+\!c_n^{(2)}b_n^{(1)}
\right) 
\right\},
\label{reflector by modes}
\end{eqnarray}
where the oscillators 
$\alpha_n^{\mu},b_n$ and $c_n$ act on the bras  
according to their superscripts.  
The bra $\langle k|$ is a grassmann-odd vector 
with the ghost number $-1$.   
Thus the reflector is a grassmann-odd vector 
and has the ghost number equal to one.

As usual in the description of many body system,  
we identify 
${\cal H}^{(1)} \otimes {\cal H}^{(2)}$ 
with
${\cal H}^{(2)} \otimes {\cal H}^{(1)}$ 
by imposing the relation,  
\begin{eqnarray}
|A\rangle_1|B\rangle_2 
=(-)^{G(A)G(B)}|B\rangle_2|A\rangle_1.
\label{equiv relation tensor H} 
\end{eqnarray}
This identification is used implicitly  
throughout this paper. 
We can see that the reflector is symmetric 
under the exchange of open-string indices.  
\begin{eqnarray}
&& 
\Bll \omega_{12}\Br=\Bll \omega_{21}\Br.
\label{symmetry of reflector}
\end{eqnarray}
This actually means the identity, 
$\Bll \omega_{12}\Br \bigl. A\brr_1 \bigl.B\brr_2
\!=\!\Bll \omega_{21}\Br \bigl.A\brr_1\bl B\brr_2$ 
for any two states.  By the equivalence relation 
(\ref{equiv relation tensor H})  
we can identify 
$\langle \omega_{21}|A\rangle_1|B\rangle_2$ 
with 
$(-)^{G(A)G(B)}\langle \omega_{21}|B\rangle_2|A\rangle_1$.  
This turns out to be $(-)^{G(A)G(B)}\langle B^T|A\rangle $. 
Thus eq.(\ref{symmetry of reflector}) means  
$\langle A^T|B \rangle$$=(-)^{G(A)G(B)}\langle B^T|A\rangle$.

\subsubsection*{Inverse reflector $\Bl S_{12}\Brr$}
The inverse reflector $\Bl S_{12}\Brr$ 
$\in {\cal H}^{\otimes 2}$ is introduced as the 
BPZ-conjugate of the reflector. 
It enjoys the same properties as (\ref{T and reflector}) 
and (\ref{T and ket}),  
\begin{eqnarray}
&& 
O^{(1)}\Bl S_{12}\Brr 
=(O^{T})^{(2)}\Bl S_{12} \Brr  
~~~~(\mbox{$O$ is arbitrary operator} ), 
\label{T and inverse reflector}
\\
&&
{}_1\bll k \bigr. \Bl S_{12} \Brr= \bl k \brr_2.
\label{T and bra}
\end{eqnarray} 
The first equation implies the BRST invariance of 
the inverse reflector, 
\begin{eqnarray}
\Bigl(Q^{(1)}\!+\!Q^{(2)}\Bigr)
\Bl S_{12}\Brr =0. 
\label{BRST invariance of inverse reflector}
\end{eqnarray}
If we regard the inverse reflector as an element of 
{\it Hom}$(({\cal H}^*)^{(1)},{\cal H}^{(2)})$, 
it gives 
\begin{eqnarray}
{}_1\bll A^T \bigr. \Bl S_{12} \Brr 
=
(-)^{G(A)}\bl A \brr_2.
\label{conjugate and inverse reflector}
\end{eqnarray}
This can be shown 
by using (\ref{T and inverse reflector}) 
and (\ref{T and bra}) 
similarly to the case of 
eq.(\ref{conjugate and reflector}).

Oscillator representation of the inverse reflector becomes   
\begin{eqnarray}
\Bl S_{12}\Brr &=&
\prod_{n=1}^{\infty}
\exp 
\left\{
\frac{(-)^{n+1}}{n}
\sum_{\mu}
\alpha_{-n}^{\mu (1)}\alpha_{-n \mu}^{(2)}
\!+\!
(-)^n
\left(
c_{-n}^{(1)}b_{-n}^{(2)}\!+\!c_{-n}^{(2)}b_{-n}^{(1)}
\right)
\right\} 
\nonumber \\
&&~~~~~~~~~~~~
\times 
\sum_k
\left(
c_0^{(1)}+c_0^{(2)}
\right)
c_1^{(1)}|k \rangle_1 \otimes
c_1^{(2)}|-k\rangle_2 .
\label{inverse reflector by modes}
\end{eqnarray}
If one takes the conjugation of eq.(\ref{inverse reflector by modes}), 
one finds that it becomes the representation (\ref{reflector by modes}).    
The ket  $|k\rangle$ is a grassmann-even vector 
with no ghost number. 
Hence the inverse reflector is a grassmann-odd vector and has 
the ghost number equal to three. 
We also see that it is anti-symmetric under the exchange of 
open-string indices. 
\begin{eqnarray} 
\Bl S_{12}\Brr = - \Bl S_{21} \Brr.  
\label{asymmetry of inverse reflector}
\end{eqnarray}

Let us explain why the BPZ-conjugate of the reflector 
is called {\it inverse} reflector.  
Take arbitrary state $A \in {\cal H}^{(1)}$.
We examine the state 
$\Bll \omega_{12}\Br \Bigl. S_{23}\Brr \bl A \brr_1$
$\in {\cal H}^{(3)}$.
It turns out to be 
$A \in {\cal H}^{(3)}$ by eqs.(\ref{conjugate and reflector}) 
and (\ref{conjugate and inverse reflector}) 
as follows.
\begin{eqnarray}
\Bll \omega_{12}\Br \Bigl. S_{23} \Brr 
\bl A \brr_1
&=& 
(-)^{G(A)}\Bll \omega_{12} \Br 
\bigl. A \brr_1 
\Bl S_{23}\Brr  
\nonumber \\
&=& 
(-)^{G(A)}~_{2}\bll  A^T \bigr. 
\Bl S_{23} \Brr 
\nonumber \\
&=& 
\bl A \brr_3.  
\end{eqnarray}
If one regards $\Bll \omega_{12}\Br \Bigl. S_{23} \Brr$ 
as an element of {\it Hom}$({\cal H}^{(1)},{\cal H}^{(3)})$,  
it can be expressed in a convenient form,    
\begin{eqnarray}
\Bll \omega_{12}\Br 
\Bigl. S_{23} \Brr  ={}_3 P_1, 
\label{omega-s=1}
\end{eqnarray}
where ${}_3P_1$ is the identity which maps 
$|A \rangle_1$ to $|A \rangle_3$.

\subsubsection*{Symmetric $3$-vertex $\Bll 1~2~3 \Br$}
\begin{figure}
\psfrag{f1V1}{$f_1(V_1)$}
\psfrag{f2V2}{$f_2(V_2)$}
\psfrag{f3V3}{$f_3(V_3)$}
\psfrag{f1(0)}{$f_1(0)$}
\psfrag{f2(0)}{$f_2(0)$}
\psfrag{f3(0)}{$f_3(0)$}
\begin{center}
\includegraphics[height=7cm]{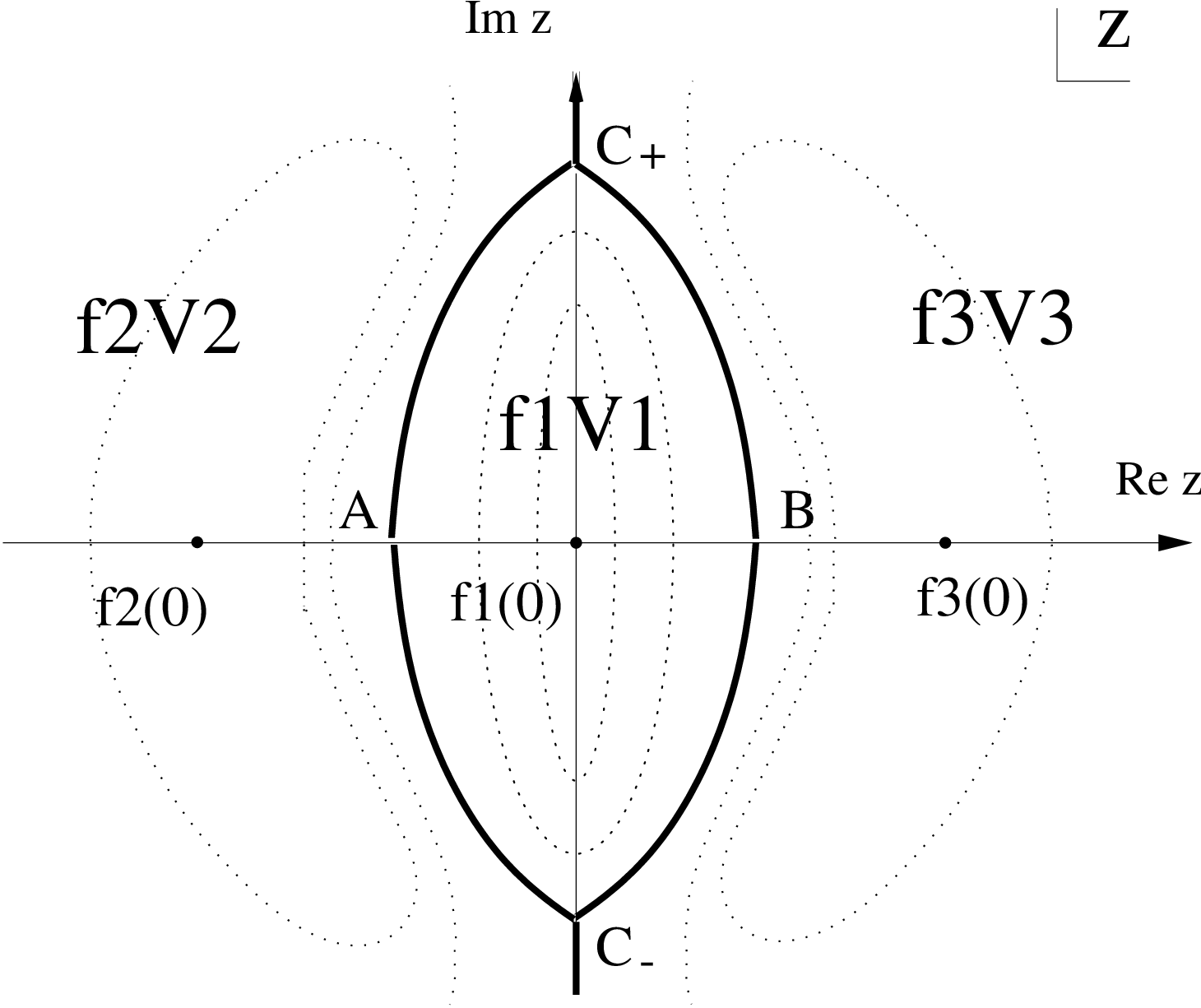}
\caption{
{\small 
Images of $f_i(V_i)$ on the $z$-plane. 
They exactly cover the $z$-plane.
Their boundaries are the bold solid lines. 
$A,B=\pm 1/\sqrt{3}$.
$C_{\pm}=\pm i$.
$f_1(0)=0,f_2(0)=-\sqrt{3},f_3(0)=\sqrt{3}$.}}
\label{fiz}
\end{center}
\end{figure}

\begin{figure}
\psfrag{f1(V1)}{$f_1(V_1)$}
\psfrag{f2(V2)}{$f_2(V_2)$}
\psfrag{f3(V3)}{$f_3(V_3)$}
\psfrag{f1(0)}{$f_1(0)$}
\psfrag{f2(0)}{$f_2(0)$}
\psfrag{f3(0)}{$f_3(0)$}
\begin{center}
\includegraphics[height=7cm]{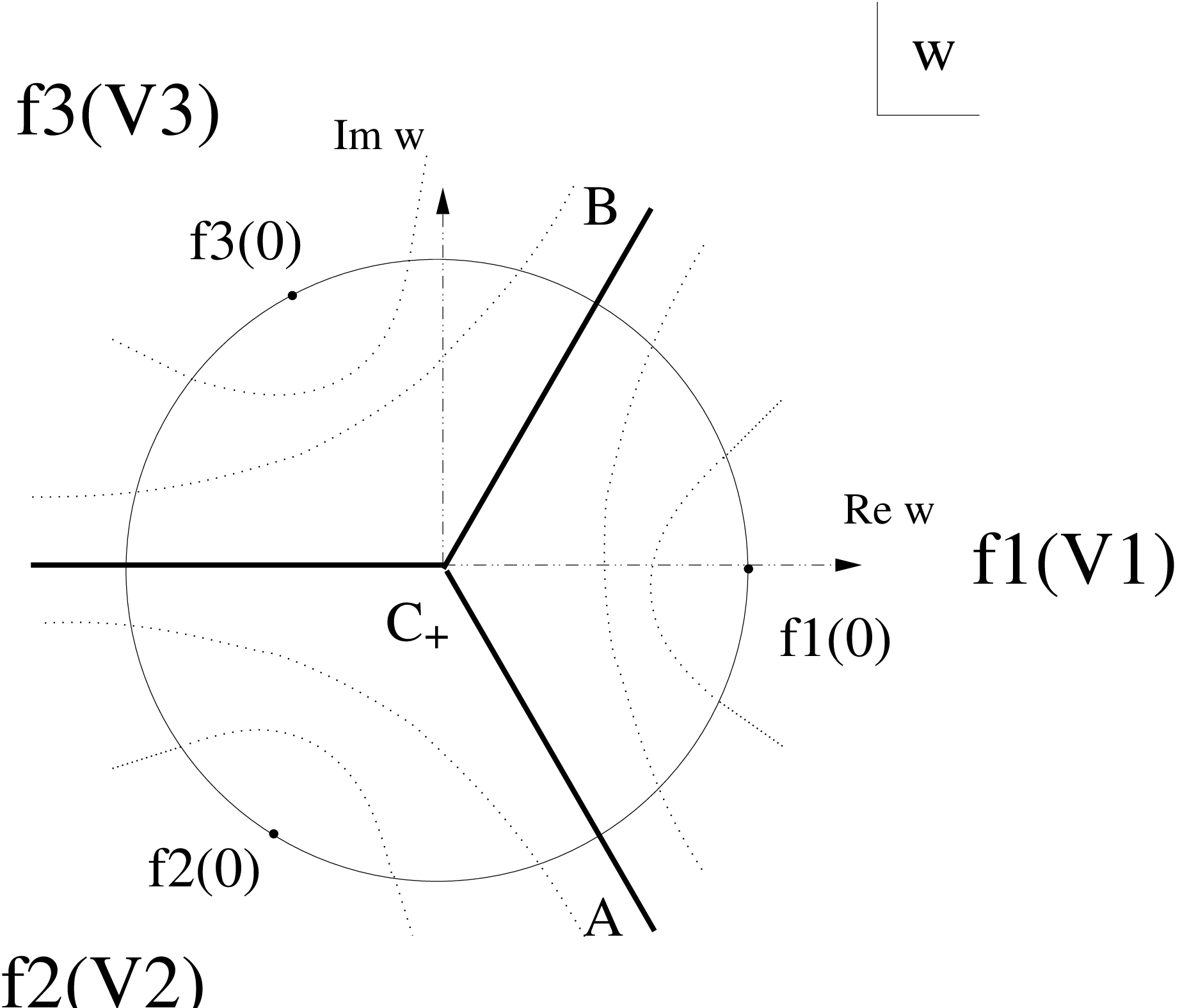}
\caption{
{\small 
Images of $f_i(V_i)$ on the $w$-plane. 
They exactly cover the $w$-plane. 
Their boundaries are the bold solid lines. 
$\textstyle{A,B=e^{\pm \frac{\pi i}{3}}}$. 
$\textstyle{f_1(0)=1,f_2(0)=e^{\frac{2\pi i}{3}}, 
f_3(0)=e^{\frac{4 \pi i}{3}}}$.}}
\label{fiu}
\end{center}
\end{figure}

Let $V_i$ for $i\!=\!1,2,3$ be unit disks 
$|v_i|\!\leq\! 1$ on the $v_i$-plane.  
Let $f_i$ for $i\!=\!1,2,3$
be holomorphic maps from $V_i $ to the $z$-plane 
of the following forms.    
\begin{eqnarray}
f_1=F,~~~
f_2=S\!\circ\!F,~~~
f_3=S\!\circ\!S\!\circ\!F, 
\label{maps of 3 vertex in z}
\end{eqnarray}
where $F$ and $S$ are meromorphic functions given by  
\begin{eqnarray}
F(v)= 
i \frac{1-\left( \frac{1+iv}{1-iv}\right)^{2/3}}
{1+\left( \frac{1+iv}{1-iv}\right)^{2/3}}, 
~~~~~
S(z)= 
\frac{\frac{1}{2}z-\frac{\sqrt{3}}{2}}
{\frac{\sqrt{3}}{2}z+\frac{1}{2}}. 
\label{F and S in z}
\end{eqnarray}
$F(v)$ has branch points at $v\!=\!i,\infty$. 
For the description we need three Riemann sheets. 
It is understood in (\ref{F and S in z}) that 
the cut of $F$ is taken so that  unit disk 
$|v|\!\leq\!1$ is on a single sheet. 
$S(z)$ is the projective action of 
$\textstyle{
\left(\begin{array}{cc}
\frac{1}{2} & -\frac{\sqrt{3}}{2} \\
\frac{\sqrt{3}}{2} & \frac{1}{2} 
\end{array}\right)}$.  
$\Bigl\{1,S,S^2\Bigr\}$ is a ${\bf Z}_3$ subgroup 
of $SL_2({\bf R})$. 
The images $f_i(V_i)$ are depicted in Figure \ref{fiz}.
%
%
In open-string field theory 
it may be convenient to use the $w$- and $u$-planes 
instead of the $z$- and $v$-planes by   
\begin{eqnarray}
z \mapsto w= \frac{1+iz}{1-iz},~~~~~
v \mapsto u= \frac{1+iv}{1-iv}.
\label{wu-planes}
\end{eqnarray}
The unit disks and upper half-planes 
are mapped respectively to 
the right half-planes and unit disks 
by these maps. 
The unit disk $V_i$ on the $v_i$-plane 
is mapped to the right half-plane on the $u_i$-plane.
If one regards $f_i$ as holomorphic maps 
from the right half-plane (on the $u_i$-plane) 
to the $w$-plane, $F$ and $S$ in (\ref{F and S in z}) 
are given by \footnote{$F$ in (\ref{F and S in w}) shows 
that the cut can be taken along the negative real line 
in the $u$-plane. 
In the $v$-plane it is the line on the imaginary axis 
starting at $i$. } 
\begin{eqnarray}
F(u) 
= u^{2/3},~~~~~~ 
S(w) 
= e^{-\frac{2 \pi i}{3}}w.  
\label{F and S in w}
\end{eqnarray}
The images $f_i(V_i)$ on the $w$-plane are depicted in 
Figure \ref{fiu}. 
%
%
%

Let ${\cal H}^{(i)}$ for $i\!=\!1,2,3$ 
be the open-string Hilbert spaces 
attached to the coordinate patches $(V_i,v_i)$.   
Open-string trivalent vertex $\Bll 1~2~3 \Br$ is a vector of 
$({\cal H}^{(1)}\otimes{\cal H}^{(2)} \otimes {\cal H}^{(3)})^*$.  
We formulate this vector along the line given in \cite{LPP}. 
Let $U$ and $V$ be respectively  unit disks  
$|z|\!\leq\!1$ and $|v|\!\leq\!1$. 
Let ${\cal O}\Bigl[(U,z)\Bigr]$ 
and ${\cal O}\Bigl[(V,v)\Bigr]$ 
be the sets of local field operators 
associated with the coordinates patches $(U,z)$ and $(V,v)$. 
Let ${\cal H}^{(U,z)}$ and ${\cal H}^{(V,v)}$ 
be the Hilbert spaces attached to these coordinates patches. 
We regard $F$ in (\ref{F and S in z}) as a holomorphic 
map from $V$ to the $z$-plane. 
This holomorphic map induces a linear map from    
${\cal H}^{(V,v)}$ to ${\cal H}^{(U,z)}$.    
Since $F$ is biholomorphic at $F(0)\!=\!0$, 
the vacuum state of ${\cal H}^{(V,v)}$ should be mapped to 
the vacuum state of ${\cal H}^{(U,z)}$.  
For a primary field $\varphi$, the relation between 
$\varphi^U(z)$ 
$\in{\cal O}\Bigl[(U,z)\Bigr]$ 
and 
$\varphi^V(v)$ 
$\in{\cal O}\Bigl[(V,v)\Bigr]$  
is given by 
\begin{eqnarray}
\varphi^U(z)~(dz)^{\Delta}=
\varphi^V(v)~(dv)^{\Delta}. 
\label{primary varphi 2}
\end{eqnarray}
Analogously to the map $T$ in the BPZ conjugation,  
this relation is used to introduce  
a linear map between the operators acting on 
${\cal H}^{(U,z)}$ and ${\cal H}^{(V,v)}$.  
We write it as $O^{(V,v)} \mapsto F[O]^{(U,z)}$.
Thus we obtain the induced map, 
which we also call $F$, 
as follows. 
\begin{eqnarray}
\begin{array}{ccc}
{\cal H}^{(V,v)} & 
\stackrel{F}
{-\!\!\!-\!\!\!-\!\!\!-\!\!\! \longrightarrow} &
{\cal H}^{(U,z)} 
\\
|A\rangle_{(V,v)}
\!=\!
O_A^{(V,v)}|0\rangle_{(V,v)} 
& \longmapsto &
|F[A]\rangle_{(U,z)}
\!\equiv \! 
F[O_A]^{(U,z)}|0\rangle_{(U,z)}. 
\end{array}
\label{def of F}
\end{eqnarray}
Using the correspondence between states and 
local field operators,  
the induced map (\ref{def of F}) is equivalently described  
as a linear map between   
${\cal O}\Bigl[(V,v)\Bigr]$ 
and 
${\cal O}\Bigl[(U,z)\Bigr]$,   
\begin{eqnarray}
\begin{array}{ccc}
{\cal O}\Bigl[(V,v)\Bigr] &
\stackrel{F}
{-\!\!\!-\!\!\!-\!\!\!-\!\!\! \longrightarrow} &
{\cal O}\Bigl[(U,z)\Bigr] \\
\varphi_A^V 
& \longmapsto &
F[\varphi_A]^U
\equiv \varphi^U_{F[A]}.
\end{array} 
\label{def of F op}
\end{eqnarray}

Let $U_i$ for $i\!=\!1,2,3$ 
be unit disks on the $z$-plane 
centered respectively at $z\!=\!S^{i-1}(0)$.  
We put $z_i\!\equiv\!z\!-\!S^{i-1}(0)$. 
Let ${\cal O}\Bigl[(U_i,z_i)\Bigr]$ be the sets of local field 
operators associated with the coordinate patches $(U_i,z_i)$. 
We regard $S^{1-i}$ as a holomorphic map from $U_i$ to $U$. 
Following the same argument as above,  
we obtain linear maps $S^{1-i}$ between 
${\cal O}\Bigl[(U_i,z_i)\Bigr]$ 
and 
${\cal O}\Bigl[(U,z)\Bigr]$, 
\begin{eqnarray}
\begin{array}{ccc}
{\cal O}\Bigl[(U_i,z_i)\Bigr] &
\stackrel{S^{1-i}}
{-\!\!\!-\!\!\!-\!\!\!-\!\!\! \longrightarrow} &
{\cal O}\left[(U,z)\right] \\
\varphi_A^{U_i} & \longmapsto &
S^{1-i}[\varphi_A]^U
\equiv \varphi^U_{S^{1-i}[A]}.
\end{array} 
\label{def of S op}
\end{eqnarray}

\begin{figure}[t]
\psfrag{Bll123Br}{$=\Bll 1~2~3 \Br$}
\begin{center}
\includegraphics[height=4cm]{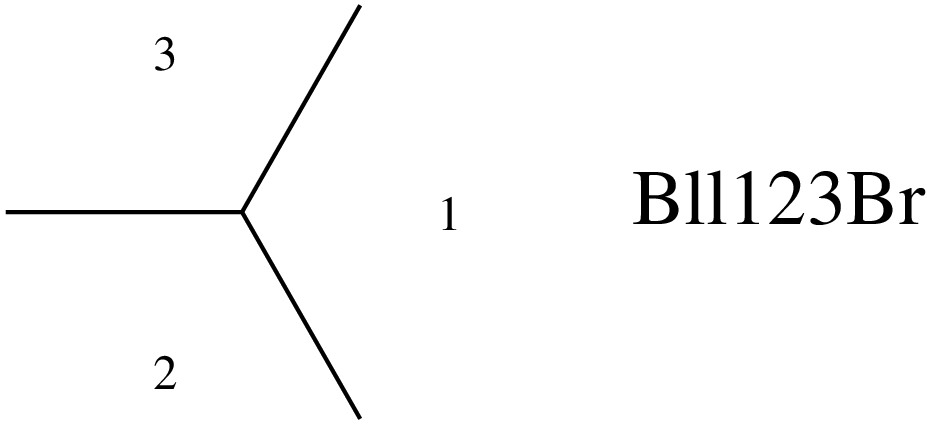}
\caption{
{\small 
Witten's trivalent vertex.
The indices $1,2$ and $3$ label three open-strings.}}
\label{Witten vertex}
\end{center}
\end{figure}

We can now introduce open-string trivalent vertex. 
For each holomorphic map $f_i$ in 
(\ref{maps of 3 vertex in z}),  
we consider $S^{1-i}F$, which is a combination of the maps 
(\ref{def of F op}) and (\ref{def of S op}). It is understood 
as a map from ${\cal O}\Bigl[(V_i,v_i)\Bigr]$ to 
${\cal O}\Bigl[(U,z)\Bigr]$ as follows. 
\begin{eqnarray} 
\begin{array}{ccccc}
{\cal O}\Bigl[(V_i,v_i)\Bigr]&
\stackrel{F}
{-\!\!\!-\!\!\!\longrightarrow}&
{\cal O}\Bigl[(U_i,z_i)\Bigr]& 
\stackrel{S^{1-i}}
{-\!\!\!-\!\!\!\longrightarrow}&
{\cal O}\Bigl[(U,z)\Bigr].
\end{array}
\label{def of f op}
\end{eqnarray}
The trivalent vertex 
$\Bll 1~2~3 \Br$ is given \cite{LPP} by 
\begin{eqnarray}
\Bll 1~2~3 \Br  
\bigl. A \brr_1 \bl B \brr_2 \bl C \brr_3 
&=& 
\Bll  
F[\varphi_A](0)~S^2F[\varphi_B](-\sqrt{3})~
SF[\varphi_C](\sqrt{3})
\Brr  ,
\label{LPP vertex}
\end{eqnarray}
where the RHS is the three point function on the $z$-plane.
One often depicts the trivalent vertex as Figure \ref{Witten vertex}. 
%
%

The most significant property of the trivalent vertex is the 
BRST invariance,  
\begin{eqnarray}
\Bll 1~2~3 \Br
\left(
Q^{(1)}\!+\!Q^{(2)}\!+\!Q^{(3)}
\right)
=0.  
\label{BRST invariance of Witten vertex}
\end{eqnarray}
Let us show eq.(\ref{BRST invariance of Witten vertex}). 
We first compute 
$\Bll  1~2~3 \Br 
\left(Q^{(1)}\bl A \brr_1\right)
\bl B\brr_2 \bl C\brr_3$ 
as follows. 
\begin{eqnarray}
\Bll 1~2~3 \Br 
\left(Q^{(1)}\bl A \brr_1\right)
\bl B\brr_2 \bl C \brr_3 
&=&
\Bll
F[\varphi_{QA}]~S^2F[\varphi_B]~SF[\varphi_C] 
\Brr 
\nonumber \\
&=&
\Bll 
F[\delta_{B\!R\!S}\varphi_{A}]~
S^2F[\varphi_B]~
SF[\varphi_C] 
\Brr 
\nonumber \\
&=&
\Bll
\delta_{B\!R\!S} 
\Bigl(F[\varphi_{A}]\Bigr)~
S^2F[\varphi_B]~
SF[\varphi_C] 
\Brr , 
\label{proof BRST inv of witten vertex 1}
\end{eqnarray}
where 
$\delta_{B\!R\!S}\varphi_A$ 
is the BRST transform of $\varphi_A$ 
and is equal to $\varphi_{QA}$. 
We also use the relation 
$F[\delta_{B\!R\!S}\varphi]\!=\!
\delta_{B\!R\!S}F[\varphi]$.  
It is ensured by the scalar property of the BRST transformation. 
If we take the Hamiltonian interpretation of the correlation function 
and use the BRST invariance of the vacua, 
we can further compute eq.(\ref{proof BRST inv of witten vertex 1}) 
as follows. 
\begin{eqnarray}
&&
\Bll 
\delta_{B\!R\!S} 
\Bigl(F[\varphi_{A}]\Bigr)~
S^2F[\varphi_B]~
SF[\varphi_C] 
\Brr 
\nonumber \\
&&=
\bll 0 \br 
\delta_{B\!R\!S} 
\Bigl(F[\varphi_{A}]\Bigr)~
S^2F[\varphi_B]~
SF[\varphi_C]
\bl 0 \brr 
\nonumber \\
&& 
=(-)^{G(A)+1}
\bll 0 \br  
F[\varphi_A]~
\delta_{B\!R\!S}
\Bigl(S^2F[\varphi_B]\Bigr)~
FS[\varphi_C]
\bl 0 \brr 
\nonumber \\
&&~~~~~~~~~~~~~~~~~~~~~~~~ 
+
(-)^{G(A)+G(B)+1}
\bll 0 \br  
F[\varphi_A]~
S^2F[\varphi_B]~
\delta_{B\!R\!S}
\Bigl(SF[\varphi_C]\Bigr)
\bl 0 \brr  
\nonumber \\
&&
=(-)^{G(A)+1}
\Bll 
F[\varphi_A]~
S^2F[\varphi_{QB}]~
SF[\varphi_C]
\Brr  
\nonumber \\
&&~~~~~~~~~~~~~~~~~~~~~~~~
+ 
(-)^{G(A)+G(B)+1}
\Bll  
F[\varphi_A]~
S^2F[\varphi_{B}]~
SF[\varphi_{QC}]
\Brr  ,
\label{proof BRST inv of witten vertex 2}
\end{eqnarray}
which turns out to be 
$(-)\Bll 1~2~3 \Br 
\left(
Q^{(2)}\!+\!Q^{(3)}
\right)
\bl A \brr_1 \bl B \brr_2 \bl C \brr_3$. 
Thus we obtain the BRST invariance 
(\ref{BRST invariance of Witten vertex}) 
of the $3$-vertex.

The oscillator representation of the $3$-vertex has the form,  
\begin{eqnarray}
&& 
\Bll 1~2~3 \Br 
\nonumber \\
&&= 
\sum_{\sum_{i}^3 k_i=0}
{}_{1}
\bll k_1 \br c_{-1}^{(1)}c_0^{(1)}
\otimes 
{}_{2}  
\bll  k_2 \br c_{-1}^{(2)}c_0^{(2)}
\otimes 
{}_{3}  
\bll  k_3 \br c_{-1}^{(3)}c_0^{(3)} 
\nonumber \\
&&~~~~~~~
\times 
\exp
\left\{
\frac{1}{2}\sum_{i,j=1}^{3}
\sum_{n,m=0}^{\infty}
\sum_{\mu}
(-)^{n+m}N_{nm}^{ij}
\alpha_{n}^{\mu (i)}\alpha_{m \mu}^{(j)}
+
\sum_{i,j=1}^{3}
\sum_{n \geq 0,m \geq 1}^{\infty}
(-)^{n+m+1}N_{c~nm}^{ij}
b_{n}^{(i)}c_{m}^{(j)}
\right\}, 
\nonumber \\
&&
\label{3 vertex by modes}
\end{eqnarray}
where $N_{nm}^{ij}$ and $N_{c~nm}^{ij}$ are the Neumann coefficients 
given by the Fourier components of the two-points functions 
of $X$ and $(b,c)$ in a suitable coordinate. 
Their explicit forms can be found in \cite{Suehiro,LPP}.  
The oscillator representation clearly shows that 
the $3$-vertex is a grassmann-odd vector and 
has the ghost number equal to three. 
In particular it is symmetric under the cyclic permutation 
of open-string indices,   
\begin{eqnarray}
\Bll 1~2~3 \Br =\Bll 2~3~1 \Br.  
\label{cyclic symm of LPP vertex}
\end{eqnarray}
This actually means 
$\Bll 1~2~3 \Br 
\bigl. A \brr_1 \br B \brr_2 \br C \brr_3$
$=\Bll 2~3~1 \Br \bigl. A \brr_1 \bl B \brr_2 \bl C \brr_3$ 
for any three states. 
The RHS of this equation is identified with    
$(-)^{G(A)(G(B)+G(C))}$
$\Bll 2~3~1 \Br \bigl. B \brr_2 \bl C \brr_3 \bl A \brr_1$. 
It is equal to 
$(-)^{G(A)(G(B)+G(C))}$
$\Bll 1~2~3 \Br \bigl. B \brr_1 \bl C \brr_2 \bl A \brr_3$.  
Thus eq.(\ref{cyclic symm of LPP vertex}) means  
\begin{eqnarray}
\Bll 1~2~3 \Br 
\Bigl\{
\bl A \brr_1 \bl B \brr_2 \bl C \brr_3
-
(-)^{G(A)(G(B)+G(C))}
\bl B \brr_1 \bl C \brr_2 \bl A \brr_3 
\Bigr\}=0
\end{eqnarray} 
for any three states.

\subsubsection*{Witten's $\Phi^3$-action of 
open-string field theory}
Open-string field $\Phi$ is a vector of the open-string 
Hilbert space ${\cal H}$. 
It is grassmann-odd and has the ghost number $G(\Phi)\!=\!1$. 
The cubic action of open-string field theory is defined by 
\begin{definition}
[Cubic action of open-string field theory]
\label{def micro action}
\begin{eqnarray}
S^{cubic}\Bigl[\Phi\Bigr]
= 
\frac{1}{2}
\Bll \omega_{12}\Br 
\bigl. \Phi \brr_1 
\left( 
Q^{(2)}\bl \Phi \brr_2 
\right)
+
\frac{1}{3}
\Bll 1~2~3 \Br 
\bigl. \Phi \brr_1 \bl \Phi \brr_2 \bl \Phi \brr_3. 
\label{microscopic action}
\end{eqnarray}
\end{definition}
Selection rules by the ghost number conservation imply 
that each term in the action does not vanish for a generic $\Phi$.

Guiding principle for the construction 
\cite{Witten NCG} of the above action 
was open-string gauge algebra $\left(Q,\star \right)$. 
Here $Q$ is the BRST charge.
$\star$ is a non-commutative associative algebra on ${\cal H}$, 
\begin{eqnarray}
|A \star B \rangle_1  \equiv 
\Bll 0~2~3 \Br \Bigl. S_{01} \Brr 
\bl A \brr_2 \bl B \brr_3, 
\label{star product}
\end{eqnarray} 
where $\star$ is understood  
as a map from ${\cal H}^{(2)}\!\times\!{\cal H}^{(3)}$ 
to ${\cal H}^{(1)}$. 
The ghost number of the RHS of eq.(\ref{star product}) 
is equal to $G(A)+G(B)$. 
Hence $\star$ preserves the ghost number, 
$G(A \star B)\!=\!G(A)+G(B)$. 
Non-commutativity, $A \star B \neq B \star A$,  
clearly has its origin in the use of the $3$-vertex. 
Associativity of the algebra,  
\begin{eqnarray}
(A \star B)\star C = A \star (B \star C), 
\label{associativity of star}
\end{eqnarray}
follows from the GGRT theorem \cite{LPP},\cite{Kugo-Takahashi}.  
It states in this particular case that 
$4$-vertex $\Bll 1~2~3~4 \Br$, which is 
introduced by  
$\Bll 1~2~3~4 \Br \equiv 
\Bll 1~2~a \Br  
\Bll a'~3~4 \Br \Bigl. S_{a'a} \Brr$,  
becomes symmetric under the cyclic rotation 
$(1,\!2,\!3,\!4)\mapsto (4,\!1,\!2,\!3)$. 
Actually one can evaluate the RHS 
of eq.(\ref{associativity of star}) as follows.  
\begin{eqnarray}
|A \star (B \star C)\rangle_5 
&=& 
\Bll  1~2~3 \Br \Bigl. S_{15} \Brr 
\bl A \brr_2
\bl B \star C \brr_3 
\nonumber \\ 
&=& 
\Bll 1~2~3 \Br \Bigl. S_{15} \Brr  
\bl A \brr_2 
\left(
\Bll 4~6~7 \Br \Bigl. S_{43} \Brr  
\bl B \brr_6 \bl C \brr_7 
\right) 
\nonumber \\ 
&=& 
\left(
\Bll 1~2~3 \Br \Bll 4~6~7 \Br \Bigl. S_{43} \Brr 
\right) 
\Bl S_{15} \Brr  
\bl A \brr_2 \bl B \brr_6 \bl C \brr_7  
\nonumber \\ 
&=& 
\Bll 1~2~6~7 \Br \Bigl. S_{15} \Brr  
\bl A \brr_2 \bl B \brr_6  \bl C \brr_7 
\nonumber \\ 
&=& 
\Bll 1~2~3~4 \Br \Bigl. S_{15} \Brr  
\bl A \brr_2 \bl B \brr_3 \bl C \brr_4. 
\nonumber 
\end{eqnarray}
A similar computation shows 
$|(A \star B)\star C \rangle_5=$
$\Bll 4~1~2~3 \Br \Bigl. S_{15} \Brr 
\bl A \brr_2 \bl B \brr_3 \bl C \brr_4$. 
Due to the cyclic symmetry of the above $4$-vertex  
these two become equal. The BRST charge is another important 
constituent of the gauge algebra. It is 
nilpotent and acts as the derivation 
on the $\star$-algebra,  
\begin{eqnarray}
Q\Bigl (A \star B \Bigr)
=QA \star B + (-)^{G(A)}A \star QB. 
\label{Q and star}
\end{eqnarray}
This can be seen from the BRST invariance of the $3$-vertex 
as follows. 
\begin{eqnarray}
Q^{(1)}|A \star B \rangle_1  
&=&  
Q^{(1)}\Bll 0~2~3 \Br \Bigl. S_{01} \Brr 
\bl A \brr_2 \bl B \brr_3 
\nonumber \\
&=& 
\Bll 0~2~3 \Br 
\left((-)Q^{(1)}\Bl S_{01} \Brr \right)
\bl A \brr_2 \bl B \brr_3
\nonumber \\
&=& 
\Bll 0~2~3 \Br 
\left( Q^{(0)}\Bl S_{01}\Brr \right)
\bl A \brr_2 \bl B \brr_3 
\nonumber \\
&=& 
\left( 
\Bll 0~2~3 \Br (-)\left(Q^{(2)}+Q^{(3)}\right)
\right)
\Bl S_{01} \Brr 
\bl A \brr_2 \bl B \brr_3 
\nonumber \\
&=& 
\Bll 0~2~3 \Br \Bigl. S_{01} \Brr 
\Bigl( Q^{(2)}\bl A \brr_2 \Bigr)
\bl B \brr_3
+(-)^{G(A)}
\Bll 0~2~3 \Br \Bigl. S_{01} \Brr 
\bl A \brr_2 
\Bigl( Q^{(3)} \br B \brr_3 \Bigr) 
\nonumber \\
&=& 
|QA \star B\rangle_1+(-)^{G(A)}|A \star QB \rangle_1. 
\end{eqnarray}

The action (\ref{microscopic action}) was obtained   
\cite{Witten NCG} originally 
in a form by which role of the open-string 
gauge algebra becomes manifest. 
Let us rewrite the cubic term in the action as follows. 
\begin{eqnarray}
\Bll 1~2~3 \Br 
\bigl. \Phi \brr_1 \bl \Phi \brr_2 \bl \Phi \brr_3
&=& 
\Bll 1~2~3 \Br 
\Bigl(
{}_1P_a \bl \Phi \brr_a
\Bigr)
\bl \Phi \brr_2 \bl \Phi \brr_3
\nonumber \\
&=& 
\Bll 1~2~3 \Br  
\Bll \omega_{ab} \Br \Bigl. S_{b1} \Brr  
\bl \Phi \brr_a \bl \Phi \brr_2 \bl \Phi \brr_3 
\nonumber \\
&=& 
\Bll \omega_{ab}\Br 
\bigl. \Phi \brr_a 
\Bll  1~2~3 \Br \Bigl. S_{1b} \Brr  
\bl \Phi \brr_2 \bl \Phi \brr_3 
\nonumber \\
&=& 
\Bll \omega_{12} \Br 
\bigl. \Phi \brr_1 
\bl \Phi \star \Phi \brr_2. 
\end{eqnarray}
By using this expression of the cubic term   
we can write down the action (\ref{microscopic action}) 
into the following form. 
\begin{eqnarray}
S^{cubic}\Bigl[\Phi \Bigr]=
\frac{1}{2}\Bll \omega_{12}\Br \bigl. \Phi \brr_1 
\left\{ 
Q^{(2)}\bl \Phi \brr_2
+ 
\frac{2}{3}
\bl \Phi \star \Phi \brr_2
\right\}. 
\label{cubic action CS}
\end{eqnarray}
Analogy with the Chern-Simons gauge theory is very suggestive.
The action enjoys the following gauge symmetry.  
\begin{eqnarray}
\delta_{\rho}\Phi=Q \rho+\Phi \star \rho-\rho \star \Phi, 
\end{eqnarray}
where $\rho$ is arbitrary vector of ${\cal H}$ 
with $G(\rho)\!=\!0$. The invariance of the action  
can be shown by using the remarks noted 
in the previous paragraph.


\section{Low Energy Theory Vertices}
Let us consider a second-quantization of open-string field theory. 
Schematically it is performed by the path-integral, 
\begin{eqnarray}
\int\! D\Phi \exp \left\{-S^{cubic}[\Phi]\right\}. 
\end{eqnarray}
The path-integral may be computed by treating  
the cubic term as a perturbation. All connected vacuum 
Feynman graphs contribute to the path-integral. 
Vacuum Feynman graph consists of the trivalent vertices 
connected by the open-string propagators 
without any external strings.  
The propagator is determined from the quadratic part 
of the action (\ref{microscopic action}). 
In the Siegel gauge $b_0\Phi\!=\!0$ it formally turns out to be  
$b_0 \frac{1}{L_0}$. 
This causes the short- and long-distance divergences 
as world-sheet theory. In this paper we hope to present  
a renormalization group analysis for the short-distance 
divergence.

\begin{figure}
\begin{center}
\includegraphics[height=5cm]{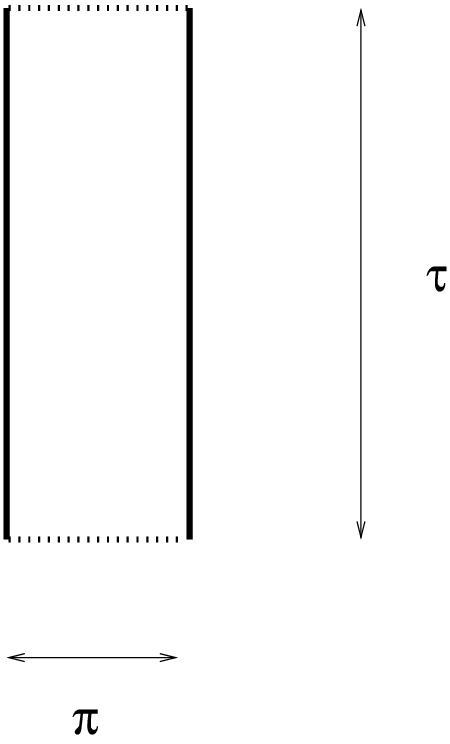}
\caption{
{\small 
Open-string strip. $\tau (>0)$ is the Schwinger imaginary time.
The width of open-string is set to $\pi$.}}
\label{strip}
\end{center}
\end{figure}

It is convenient to use the Schwinger representation 
of the propagator, 
\begin{eqnarray}
b_0\frac{1}{L_0}=\int_0^{\infty}\!d\tau  
~b_0 \exp \Bigl\{-\tau L_0\Bigr\}.
\label{propagator}
\end{eqnarray}
One can interpret $e^{-\tau L_0}$ as the evolution operator 
for open-string. It is realized by a strip of length $\tau$ 
(Figure \ref{strip}). 
%
%
Open-string diagrams are metrized trivalent ribbon graphs.   
Metric of the graph is given by 
the (imaginary time) parameters $\tau$. 
To control the short-distance divergence  
we introduce the cut-off scale parameter $\zeta$ $(>0)$
and use the regularized propagator given by 
\begin{eqnarray}
\left( b_0 \frac{1}{L_0}\right)^{reg}=
\int_{2\zeta}^{\infty}\!d\tau 
~b_0 \exp \Bigl\{-\tau L_0 \Bigr\}.
\label{reg propagator}
\end{eqnarray}

\begin{figure}[t]
\begin{center}
\includegraphics[height=7cm]{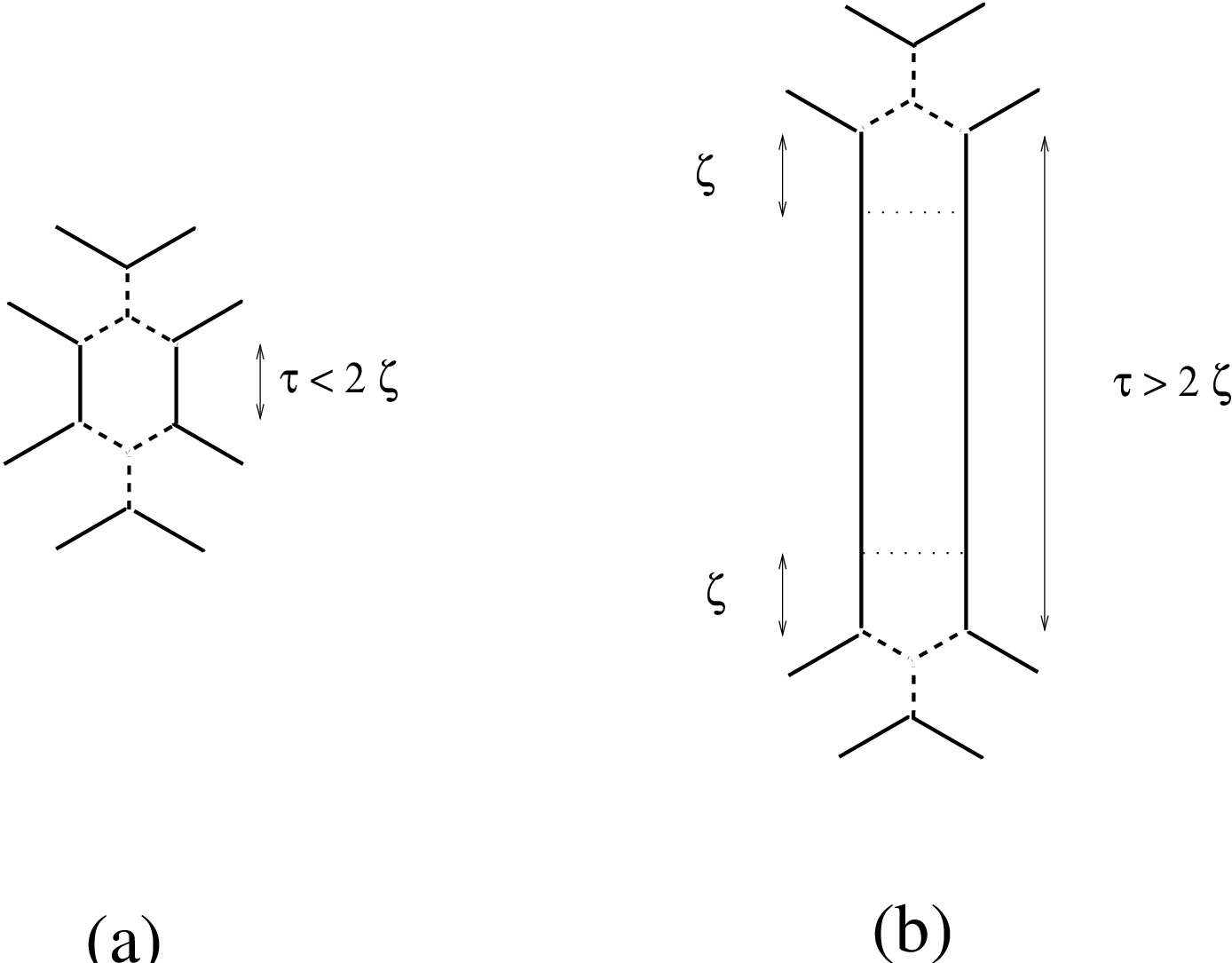}
\caption{{\small  
An open-string diagram which contains $(a)$ can not 
be reproduced from the trivalent vertices at the cut-off 
scale $\zeta$.  On the other hand, 
when all internal strips are $(b)$ the diagram can 
be reproduced.}}
\label{example}
\end{center}
\end{figure}

In presence of the cut-off scale $\zeta$, 
any open-string diagram which has at least one 
internal strip of length less than $2\zeta$ can not 
be reproduced from the trivalent vertices 
by using the regularized propagator. 
This is because $\tau$ in eq.(\ref{reg propagator}) is 
greater than $2\zeta$. 
Any internal strip, which connects the trivalent vertices, 
must have length greater than $2\zeta$. 
On the other hand, when all the internal strips 
have length more than $2\zeta$, it can be 
reproduced from the trivalent vertices in this cut-off theory  
(Figure \ref{example}). 
%
%
If we wish to take account of all the diagrams 
even in presence of the cut-off scale $\zeta$, 
we need to introduce an action which contains  
higher interactions beyond the cubic term. 
These interactions are obtained from graphs 
which are {\it one-particle irreducible} 
with respect to the regularized propagator 
(\ref{reg propagator}). They are those graphs 
all internal strips of which have length less than 
$2 \zeta$. 
If we take the perspective of renormalization group 
$ala$ Wilson \cite{Wilson-Kogut} or Polchinski \cite{Polchinski},  
the action is nothing but a low energy action 
obtained by integrating out all the contributions 
from length scale less than $\zeta$.  
We denote the low energy action at the cut-off scale $\zeta$ 
by $S_{eff}^{\zeta}[\Phi]$. 
It becomes the microscopic action 
$S^{cubic}[\Phi]$ as $\zeta$ goes to $0$, 
while, as $\zeta$ goes to $\infty$, 
it provides a macroscopic description.

Contribution from open-string loops might play an important role   
in the low energy description, 
particularly related with duality between 
open- and closed-strings \cite{Verlinde}.  
Nevertheless, at the present stage, we do not 
have a systematic machinery enough to handle with it. 
In this paper we study the classical part of 
$S_{eff}^{\zeta}[\Phi]$ which does not include the loop effect.

\subsection{Low energy theory vertices ; examples}
Since our consideration is limited to the classical case,
we call connected open-string diagrams which are tree 
simply as open-string diagrams. 
External open-strings of the diagram are clockwise-ordered.  
Open-string diagram is 
a metrized connected trivalent tree ribbon graph 
with clockwise-ordered external ribbons. 
Open-string $n$-diagram is 
a metrized connected trivalent tree ribbon graph 
with clockwise-ordered $n$ external ribbons.

Let ${\cal M}_n^{\partial}$  
be the set of clockwise-ordered $n$ different points 
$(z_1,\cdots,z_n)$ on the boundary of two-disk $D$ 
(or the upper half-plane) divided 
by the standard action\footnote{
$(z_1,\cdots,z_n)$ $\mapsto$ 
$\left(\frac{az_1+b}{cz_1+d},\cdots,\frac{az_n+b}{cz_n+d} \right)$ 
where 
$\textstyle{
\left( \begin{array}{cc}
a & b \\
c & d 
\end{array}\right) \in SL_2({\bf R})}$.} 
of $SL_2({\bf R})$. 
($n\!\geq\!3$.)
The dimension of ${\cal M}_n^{\partial}$ is $n-3$.
The set of metrized connected trivalent tree ribbon graphs 
with clockwise-ordered $n$ external ribbons is identified with 
${\cal M}^{\partial}_n$. 
Each ribbon graph has $n-3$ internal strips. 
Their length play the role of local coordinates 
of ${\cal M}_n^{\partial}$.  
Infinities of ${\cal M}_n^{\partial}$ are the configurations 
in which some of $z_i$ coincide with one another on $\partial D$. 
Stable compactification of ${\cal M}_n^{\partial}$ 
is given, under the above identification, 
by adding the trivalent tree ribbon graphs 
at least one internal strip of which has infinite length.
We denote this compactification by ${\cal CM}_n^{\partial}$.   
Topologically ${\cal CM}_n^{\partial}$ becomes a $(n\!-\!3)$-dimensional 
ball $B_{n-3}$. We fix orientation of ${\cal CM}_n^{\partial}$ 
by the standard orientation of $B_{n-3}$.

If we have an open-string $n$-diagram,  
we can obtain another one by permuting the labels of 
external open-strings from $(1,2,\cdots,n\!-\!1,n)$ 
to $(2,3,\cdots,n,1)$.
In general, it is different from the original diagram. 
This permutation of open-string indices 
gives an automorphism of ${\cal CM}_n^{\partial}$. 
We denote this automorphism by $s$. 
Clearly $s$ generates ${\bf Z}_n$ (or its subgroup) 
action on ${\cal CM}_n^{\partial}$. 
As we will see in the subsequent discussions,   
the automorphism $s$ is related with asymmetry of open-string vertex 
under the cyclic-permutation of open-string indices.

\subsubsection{Open-string $4$-vertex~ 
$\Bll 1~2~3~4\!:\!\zeta \Br$}
\begin{figure}
\begin{center}
\psfrag{CM4}{${\cal CM}_4^{\partial}$}
\psfrag{+infty}{$+\infty$}
\psfrag{-infty}{$-\infty$}
\psfrag{xgeq0}{$x \geq 0$}
\psfrag{x<0}{$x < 0$}
\psfrag{x}{$x$}
\psfrag{axa}{$|x|$}
\includegraphics[height=7cm]{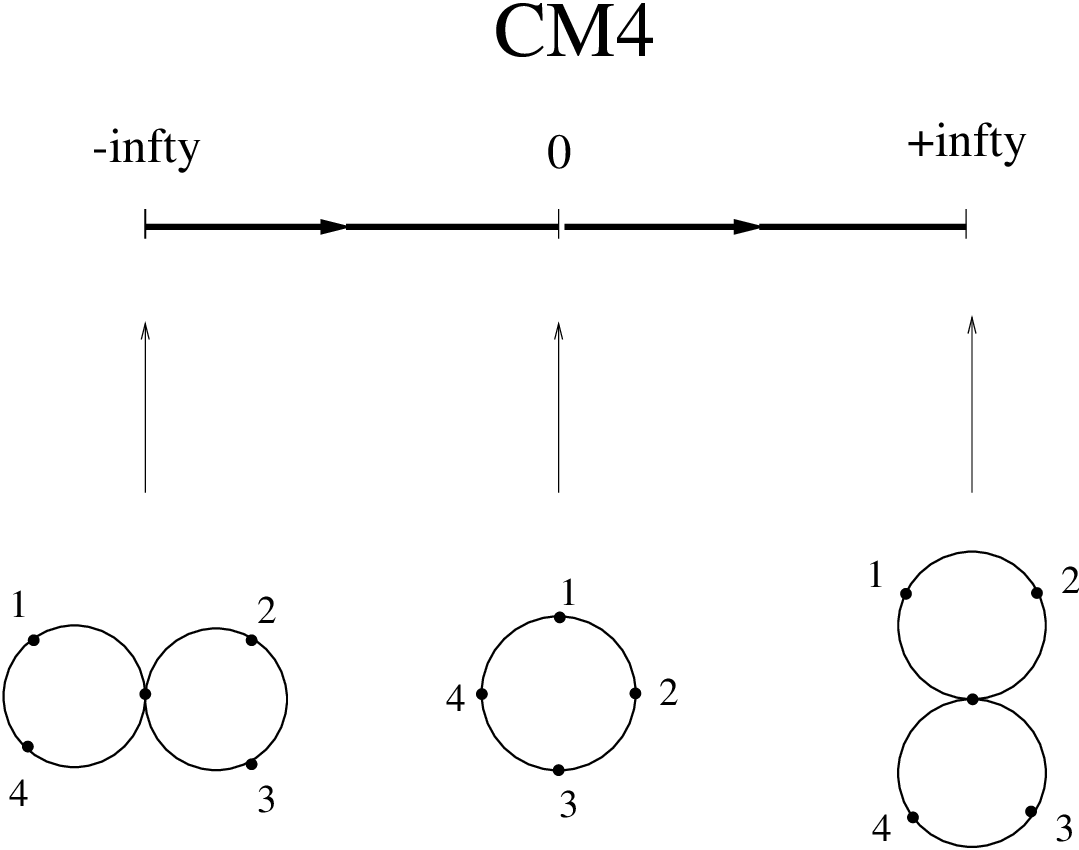} 
\includegraphics[height=7cm]{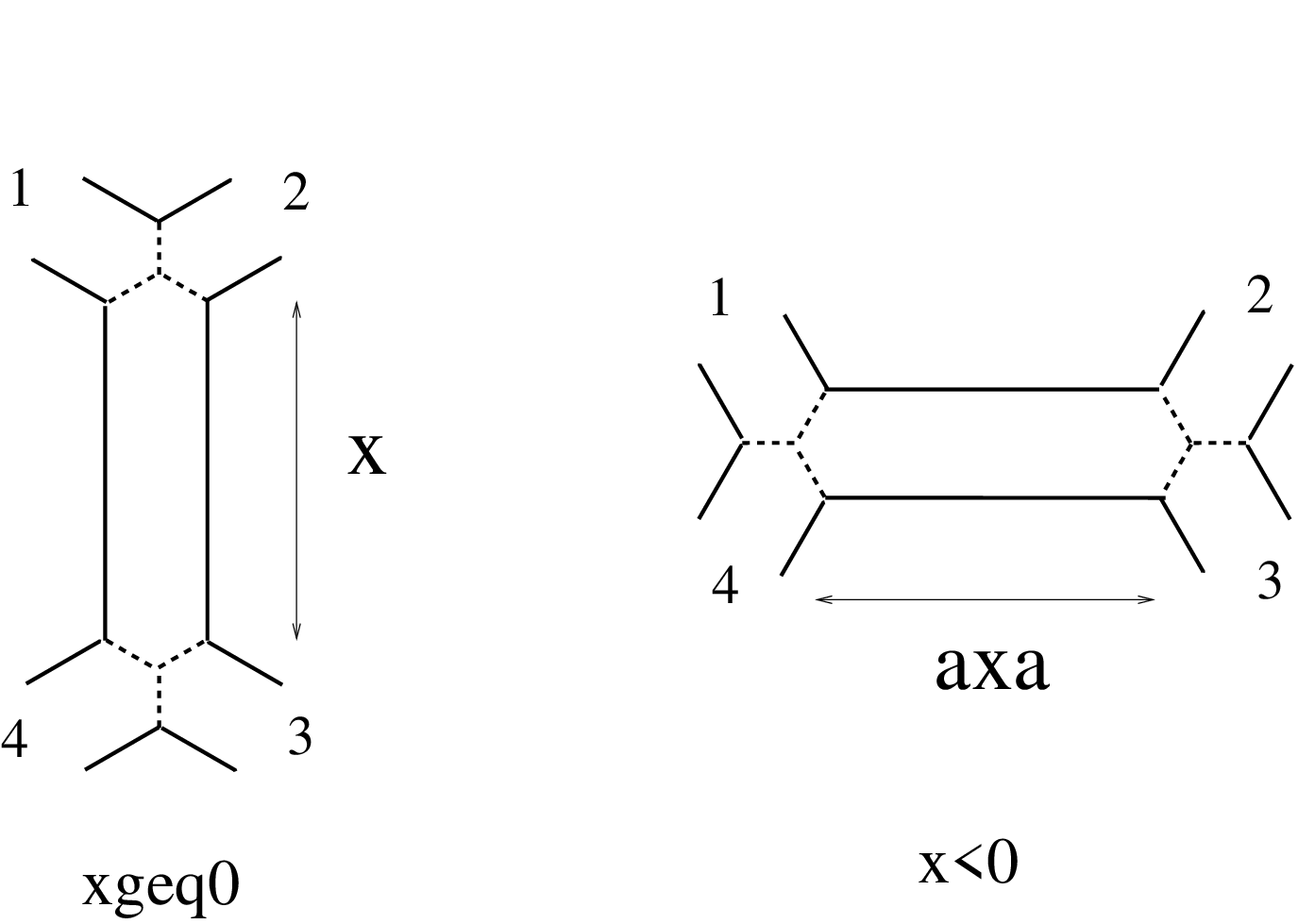} 
\caption{{\small 
${\cal CM}^{\partial}_4=[-\infty,+\infty]$. 
The configurations of four points on $\partial D$ 
at $x\!=\!0,\pm \infty$ are depicted in the second line.
Open-string diagram at $x \in [-\infty,+\infty]$ 
is on the third line. The $s,t$-channels are respectively 
$x>0$ and $x<0$.}}
\label{figCM4}
\end{center}
\end{figure}

The compactification ${\cal CM}_4^{\partial}$ is $[-\infty,\infty]$. 
We identify the $s$-channel of four open-string interaction 
with $[0,\infty]$ and the $t$-channel with $[-\infty,0]$.  
(Figure \ref{figCM4}). 
%
%
Length of the internal strip is $x$ in the $s$-channel 
while it is $|x|$ in the $t$-channel.

We can construct a state $\langle x |$ $\in ({\cal H}^{\otimes 4})^*$ 
for each $x \in {\cal CM}_4^{\partial}$ 
by applying the Feymman rule to the corresponding diagram. 
Explicitly we define $\langle x |^{(1234)}$ as follows.  
\begin{eqnarray}
\langle x |^{(1234)}=
\left\{
\begin{array}{cr}
\Bll 1~2~a \Br \Bll a'~3~4 \Br 
\left(  b_0e^{-xL_0}\Bl S_{a'a}\Brr \right)
& \mbox{for}~ x>0 ,
\\
0
& \mbox{for}~ x=0 ,
\\
\Bll 2~3~a \Br \Bll a'~4~1 \Br  
\left( (- b_0)e^{xL_0}\Bl S_{a'a}\Brr \right)
& \mbox{for}~ x<0.
\end{array}
\right.
\label{state x}
\end{eqnarray}
We need not specify insertions of $b_0$ and $L_0$ 
on the inverse reflector $\Bl S_{a'a}\Brr$ 
because either way gives the same by the BPZ conjugation. 
The reason why we attach the superscript $(1234)$ 
to the state $\langle x |$ will become clear soon. 
The state $\langle x |^{(1234)}$ is grassmann-even  
and has the ghost number four. 
We have a  
$({\cal H}^{\otimes 4})^*$-valued one-form 
$\langle \Omega |^{(1234)}$ 
on ${\cal CM}^{\partial}_4$. 
It is defined by 
\begin{eqnarray}
\Omega(x)^{(1234)} = dx \langle x |^{(1234)}.
\label{1-form omega} 
\end{eqnarray}

Let us consider an effect of the cyclic permutation 
of open-string indices. 
Another state can be obtained 
if we permute the open-string indices  
from $(1,2,3,4)$ to $(2,3,4,1)$ 
in the RHS of eq.(\ref{state x}).
We denote this new state by $\langle x |^{(2341)}$. 
We also let  
$\langle \Omega |^{(2341)}$
be the one-form $dx \langle x |^{(2341)}$.  
Two states $\langle x |^{(1234)}$ and $\langle x |^{(2341)}$ 
are related with each other by 
$\langle -x|^{(1234)} = -\langle x |^{(2341)}$. 
This follows from 
the anti-symmetry of the inverse reflector and 
the cyclic symmetry of the trivalent vertex 
besides its odd-grassmannity.
The permutation exchanges the $s$- and $t$-channels.  
Thus the automorphism $s$ generates a ${\bf Z}_2$-action 
on ${\cal CM}_4^{\partial}$. 
It is given by $s(x)\!=\!-x$. 
Therefore $\langle \Omega |^{(2341)}$ 
is the pull-back of $\langle \Omega |^{(1234)}$ by $s$.
\begin{eqnarray}
\Omega^{(2341)}=s^*\Omega^{(1234)}. 
\label{symmetry of 1-form omega}
\end{eqnarray}

\begin{figure}[t]
\begin{center}
\includegraphics[height=5cm]{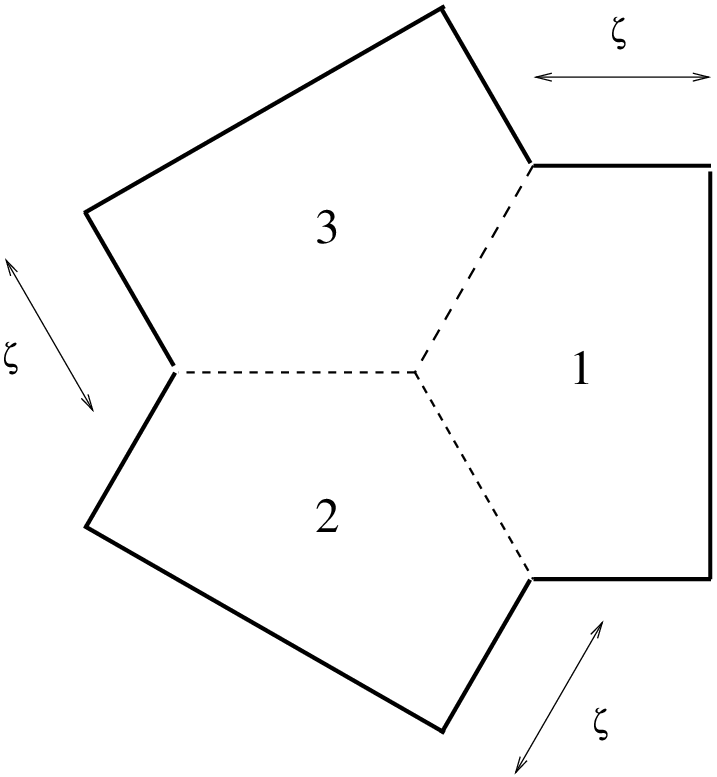}
\caption{
{\small Open-string trivalent vertex at the cut-off scale 
$\zeta$.}}
\label{3vertex at zeta}
\end{center}
\end{figure}

We now examine the low energy description at the cut-off 
scale $\zeta$. 
We put ${\cal V}_4(\zeta)\!\equiv\!(-2\zeta,2\zeta)$. 
Open-string diagrams which belong to ${\cal V}_4(\zeta)$
can not be reproduced from the trivalent vertices 
at this length scale. 
Any state $\langle x |$ for 
$x \in {\cal V}_4(\zeta)$ 
must appear in 
$S^{\zeta}_{eff}[\Phi]$ 
as a part of an interaction vertex.  
It will be a form such as  
$\frac{1}{4}\int_{-2\zeta}^{2\zeta}
\langle \Omega |^{(1234)}
|\Phi \rangle_1 |\Phi \rangle_2 |\Phi \rangle_3 |\Phi \rangle_4$. 
To specify the correct form,  
we first consider the states $\langle \pm 2\zeta |$. 
They are expected to be obtained simply 
by joining two $3$-vertices at one open-string,  
since $x\!=\!\pm 2\zeta$ are the boundaries of ${\cal V}_4(\zeta)$. 
For this to be possible we need to modify 
slightly the trivalent vertex. 
\begin{definition}[Open-string $3$-vertex at $\zeta$]
\label{def of modified 3-vertex}
Open-string $3$-vertex at the cut-off scale $\zeta$ is 
defined by 
\begin{eqnarray}
\Bll 1~2~3\!:\!\zeta \Br  
= 
\Bll 1~2~3 \Br 
\prod_{i=1,2,3}e^{-\zeta L_0^{(i)}}.
\label{3 open-string vertex}
\end{eqnarray}
\end{definition}
The modified $3$-vertex 
(Figure \ref{3vertex at zeta}) 
%
%
is a grassmann-odd vector which satisfies 
the BRST invariance and the cyclic symmetry 
in the same manner as the original one. 
In (\ref{3 open-string vertex})
we attach the propagator of length $\zeta$ 
to each (not a specific) external open-string  
in order to keep the cyclic symmetry of $3$-vertex. 
States obtained by joining two modified $3$-vertices 
are not $\langle \pm 2\zeta|$ but 
$\langle \pm 2\zeta |\prod_{i=1}^4e^{-\zeta L_0^{(i)}}$. 
Appearance of the propagators of length $\zeta$ is a common 
phenomenon in the low energy description. 
Let ${\cal V}_n(\zeta)$ be the set of 
open-string $n$-diagrams all internal strips 
of which have length less than $2\zeta$. 
They are by no means obtained from lower diagrams 
at the cut-off scale $\zeta$.
Open-string diagram at the boundary of ${\cal V}_n(\zeta)$ 
has at least one internal strip of length equal to $2\zeta$.
Let us take any two open-string diagrams 
which belong respectively to  
${\cal V}_{n_1+1}(\zeta)$ and ${\cal V}_{n_2+1}(\zeta)$,     
and join them at one external string. 
As the result we obtain a $(n_1\!+\!n_2)$-diagram. 
It must be at the boundary of ${\cal V}_{n_1+n_2}(\zeta)$. 
For this to be possible, each external strip 
needs to have length $\zeta$.  
This shows that we must insert $e^{-\zeta L_0}$  
to each external open-string 
in order to obtain the vertices at the cut-off scale $\zeta$. 
\begin{definition}[Open-string $4$-vertex at $\zeta$]
\label{def of 4-vertex}
Open-string $4$-vertex at the cut-off scale $\zeta$ is 
defined by an integration of the $({\cal H}^{\otimes 4})^*$-valued 
one-form $\langle \Omega |$ (\ref{1-form omega}) over 
${\cal V}_4(\zeta)$ multiplied by the external propagators
of length $\zeta$, 
\begin{eqnarray}
\Bll 1~2~3~4\!:\!\zeta \Br  
= 
\int_{-2\zeta}^{2\zeta}
\langle \Omega |^{(1234)}
\prod_{i=1}^{4}e^{-\zeta L_0^{(i)}}. 
\label{4 open-string vertex}
\end{eqnarray}
Indices in the $4$-vertex are understood to label 
the Hilbert spaces attached to clockwise-ordered 
four open-strings on $\partial D$. 
Thereby the vertex is regarded as a vector of 
$({\cal H}^{(1)}\otimes {\cal H}^{(2)}\otimes 
{\cal H}^{(3)}\otimes {\cal H}^{(4)})^*$. 
\end{definition}
Open-string $4$-vertex (\ref{4 open-string vertex}) 
is grassmann-even and has the ghost number equal to four. 
When we use the one-form $\Omega^{(2341)}$ 
instead of $\Omega^{(1234)}$ in the above definition,  
we obtain another $4$-vertex 
$\Bll 2~3~4~1\!:\!\zeta \Br$.  
But, owing to the relation (\ref{symmetry of 1-form omega}), 
these two are shown to be identical. 
\begin{eqnarray}
\Bll 2~3~4~1\!:\!\zeta \Br
&=& 
\int_{-2\zeta}^{2\zeta}
\langle \Omega |^{(2341)}
\prod_{i=1}^{4}e^{-\zeta L_0^{(i)}}
\nonumber \\
&=&
\int_{-2\zeta}^{2\zeta}
s^* \langle \Omega |^{(1234)}
\prod_{i=1}^{4}e^{-\zeta L_0^{(i)}}
\nonumber \\
&=&
-\int_{-2\zeta}^{2\zeta}
\langle \Omega |^{(1234)}
\prod_{i=1}^{4}e^{-\zeta L_0^{(i)}}
\nonumber \\
&=&
-\Bll 1~2~3~4\!:\!\zeta \Br .  
\label{asymmetry of 4 vertex}
\end{eqnarray}
We say this as asymmetry of the open-string $4$-vertex 
under the cyclic permutation. 
The factor $(-)$ in (\ref{asymmetry of 4 vertex}) 
originates in the fact that the automorphism $s$ 
does not preserve the orientation 
of ${\cal V}_4(\zeta)$.

The $4$-vertex is not invariant under the action of 
the BRST charge. Instead we have : 
\begin{proposition}
[$Q$-action on $4$-vertex]
\label{prop Q-action on 4-vertex}
Action of the BRST charge on the $4$-vertex 
(\ref{4 open-string vertex}) becomes as follows. 
\begin{eqnarray}
\Bll 1~2~3~4\!:\!\zeta \Br 
\Bigl(\sum_{i=1}^4Q^{(i)}\Bigr)
=
\Bll 1~2~a\!:\!\zeta \Br
\Bll a'~3~4\!:\!\zeta \Br 
\Bigl. S_{a'a}\Brr  
-
\Bll 2~3~a\!:\!\zeta \Br 
\Bll a'~4~1\!:\!\zeta \Br
\bigl. S_{a'a}\Brr . 
\label{Q action on 4-vertex}
\end{eqnarray}
\end{proposition}
Eq.(\ref{Q action on 4-vertex}) has the following geometrical 
interpretation. 
Orientation of ${\cal V}_4(\zeta)$ is given, 
as used in eq.(\ref{4 open-string vertex}), 
by the orientation of 
${\cal CM}_4^{\partial}\!=\![-\infty,\infty]$. 
Therefore we have 
\begin{eqnarray}
\partial {\cal V}_4(\zeta) 
&=& 
\Bigl\{2\zeta \Bigr\}-
\Bigl\{ -2\zeta \Bigr\},
\label{boundary of V4} 
\\
\partial 
\Bigl\{ 
\pm 2\zeta 
\Bigr\}
&=& 0, 
\label{boundary of V3}
\end{eqnarray}
where $\partial$ is the boundary operator.
Similarity between eqs.(\ref{Q action on 4-vertex}) 
and (\ref{boundary of V4}) including their signatures 
should be emphasized. 
Each term in the RHS of eq.(\ref{Q action on 4-vertex}) 
is BRST-closed 
as follows from the BRST invariances 
of the $3$-vertex and the inverse reflector.  
This corresponds to eq.(\ref{boundary of V3}). 
The action of the BRST charge on the $4$-vertex 
becomes a representation of the boundary operator $\partial$.

The $4$-vertex depends on the cut-off scale parameter $\zeta$. 
It vanishes at $\zeta\!=\!0$. 
The scale dependence 
comes from ${\cal V}_4(\zeta)$ and 
$\prod e^{-\zeta L_0^{(i)}}$ 
in eq.(\ref{4 open-string vertex}).  
\begin{proposition}
[Scale dependence of $4$-vertex]
\label{scale dependence of 4-vertex}
Scale dependence of the $4$-vertex 
(\ref{4 open-string vertex}) is described by 
\begin{eqnarray}
\frac{1}{2}
\frac{d}{d\zeta}\Bll 1~2~3~4\!:\!\zeta \Br 
&=&
\Bll 1~2~a\!:\!\zeta \Br 
\Bll a'~3~4\!:\!\zeta \Br 
\left( b_0 \Bl S_{a'a}\Brr \right)
-
\Bll 2~3~a\!:\!\zeta \Br
\Bll a'~4~1\!:\!\zeta \Br 
\left( b_0 \Bl S_{a'a}\Brr \right)
\nonumber \\
&&-
\frac{1}{2}
\Bll 1~2~3~4\!:\!\zeta \Br
\Bigl(\sum_{i=1}^4L_0^{(i)}\Bigr).
\label{cut-off dependence of 4 vertex}
\end{eqnarray}
\end{proposition}

~

Now we prove Propositions 
\ref{prop Q-action on 4-vertex} and 
\ref{scale dependence of 4-vertex}. 
The following lemma becomes useful in the proof of 
Proposition \ref{prop Q-action on 4-vertex}. 
\begin{lemma}
\label{lemma for Qx}
The BRST charge acts on the state $\langle x|^{(1234)}$ as  
\begin{eqnarray}
\langle x |^{(1234)}
\Bigl(
\sum_{i=1}^4 Q^{(i)}
\Bigr)
=
\left\{
\begin{array}{lc}
\partial_x 
\left[
\Bll 1~2~a \Br \Bll a'~3~4 \Br
\left(e^{-xL_0}\Bl S_{a'a}\Brr \right)
\right] 
& 
~~~\mbox{for}~~x>0,
\\
~
&
~
\\
\partial_x 
\left[
\Bll 2~3~a \Br \Bll a'~4~1 \Br
\left(e^{xL_0}\Bl S_{a'a}\Brr \right)
\right] 
& 
~~~\mbox{for}~~x<0.
\label{Q on x}
\end{array}
\right. 
\end{eqnarray}
\end{lemma} 

~

\noindent 
\underline{{\it Proof of Lemma \ref{lemma for Qx}}}~: 
We show eq.(\ref{Q on x}) for the case of $x >0$. 
We rewrite the LHS of eq.(\ref{Q on x}) as follows.  
\begin{eqnarray}
&& \langle x |^{(1234)}
\Bigl(\sum_iQ^{(i)}\Bigr) 
\nonumber \\
&&=
\Bigl\{
\Bll 1~2~a \Br \Bll a'~3~4 \Br 
\Bigl(\sum_iQ^{(i)}\Bigr)
\Bigr\}
\left(b_0e^{-xL_0}\Bl S_{a'a}\Brr \right)
\nonumber \\
&&=
\left\{
\left(
  - \Bll 1~2~a \Br 
         \left(Q^{(1)}\!+\!Q^{(2)}\right)
\right)
  \Bll a'~3~4 \Br 
+
  \Bll 1~2~a \Br
  \left(
     \Bll a'~3~4 \Br 
          \left(Q^{(3)}\!+\!Q^{(4)}\right)
  \right)
\right\}
\left( b_0e^{-xL_0}\Bl S_{a'a}\Brr \right). 
\nonumber \\
\label{proof lemma Qx 1}
\end{eqnarray}
Eq.(\ref{proof lemma Qx 1}) can be further computed by using 
the BRST invariance of the $3$-vertices.
\begin{eqnarray}
Eq.(\ref{proof lemma Qx 1})
&=&
\Bigl\{
   \left(
      \Bll 1~2~a \Br Q^{(a)}
   \right)
   \Bll a'~3~4 \Br 
-
   \Bll 1~2~a \Br 
   \left(
      \Bll a'~3~4 \Br Q^{(a')}
   \right)
\Bigr\}
\left( b_0e^{-xL_0}\Bl S_{a'a}\Brr \right)
\nonumber \\
&=&
- \Bll 1~2~a \Br \Bll a'~3~4 \Br
\Bigl\{
      \left(Q^{(a)}\!+\!Q^{(a')} \right)
      \left( b_0e^{-xL_0}\Bl S_{a'a}\Brr \right)
\Bigr\}.
\label{proof lemma Qx 2}
\end{eqnarray}
Then, by using the BRST invariance of the inverse reflector 
and the anti-commutation relation, $\{ Q,b_0 \}\!=\!L_0$, 
eq.(\ref{proof lemma Qx 2}) becomes as follows. 
\begin{eqnarray}
Eq.(\ref{proof lemma Qx 2})
&=& 
- \Bll 1~2~a \Br \Bll a'~3~4 \Br
\left( 
      L_0e^{-xL_0}\Bl S_{a'a}\Brr 
\right)
\nonumber \\
&=&
\partial_x 
\left[
   \Bll 1~2~a \Br \Bll a'~3~4 \Br
   \left(e^{-xL_0}\Bl S_{a'a}\Brr \right)
\right]. 
\end{eqnarray}
This is the RHS of eq.(\ref{Q on x}) for the case of 
$x\!\geq\!0$.\P 

~

\noindent 
\underline{{\it Proof of Proposition \ref{prop Q-action on 4-vertex}}}~:  
The proof is done by a bunch of simple calculations. 
We compute the LHS of eq.(\ref{Q action on 4-vertex}) 
by using Lemma \ref{lemma for Qx}. 
\begin{eqnarray}
&& 
\Bll 1~2~3~4\!:\!\zeta \Br
\Bigl(
\sum_iQ^{(i)}
\Bigr)
\nonumber \\
&&=
\int_0^{2\zeta} 
\langle \Omega|^{(1234)}
\Bigl(\sum_jQ^{(j)}\Bigr)
\prod_j e^{-\zeta L_0^{(i)}} 
+
\int^0_{-2\zeta} 
\langle \Omega|^{(1234)}
\Bigl(\sum_jQ^{(j)}\Bigr)
\prod_j e^{-\zeta L_0^{(i)}} 
\nonumber \\
&&=
\left\{
\Bll 1~2~a \Br \Bll a'~3~4 \Br
\left(e^{-2\zeta L_0}\Bl S_{a'a}\Brr \right)
-
\Bll 1~2~a \Br \Bll a'~3~4 \Br 
\Bigl. S_{a'a}\Brr 
\right\}
\prod_ie^{-\zeta L_0^{(i)}}
\nonumber \\
&&~~~~~~~
+
\left\{
\Bll 2~3~a \Br \Bll a'~4~1 \Br 
\Bigl. S_{a'a}\Brr  
-
\Bll 2~3~a \Br \Bll a'~4~1 \Br
\left(e^{-2\zeta L_0}\Bl S_{a'a}\Brr \right) 
\right\}
\prod_ie^{-\zeta L_0^{(i)}}
\label{proof Q-action on 4-vertex 1}
\end{eqnarray}
Then we arrange eq.(\ref{proof Q-action on 4-vertex 1}) 
as follows. 
\begin{eqnarray}
Eq.(\ref{proof Q-action on 4-vertex 1})
&=& 
\Bll 1~2~a\!:\!\zeta \Br \Bll a'~3~4\!:\!\zeta \Br 
\Bigl. S_{a'a}\Brr  
-
\Bll 2~3~a\!:\!\zeta \Br \Bll a'~4~1\!:\!\zeta \Br 
\Bigl. S_{a'a}\Brr 
\nonumber \\
&&~~~
-
\Bigl(
\Bll 1~2~a \Br \Bll a'~3~4 \Br 
\Bigl. S_{a'a}\Brr -
\Bll 2~3~a \Br \Bll a'~4~1 \Br 
\Bigl. S_{a'a}\Brr 
\Bigr)
\prod_ie^{-\zeta L_0^{(i)}}  
\nonumber \\
&=& 
\Bll 1~2~a\!:\!\zeta \Br \Bll a'~3~4\!:\!\zeta \Br 
\Bigl. S_{a'a}\Brr 
-
\Bll 2~3~a\!:\!\zeta \Br \Bll a'~4~1\!:\!\zeta \Br 
\Bigl. S_{a'a}\Brr  
\nonumber \\
&&~~~
-
\Bigl(
\Bll 1~2~~3~4 \Br -\Bll 2~3~4~1 \Br 
\Bigr)
\prod_ie^{-\zeta L_0^{(i)}}. 
\end{eqnarray}
The last term vanishes identically because of the cyclic 
symmetry of $\Bll 1~2~3~4 \Br$ \cite{LPP}. (We remind the reader that 
$\Bll 1~2~3~4 \Br$ and our $4$-vertex $\Bll 1~2~3~4\!:\!\zeta \Br$ 
are different.) 
Therefore we obtain the RHS of eq.(\ref{Q action on 4-vertex}).\P

~

\noindent 
\underline{{\it Proof of Proposition 
\ref{scale dependence of 4-vertex}}}~: 
The first-order variation of the $4$-vertex 
with respect to the cut-off scale $\zeta$ 
can be computed as follows : 
\begin{eqnarray}
\delta \Bll 1~2~3~4\!:\!\zeta \Br  
&=& 
\delta 
\left[
   \int_{{\cal V}_4(\zeta)}
         \langle \Omega |^{(1234)}
   \prod_{i=1}^{4}e^{-\zeta L_0^{(i)}}
\right] 
\nonumber \\
&=& 
\delta 
\left[
   \int_{{\cal V}_4(\zeta)}
         \langle \Omega |^{(1234)}
\right]
   \prod_{i=1}^{4}e^{-\zeta L_0^{(i)}}
+\int_{{\cal V}_4(\zeta)}
         \langle \Omega |^{(1234)}
\delta 
  \left[
   \prod_{i=1}^{4}e^{-\zeta L_0^{(i)}}
  \right] 
\nonumber \\
&=& 
2 \delta \zeta 
\left\{
\Bigl( 
\langle 2\zeta |^{(1234)}+
\langle -2\zeta |^{(1234)} 
\Bigr)
\prod_{i=1}^{4}e^{-\zeta L_0^{(i)}}
-\frac{1}{2}
\Bll 1~2~3~4\!:\!\zeta \Br 
\Bigl(\sum_{i=1}^4L_0^{(i)}\Bigr)
\right\}. 
\nonumber \\
\label{proof scale dependence 4-vertex 1}
\end{eqnarray}
It is easy to see that eq.(\ref{proof scale dependence 4-vertex 1}) 
precisely gives the RHS of 
eq.(\ref{cut-off dependence of 4 vertex}).\P 


\subsubsection
{Open-string $5$-vertex~ $\Bll 1~2~3~4~5\!:\!\zeta \Br$}
\begin{figure}
\psfrag{calUi}{${\cal U}_i$}
\psfrag{calU1}{${\cal U}_1$}
\psfrag{calU2}{${\cal U}_2$}
\psfrag{calU3}{${\cal U}_3$}
\psfrag{calU4}{${\cal U}_4$}
\psfrag{calU5}{${\cal U}_5$}
\psfrag{CM5}{${\cal CM}_5^{\partial}=B_2$}
\psfrag{xi}{$x_i$}
\psfrag{yi}{$y_i$}
\psfrag{(xi,yi)}{$(x_i,y_i)$}
\psfrag{(0,0)}{$(0,0)$}
\begin{center}
\includegraphics[height=7cm]{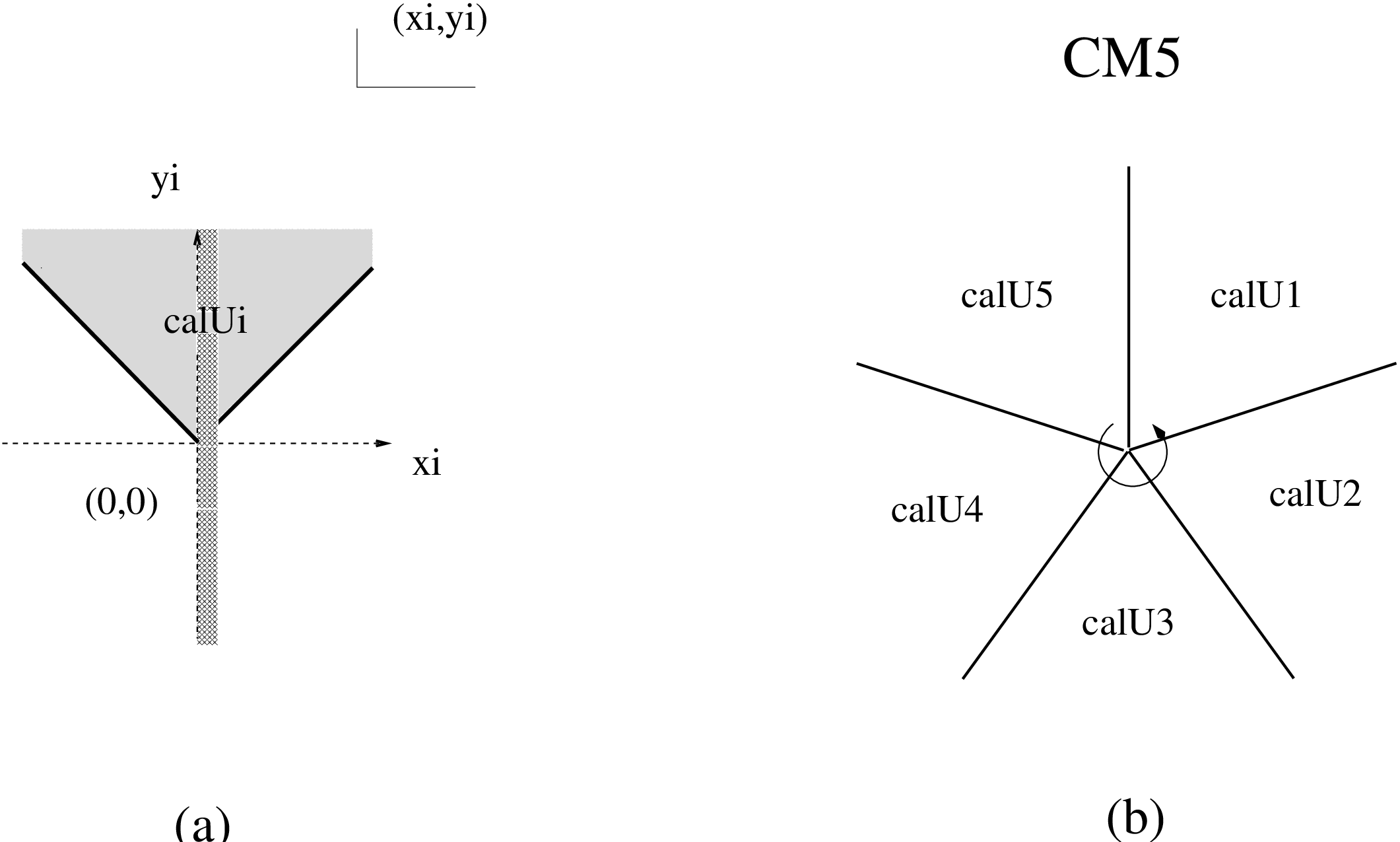}
\caption{ 
{\small 
(a) ${\cal U}_i$ is given by the shaded region on 
the $(x_i,y_i)$-plane. 
(b) The compactification ${\cal CM}_5^{\partial}$. 
It is understood as a union $\cup_{i=1}^{5}{\cal U}_i$.
The arrow denotes the orientation of 
${\cal CM}_5^{\partial}$. }}
\label{fig-ui}
\end{center}
\end{figure}

The compactification ${\cal CM}_5^{\partial}$ is a two-disk. 
We fix an identification of ${\cal CM}_5^{\partial}$ 
with the set of open-string $5$-diagrams as follows.
Let ${\cal U}_i$ 
be the fan 
$y_i \!\geq \!|x_i| \geq 0$ 
on the two-plane $(x_i,y_i)$ 
for 
$1 \!\leq \!i \!\leq \!5$.
${\cal CM}_5^{\partial}$ is understood 
as $\cup_{i=1}^5 {\cal U}_i$. 
See Figure \ref{fig-ui}. 
%
%
%
\begin{figure}
\psfrag{xigeq0}{$x_i \geq 0$}
\psfrag{xileq0}{$x_i < 0$}
\psfrag{xi}{$x_i$}
\psfrag{yi}{$y_i$}
\psfrag{|xi|}{$|x_i|$}
\begin{center}
\includegraphics[height=8cm]{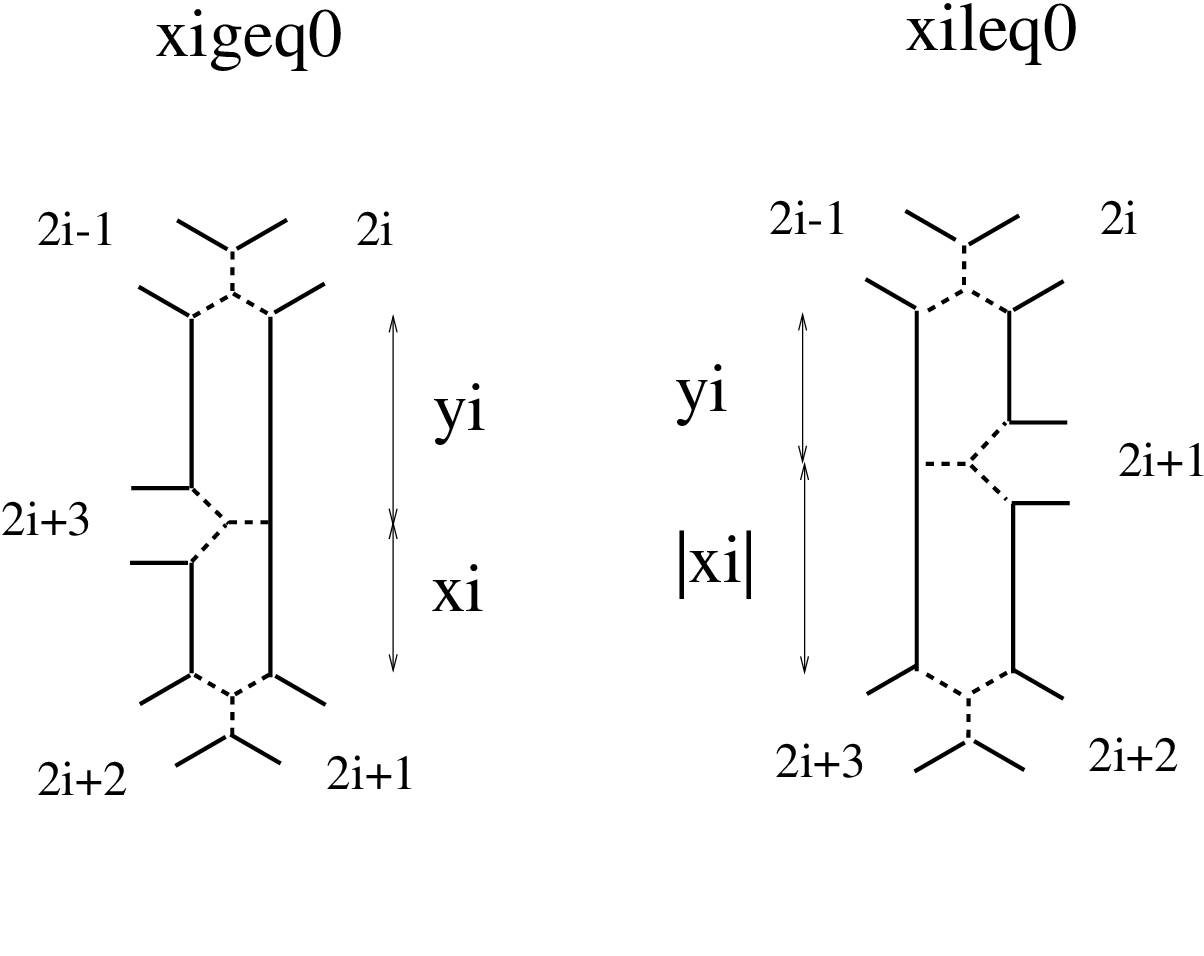}
\caption{{\small 
Open-string diagram at $(x_i,y_i)\in {\cal U}_i$.}}
\label{string-ui}
\end{center}
\end{figure}
%
%
For each $(x_i,y_i) \in {\cal U}_i$ 
we associate an open-string diagram as given in 
Figure \ref{string-ui}.   
Open-string indices in the diagrams, 
say $2i$, are understood modulo $5$. 
Length of the two internal strips are given by 
$x_i$ (or $|x_i|$) and $y_i$.   
Diagrams at the boundary $x_i\!=\!y_i$ of ${\cal U}_i$ 
coincide with those at the boundary $x_{i+1}\!=\!-y_{i+1}$ 
of ${\cal U}_{i+1}$, 
while diagrams at the other boundary $x_i\!=\!-y_i$ 
coincide with those at the boundary 
$x_{i-1}\!=\!y_{i-1}$ of ${\cal U}_{i-1}$. 
As the result we can joint these ${\cal U}_i$ 
together in the clockwise order along their boundaries 
so that they are consistent with the open-string diagrams.
Hence this gives an identification of ${\cal CM}_5^{\partial}$ 
with the set of open-string $5$-diagrams.

We have a state $\langle \Sigma|$ $\in ({\cal H}^{\otimes 5})^*$ 
for each $\Sigma \in {\cal CM}_5^{\partial}$. 
It is given by a patch-wise construction. 
We introduce a state $\Bll (x_i,y_i)\Br_i^{(12345)}$ 
$\in ({\cal H}^{\otimes 5})^*$ 
for $(x_i,y_i) \in {\cal U}_i$ 
by applying the Feynman rule 
to the corresponding open-string diagram.  
\begin{eqnarray}
&& 
\Bll
(x_i,y_i)
\Br_i^{(12345)}
\nonumber \\
&&=
\left\{
\begin{array}{cl}
  \begin{array}{l}
       \Bll 2i\!-\!1~2i~a \Br 
       \Bll b~2i\!+\!1~2i\!+\!2 \Br 
       \Bll a'~b'~2i\!+\!3 \Br 
  \\
      ~~~~~~~~~~~~~~~~~\times
        \left(b_0e^{-y_iL_0}\Bl S_{a'a}\Brr \right)
        \left(b_0e^{-x_iL_0}\Bl S_{b'b}\Brr \right) 
  \end{array}
&
   \mbox{for}~~y_i \geq x_i > \! 0, 
\\
~
&
~
\\
  \begin{array}{l}
              \Bll 2i\!-\!1~2i~a \Br
              \Bll a'~2i\!+\!1~b \Br
              \Bll b'~2i\!+\!2~2i\!+\!3 \Br
\\ 
    ~~~~~~~~~~~~~~~~~\times
              \left(b_0e^{-y_iL_0}\Bl S_{a'a}\Brr \right)
              \left((-)b_0e^{x_iL_0}\Bl S_{b'b}\Brr \right)
  \end{array}
&
\mbox{for}~~y_i \geq -x_i > 0, 
\\
~
&
~
\\
0
&
   \mbox{for}~~x_i =0, y_i \geq 0.
\end{array}
\right. 
\nonumber \\
\label{state (xi,yi)}
\end{eqnarray}
This state is grassmann-odd  
and has the ghost number equal to five. 
We then define a $({\cal H}^{\otimes 5})^*$-valued 
two-form $\bll \Omega \br^{(12345)}$. 
We put it equal to a two-form   
$dx_i \wedge dy_i \Bll(x_i,y_i)\Br_i^{(12345)}$ 
on each ${\cal U}_i$, 
\begin{eqnarray}
\Bigl. \Omega^{(12345)} \Bl_{{\cal U}_i}
=
dx_i \wedge dy_i \Bll (x_i,y_i) \Br_i^{(12345)}.
\label{2-form omega}
\end{eqnarray}

We examine an effect of the cyclic permutation 
of open-string indices. 
We can obtain another state 
by permuting the open-string indices  
from $(1,2,3,4,5)$ to $(2,3,4,5,1)$ 
in the RHS of eq.(\ref{state (xi,yi)}). 
We denote this new state by 
$\Bll (x_i,y_i) \Br_i^{(23451)}$. 
We call the corresponding two-form 
$\bll \Omega \br^{(23451)}$. 
As one can easily check,  
the state $\Bll (x_i,y_i) \Br_i^{(23451)}$ turns out 
to be $\Bll (x_i,y_i)\Br^{(12345)}_{i+3}$. 
This permutation generates a ${\bf Z}_5$-action 
on ${\cal CM}_5^{\partial}$. 
It is given by $s({\cal U}_i)={\cal U}_{i+3}$. 
Therefore $\langle \Omega |^{(23451)}$ 
is the pull-back of $\langle \Omega |^{(12345)}$ by $s$.
\begin{eqnarray}
\Omega^{(23451)}=s^*\Omega^{(12345)}. 
\label{symmetry of 2-form omega}
\end{eqnarray}

\begin{figure}[t]
\psfrag{xi}{$x_i$}
\psfrag{yi}{$y_i$}
\psfrag{(xi,yi)}{$(x_i,y_i)$}
\psfrag{(0,0)}{$(0,0)$}
\psfrag{calUizeta}{${\cal U}_i(\zeta)$}
\psfrag{calU1zeta}{${\cal U}_1(\zeta)$}
\psfrag{calU2zeta}{${\cal U}_2(\zeta)$}
\psfrag{calU3zeta}{${\cal U}_3(\zeta)$}
\psfrag{calU4zeta}{${\cal U}_4(\zeta)$}
\psfrag{calU5zeta}{${\cal U}_5(\zeta)$}
\psfrag{calV5zeta}{${\cal V}_5(\zeta)$}
\psfrag{(0,2zeta)}{$(0,2\zeta)$}
\begin{center}
\includegraphics[height=7cm]{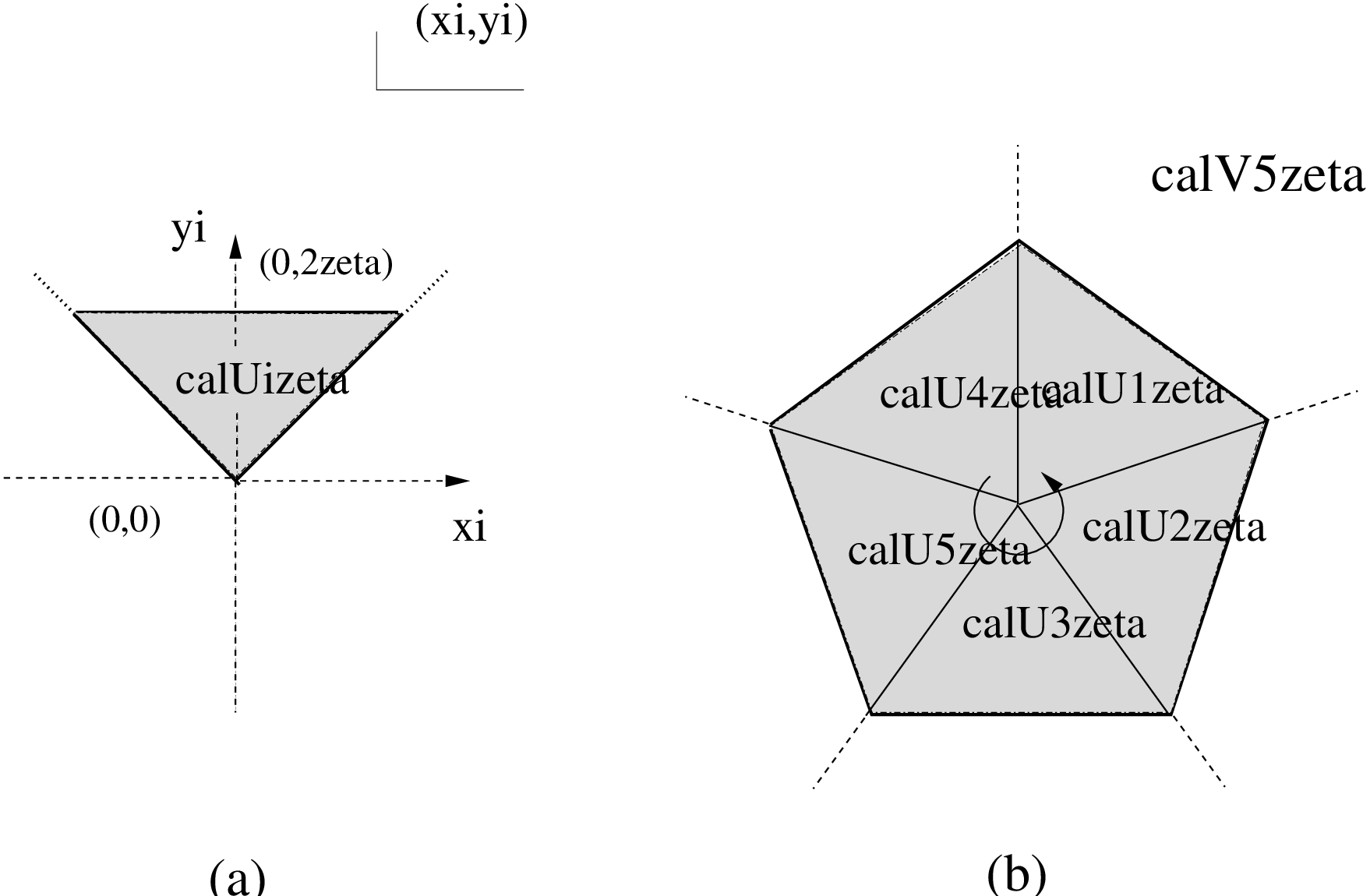}
\caption{{\small 
(a) ${\cal U}_i(\zeta)$ is the shaded region of ${\cal U}_i$. 
(b) ${\cal V}_5(\zeta)$ is a union $\cup_{i=1}^5{\cal U}_i(\zeta)$. 
It becomes the shaded region of ${\cal CM}_5^{\partial}=B_2$.
The arrow denotes the orientation of ${\cal V}_5(\zeta)$.}}
\label{fig-vi}
\end{center}
\end{figure}

We put ${\cal U}_i(\zeta)\equiv 
\Bigl\{ (x_i,y_i) \in {\cal U}_i~ \Bl~   
y_i \!<\!2\zeta \Bigr\} \Bigr.$.
Any diagram belonging to ${\cal U}_i(\zeta)$ 
can not be reproduced from lower diagrams 
at the scale $\zeta$. 
We put ${\cal V}_5(\zeta)$
$\equiv \cup_{i=1}^5 {\cal U}_i(\zeta)$.
(Figure \ref{fig-vi}.) 
%
%
The state $\Bll (x_i,y_i)\Br_i^{(12345)}$ 
must appear in $S^{\zeta}_{eff}[\Phi]$ 
as a part of an interaction vertex 
for any $(x_i,y_i)\in{\cal V}_5(\zeta)$. 
We then define open-string $5$-vertex at the scale $\zeta$ 
as follows. 
\begin{definition}[Open-string $5$-vertex at $\zeta$]
\label{def of 5-vertex}
Open-string $5$-vertex at the cut-off scale $\zeta$ is 
defined by an integration of the 
$({\cal H}^{\otimes 5})^*$-valued two-form 
$\langle \Omega |$ (\ref{2-form omega}) over 
${\cal V}_5(\zeta)$ 
multiplied by the external propagators of length $\zeta$.
\begin{eqnarray}
\Bll 1~2~3~4~5\!:\!\zeta \Br =
\int_{{\cal V}_5(\zeta)} 
\langle \Omega |^{(12345)}
\prod_{k=1}^5 e^{-\zeta L_0^{(k)}}. 
\label{5 open-string vertex}
\end{eqnarray}
Indices of the vertex are understood to label 
the Hilbert spaces ${\cal H}^{(i)}$ attached to 
clockwise-ordered five open-strings on $\partial D$. 
Thereby the vertex is regarded as a vector of 
$({\cal H}^{(1)}\otimes \cdots \otimes{\cal H}^{(5)})^*$.
\end{definition}
The $5$-vertex is grassmann-odd and has the ghost number equal to five. 
Another $5$-valent vertex might be obtained by using the two-form 
$\Omega^{(23451)}$ instead of 
$\Omega^{(12345)}$ in the definition. 
We denote it by 
$\Bll 2~3~4~5~1\!:\!\zeta \Br$.  
But they eventually become the same due to 
the relation (\ref{symmetry of 2-form omega}).  
\begin{eqnarray}
\Bll 2~3~4~5~1\!:\!\zeta \Br 
&=& 
\int_{{\cal V}_5(\zeta)}
\langle \Omega |^{(23451)}
\prod_{k=1}^{5}e^{-\zeta L_0^{(k)}}
\nonumber \\
&=&
\int_{{\cal V}_5(\zeta)}
s^* \langle \Omega |^{(12345)}
\prod_{k=1}^{5}e^{-\zeta L_0^{(k)}}
\nonumber \\
&=&
\int_{s({\cal V}_5(\zeta))={\cal V}_5(\zeta)}
\langle \Omega |^{(12345)}
\prod_{k=1}^{5}e^{-\zeta L_0^{(k)}}
\nonumber \\
&=&
\Bll 1~2~3~4~5 \!:\!\zeta \Br.  
\label{symmetry of 5-vertex}
\end{eqnarray}
We say this as symmetry of the open-string $5$-vertex 
under the cyclic permutation.

\begin{figure}[t]
\begin{center}
\includegraphics[height=4cm]{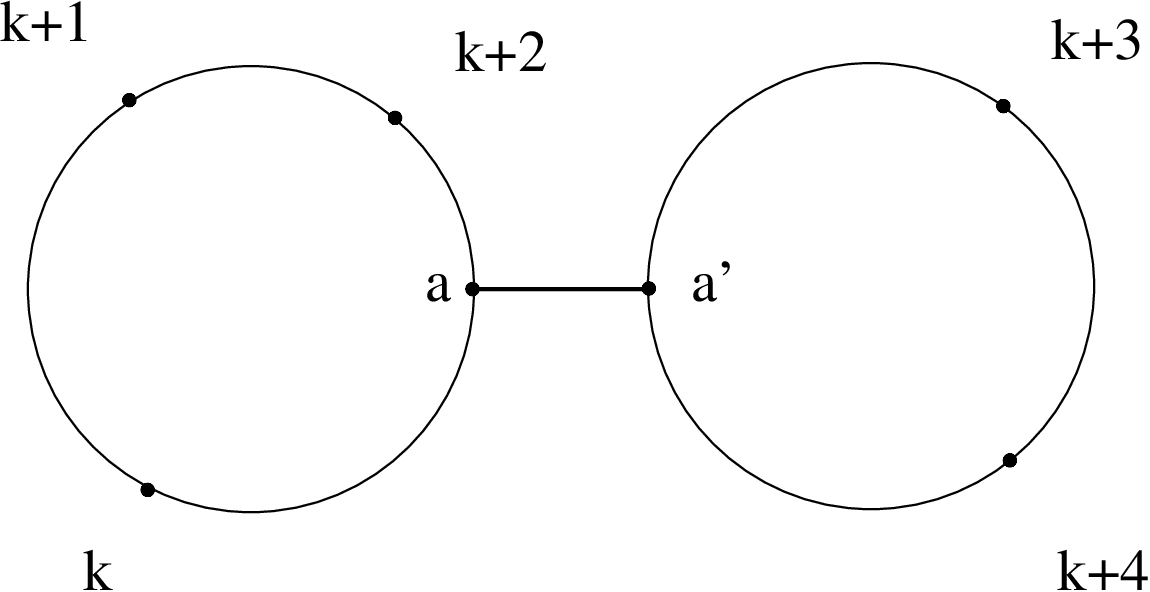}
\caption{{\small 
A typical configuration of five points 
on $\partial D$ which belongs to 
${\cal V}_{(k ~k+1~ k+2~ a)}(\zeta)\times 
{\cal V}_{(a'~k+3 ~k+4)}(\zeta)$ 
($\subset \partial {\cal V}_5(\zeta))$. 
The solid line connecting two points $a$ and $a'$ 
represents open-string strip of length equal to 
$2\zeta$.}}
\label{fig-boundaryV4}
\end{center}
\end{figure}

The $5$-vertex is not invariant under the action of 
the BRST charge. Similarly to the case of the $4$-vertex   
the BRST transform can be written down 
using the lower vertices.     
\begin{proposition}[$Q$-action on $5$-vertex]
\label{prop Q-action on 5-vertex}
Action of the BRST charge on the $5$-vertex 
(\ref{5 open-string vertex}) becomes as follows. 
\begin{eqnarray}
 \Bll 1~2~3~4~5\!:\!\zeta \Br 
\Bigl(\sum_{i=1}^5Q^{(i)}\Bigr)
=
-\sum_{k=1}^5
\Bll k~k\!+\!1~k\!+\!2~a\!:\!\zeta \Br 
\Bll a'~k\!+\!3~k\!+\!4\!:\!\zeta \Br 
\Bigl. S_{a'a}\Brr, 
\label{Q action on 5-vertex}
\end{eqnarray}
where open-string indices in the RHS 
are understood modulo $5$.
\end{proposition}
The origin of eq.(\ref{Q action on 5-vertex}) 
is also in the geometry.  
The boundary of ${\cal V}_5(\zeta)$ is a circle. 
As a set, it is a sum of 
${\cal V}_{(k ~k+1~ k+2~ a)}(\zeta)\times 
{\cal V}_{(a'~k+3 ~k+4)}(\zeta)$. 
Here we write ${\cal V}_4(\zeta)$ and  
${\cal V}_3(\zeta)$ (a single point)  
as 
${\cal V}_{(a~k ~k+1 ~k+2)}(\zeta)$ 
and 
${\cal V}_{(a'~k+3 ~k+4)}(\zeta)$
in order to show that  
external open-strings participating in 
the diagrams are labeled in the clockwise order 
respectively by $(k,k\!+\!1,k\!+\!2,a)$ 
and $(a',k\!+\!3,k\!+\!4)$.  
(Figure \ref{fig-boundaryV4}).
%
%
Orientation of ${\cal V}_5(\zeta)$ is given by 
${\cal CM}_5^{\partial}$ 
(the standard orientation of two-disk). 
Thereby the circle is oriented. 
On the other hand,  
${\cal V}_4(\zeta) \times {\cal V}_3(\zeta)$ has  
the orientation determined from  
${\cal CM}_4^{\partial} \times {\cal CM}_3^{\partial}$. 
Comparing these orientations we obtain  
\begin{eqnarray}
\partial {\cal V}_5(\zeta) 
=
-\sum_{k=1}^5
{\cal V}_{(k ~k+1 ~k+2 ~a)}(\zeta)\times 
{\cal V}_{(a'~k+3 ~k+4)}(\zeta). 
\label{dV5}
\end{eqnarray}
Looking at eq.(\ref{Q action on 5-vertex}) and 
eq.(\ref{dV5}) we find a coincidence between the BRST 
transformation  and the boundary operation. 
Actually the action of the BRST charge on the $5$-vertex 
becomes a representation of the boundary operator.

The $5$-vertex depends on the cut-off scale. 
It vanishes at $\zeta=0$.
The dependence comes from the integration region 
${\cal V}_5(\zeta)$ besides the propagators 
$\prod_ie^{-\zeta L_0^{(i)}}$ 
in eq.(\ref{5 open-string vertex}). 
\begin{proposition}
[Scale dependence of $5$-vertex]
\label{scale dependence of 5-vertex}
Scale dependence of the $5$-vertex 
(\ref{5 open-string vertex}) is described by 
\begin{eqnarray}
\frac{1}{2}
\frac{d}{d\zeta}
\Bll 1~2~3~4~5\!:\!\zeta \Br
&=& 
-\sum_{k=1}^5
\Bll k~k\!+\!1~k\!+\!2~a\!:\!\zeta \Br
\Bll a'~k\!+\!3~k\!+\!4\!:\!\zeta \Br 
\Bigl( b_0\Bl S_{a'a}\Brr \Bigr)
\nonumber \\
&&
-\frac{1}{2}
\Bll 1~2~3~4~5\!:\!\zeta \Br 
\Bigl(\sum_{k=1}^5L_0^{(k)}\Bigr). 
\label{cut-off dependence of 5-vertex}
\end{eqnarray}
\end{proposition} 

~

We give proofs of Propositions 
\ref{prop Q-action on 5-vertex} 
and 
\ref{scale dependence of 5-vertex}. 
The next lemma is useful in the proof of 
Proposition \ref{prop Q-action on 5-vertex}. 
\begin{lemma}
\label{lemma Q-action on 5-vertex}
The BRST charge acts on the state 
$\Bll (x_i,y_i) \Br_i^{(12345)}$ as follows :
\begin{eqnarray}
&& \Bll (x_i,y_i) \Br_i^{(12345)}
\Bigl(\sum_iQ^{(i)}\Bigr)
\nonumber \\
&&=
\left\{
\begin{array}{ll}
\begin{array}{l}
\Bll 2i\!-\!1~2i~a \Br 
\Bll b~2i\!+\!1~2i\!+\!2 \Br 
\Bll a'~b'~2i\!+\!3 \Br  
\\
\times 
\Bigl\{
\left(b_0e^{-y_iL_0}\Bl S_{a'a}\Brr \right)
\frac{\partial}{\partial x_i}\!
\left(e^{-x_iL_0}\Bl S_{b'b}\Brr \right)
+
\frac{\partial}{\partial y_i}\!
\left(e^{-y_iL_0}\Bl S_{a'a}\Brr \right)
\left(b_0e^{-x_iL_0}\Bl S_{b'b}\Brr \right)
\Bigr\}
\end{array}
&
\mbox{for}~~x_i >0
\\
~
&
~
\\
\begin{array}{l}
\Bll 2i\!-\!1~2i~a \Br 
\Bll a'~2i\!+\!1~b \Br  
\Bll b'~2i\!+\!2~2i\!+\!3 \Br  
\\
\times 
\Bigl\{
\left(b_0e^{-y_iL_0}\Bl S_{a'a}\Brr \right)
\frac{\partial}{\partial x_i}\! 
\left(e^{x_iL_0}\Bl S_{b'b}\Brr \right)
+ 
\frac{\partial}{\partial y_i}\!
\left(e^{-y_iL_0}\Bl S_{a'a}\Brr \right)
\left((-)b_0e^{x_iL_0}\Bl S_{b'b}\Brr \right)
\Bigr\}
\end{array}
&
\mbox{for}~~x_i < 0 
\end{array}
\right. .
\nonumber \\
&&
\label{Q on state (xi,yi)}
\end{eqnarray}
\end{lemma}
This lemma can be shown in the same manner 
as Lemma \ref{lemma for Qx}.
We omit the proof.

~

\noindent 
\underline{{\it Proof of Proposition 
\ref{prop Q-action on 5-vertex}}}~:  
We put 
$I_i\equiv$
$\displaystyle{\int_{{\cal U}_i(\zeta)}
\!dx_i\wedge\! dy_i}$ 
$\left\{
\Bll (x_i,y_i)\Br_i^{(12345)}
\Bigl(\sum_{k=1}^5Q^{(k)}\Bigr)
\right\}$. 
The LHS of eq.(\ref{Q action on 5-vertex}) is 
just $\sum_{i=1}^5I_i$. 
Due to Lemma \ref{lemma Q-action on 5-vertex} 
the integral $I_i$ reduces to an integral 
on $\partial {\cal U}_i(\zeta)$. 
The boundary $\partial {\cal U}_i(\zeta)$ has three components. 
$I_i$ becomes a sum of the three boundary integrals. 
The boundary integral along $x_i\!=\!y_i$ turns out to 
equal $(-1)\times$ the boundary integral of $I_{i+1}$ along 
$-x_{i+1}\!=\!y_{i+1}$. 
Thus their contributions to $\sum_iI_i$ cancel each other. 
(These two boundaries are identified in ${\cal V}_5(\zeta)$.)
Net contribution of $I_i$ to the sum    
becomes the boundary integral along $y_i\!=\!2\zeta$.  
Therefore we have 
\begin{eqnarray}
&& 
\Bll 1~2~3~4~5\!:\!\zeta \Br 
\Bigl(\sum_{i=1}^5Q^{(i)}\Bigr)
~~~~~~~~~ \left(=\sum_{i=1}^5I_i \right) 
\nonumber \\
&&=
\sum_{i=1}^5 
\left\{
\int_0^{2\zeta}\!dx_i
\Bll 2i\!-\!1~2i~a \Br
\Bll b~2i\!+\!1~2i\!+\!2 \Br 
\Bll a'~b'~2i\!+\!3 \Br 
\left(e^{-2\zeta L_0}\Bl S_{a'a}\Brr \right)
\left(b_0e^{-x_i L_0}\Bl S_{b'b}\Brr \right)
\right. 
\nonumber \\
&&~~~~~~~~
\left.
+\int^0_{-2\zeta}\!dx_i
\Bll 2i\!-\!1~2i~a \Br
\Bll a'~2i\!+\!1~b \Br 
\Bll b'~2i\!+\!2~2i\!+\!3 \Br
\left(e^{-2\zeta L_0}\Bl S_{a'a}\Brr \right)
\left((-)b_0e^{x_i L_0}\Bl S_{b'b}\Brr \right)
\right\} 
\nonumber \\
&&~~~~~~~~~~~~~~~~~~
\times 
\prod_{k=1}^5e^{-\zeta L_0^{(k)}} 
\nonumber \\ 
&&
=-\sum_{i=1}^5 
\Bll 2i\!-\!1~2i~a \Br 
\nonumber \\
&&~~~~~~~~
\times 
\left\{
\int_0^{2\zeta}\!dx
\Bll a'~2i\!+\!1~b \Br
\Bll b'~2i\!+\!2~2i\!+\!3 \Br 
\left(b_0e^{-xL_0}\Bl S_{b'b}\Brr \right) \right.
\nonumber \\
&&~~~~~~~~~~~~~~~
+
\int_{-2\zeta}^0
\!dx
\Bll 2i\!+\!1~2i\!+\!2~b \Br 
\Bll b'~2i\!+\!3~a' \Br 
\left((-)b_0e^{xL_0}\Bl S_{b'b}\Brr \right)
\left\}
\prod_{k=1}^5e^{-\zeta L_0^{(k)}}
\right.
\nonumber \\
&&~~~~~~~~~~~~~~~~~~~~~~~~~
\times 
\Bigl(e^{-2\zeta L_0} \Bl S_{a'a}\Brr \Bigr)
\label{proof Q-action on 5-vertex 1}
\end{eqnarray}
We rewrite eq.(\ref{proof Q-action on 5-vertex 1}) 
in terms of the vertices at the scale $\zeta$. 
It turns out to be    
\begin{eqnarray}
Eq.(\ref{proof Q-action on 5-vertex 1})
&=&
-\sum_{i=1}^5
\Bll 2i\!-\!1~2i~a\!:\!\zeta \Br
\Bll a'~2i\!+\!1~2i\!+\!2~2i\!+\!3\!:\!\zeta \Br  
\Bigl. S_{a'a}\Brr, 
\nonumber \\
&=&
-\sum_{k=1}^5
\Bll k~k\!+\!1~k\!+\!2~a\!:\!\zeta \Br 
\Bll a'~k\!+\!3~k\!+\!4\!:\!\zeta \Br 
\Bigl. S_{a'a}\Brr. 
\end{eqnarray}
This is precisely the RHS of eq.(\ref{Q action on 5-vertex}).\P

~

\noindent 
\underline{{\it Proof of Proposition 
\ref{scale dependence of 5-vertex}}}~: 
We compute the first-order variation of the $5$-vertex 
with respect to the cut-off scale parameter $\zeta$ 
as follows.   
\begin{eqnarray}
&&
\delta \Bll 1~2~3~4~5:\zeta \Br 
\nonumber \\
&&= 
\delta 
\left[
   \int_{{\cal V}_5(\zeta)}
         \langle \Omega |^{(12345)}
   \prod_{k=1}^{5}e^{-\zeta L_0^{(k)}}
\right] 
\nonumber \\
&&=
\delta 
\left[
   \int_{{\cal V}_5(\zeta)}
         \langle \Omega |^{(12345)}
\right]
   \prod_{k=1}^{5}e^{-\zeta L_0^{(k)}}
+\int_{{\cal V}_5(\zeta)}
         \langle \Omega |^{(12345)}
\delta 
  \left[
   \prod_{k=1}^{5}e^{-\zeta L_0^{(k)}}
  \right] 
\nonumber \\
&&= 
\sum_{i=1}^5  
\int_{\delta{\cal U}_i(\zeta)}\!dx_i\wedge dy_i 
\Bll (x_i,y_i)\Br_i^{(12345)} 
\prod_{k=1}^{5}e^{-\zeta L_0^{(k)}} 
-
\delta \zeta 
\Bll 1~2~3~4~5\!:\!\zeta \Br
\Bigl(\sum_{k=1}^5L_0^{(k)}\Bigr). 
\nonumber \\
\label{proof scale dependence 5-vertex 1}
\end{eqnarray} 
The first term of eq.(\ref{proof scale dependence 5-vertex 1}) 
is further evaluated as follows.  
\begin{eqnarray}
&&
\sum_i 
\int_{\delta{\cal U}_i(\zeta)}\!dx_i\wedge dy_i 
\Bll (x_i,y_i)\Br_i^{(12345)} 
\prod_{k=1}^{5}e^{-\zeta L_0^{(k)}}
\nonumber \\
&&
=2 \delta \zeta \sum_i 
\left\{
\int_0^{2\zeta}\!dx_i 
\Bll 2i\!-\!1~2i~a \Br 
\Bll b~2i\!+\!1~2i\!+\!2 \Br  
\Bll a'~b'~2i\!+\!3 \Br 
\left(b_0e^{-2\zeta  L_0}\Bl S_{a'a}\Brr \right)
\left(b_0e^{-x_i L_0}\Bl S_{b'b}\Brr \right)
\right.
\nonumber \\
&&~~~~~~~~~~~~~~+
\left. 
\int_{-2 \zeta}^0\!dx_i 
\Bll 2i\!-\!1~2i~a \Br
\Bll a'~2i\!+\!1~b \Br
\Bll b'~2i\!+\!2~2i\!+\!3 \Br
\left(b_0e^{-2\zeta L_0}\Bl S_{a'a}\Brr \right) 
\left((-)b_0e^{x_i L_0}\Bl S_{b'b}\Brr \right)
\right\}  
\nonumber \\
&&~~~~~~~~~~~~~~~~~~~~~~~~~~~\times 
\prod_{k=1}^{5}e^{-\zeta L_0^{(k)}}
\nonumber \\
&&
=-2\delta \zeta \sum_i 
\Bll 2i\!-\!1~2i~a \Br 
\nonumber \\
&&~~~~~~~~~~~~~~~~~~\times 
\left\{
\int_0^{2\zeta}\!dx
\Bll a'~2i\!+\!1~b \Br
\Bll b'~2i\!+\!2~2i\!+\!3 \Br
\left(b_0e^{-xL_0}\Bl S_{b'b}\Brr \right)
\right. 
\nonumber \\
&&~~~~~~~~~~~~~~~~~~~~~~~~~~~~~~~
\left. +
\int_{-2\zeta}^{0}\!dx
\Bll 2i\!+\!1~2i\!+\!2~b \Br
\Bll b'~2i\!+\!3~a'\Br 
\left((-)b_0e^{xL_0}\Bl S_{b'b}\Brr \right)
\right\}
\nonumber \\
&&~~~~~~~~~~~~~~~~~~~~~~~~~~~~~~~~~~~~~~~~~~~\times  
\prod_{k=1}^5e^{-\zeta L_0^{(k)}}
\left(b_0 e^{-2\zeta L_0}\Bl S_{a'a}\Brr \right).
\label{proof scale dependence 5-vertex 2}
\end{eqnarray}
We write eq.(\ref{proof scale dependence 5-vertex 2}) 
in terms of the vertices at the scale $\zeta$. 
It turns out to be 
\begin{eqnarray}
Eq.(\ref{proof scale dependence 5-vertex 2})
&=&
-2\delta \zeta \sum_i 
\Bll 2i\!-\!1~2i~a\!:\!\zeta \Br 
\Bll a'~2i\!+\!1~2i\!+\!2~2i\!+\!3\!:\!\zeta \Br 
\left( b_0 \Bl S_{a'a}\Brr \right), 
\\
&=& 
-2 \delta \zeta \sum_k
\Bll k~k\!+\!1~k\!+\!2~a\!:\!\zeta \Br
\Bll a'~k\!+\!3~k\!+\!4\!:\!\zeta \Br 
\Bigl( b_0\Bl S_{a'a}\Brr \Bigr). 
\end{eqnarray}
Hence we obtain the first term of 
the RHS of eq.(\ref{cut-off dependence of 5-vertex}).
The second term in eq.(\ref{proof scale dependence 5-vertex 1}) 
gives the second term of the RHS of 
eq.(\ref{cut-off dependence of 5-vertex}). 
\P

\subsection{Conjecture on higher open-string vertices}
We want to obtain low energy open-string $n$-valent vertices 
for the cases of $n\!\geq \!6$.  
These vertices are determined in principle from 
the sum of the graphs which are one-particle irreducible 
with respect to the regularized propagator. 
In the previous subsection explicit constructions are given 
for $n\!=\!3,4$ and $5$ cases. 
${\cal V}_3(\zeta),{\cal V}_4(\zeta)$ and ${\cal V}_5(\zeta)$ 
are the moduli of the corresponding irreducible graphs. 
The approach taken there may be generalized.

The compactification ${\cal CM}_n^{\partial}$ is $B_{n-3}$. 
Each point of ${\cal CM}_n^{\partial}$ 
represents an open-string $n$-diagram. 
We can obtain a map 
${\cal CM}_n^{\partial}\rightarrow ({\cal H}^{\otimes n})^*$
$: \Sigma \mapsto \langle \Sigma |$ 
by applying the Feynman rule to the diagrams.  
Since they are trivalent ribbon graphs, 
each diagram consists of $(n-2)$ trivalent vertices 
and $(n-3)$ internal strips. 
Two trivalent vertices are connected 
at one external open-string 
by inserting there the propagator of the form 
$b_0e^{-\tau L_0}|S_{a'a}\rangle$ in the following way. 
\begin{eqnarray}
\Bll 1~2~a\!:\!\zeta \Br 
\Bll a'~3~4:\!\zeta \Br 
\left( b_0 \Bl S_{a'a}\Brr\right) . 
\end{eqnarray}
The trivalent vertex is a grassmann-odd vector with the ghost number three 
and the propagator is a grassmann-even vector with the ghost number two. 
The state $\langle \Sigma |$ acquires  
the ghost number $n$ and the grassmannity $(-)^n$. 
We have several ways to identify 
points of ${\cal CM}_n^{\partial}$ with the diagrams. 
Different identifications are related with one another 
by the automorphisms generated by $s$. 
Let us take one of them and fix it. 
We call the state $\langle \Sigma |$ 
given under this identification 
as $\langle \Sigma |^{(1~2 \cdots n\!-\!1~n)}$.  
If one takes another identification 
related by the automorphism $s$, 
the corresponding state is called  
$\langle \Sigma |^{(2~3 \cdots n~1)}$.

To have an explicit representation of the map 
it is convenient to take a patch-wise construction. 
Let ${\cal CM}_n^{\partial}\equiv \cup_i {\cal U}_i$. 
Each ${\cal U}_i$ may be identified with a suitable cone of 
${\bf R}^{n-3}$. 
The Euclidean coordinates $(x^i_1,\cdots,x^i_{n-3})$ of ${\cal U}_i$ 
represent the metric of the underlying ribbon graph.  
The map $\Sigma \mapsto \langle \Sigma |$ is introduced   
in a patch-wise manner. We denote the state 
given at $(x^i_1,\cdots,x^i_{n-3})$  
by $\left \langle (x^i_1,\cdots,x^i_{n-3})\right |_i$.

The next step is to obtain a 
$({\cal H}^{\otimes n})^*$-valued $(n\!-\!3)$-form 
$\langle \Omega |$. 
It is given on each ${\cal U}_i$ by  
$dx^i_1\wedge \cdots \wedge dx^i_{n-3}$
$\left \langle (x^i_1,\cdots,x^i_{n-3})\right |_i$. 
Recall ${\cal V}_n(\zeta)$ is the set of open-string 
$n$-diagrams which are one-particle irreducible with respect to 
the regularized propagator (\ref{reg propagator}). 
In other word it is the set of open-string $n$-diagrams 
all internal strips of which have length less than $2\zeta$. 
${\cal V}_n(\zeta)$ also becomes a $(n\!-\!3)$-ball. 
Open-string $n$-vertex at the cut-off scale $\zeta$ 
is defined by an integration of $\Omega $ 
over the ball ${\cal V}_n(\zeta)$.  
\begin{eqnarray}
\Bll 1~2 \cdots n\!-\!1~n \!:\!\zeta \Br =
\int_{{\cal V}_n(\zeta)}
\langle \Omega |^{(1~2 \cdots n\!-\!1~n)}
\prod_{i=1}^{n}e^{-\zeta L_0^{(i)}}. 
\nonumber 
\end{eqnarray}
Indices in the vertex are understood to label 
clockwise-ordered $n$ open-strings on $\partial D$. 
The vertex is regarded as a vector of 
$({\cal H}^{(1)}\otimes \cdots \otimes {\cal H}^{(n)})^*$. 
It has the ghost number $n$ and the grassmannity $(-)^n$. 
The automorphisms of ${\cal CM}_n^{\partial}$ 
will provide symmetry of the vertex. 
Depending on choice of the identifications,   
we have several $\Omega$. 
They are expected to be related with one another by the 
automorphisms.  
\begin{eqnarray}
s^*\Omega^{(1~2 \cdots n-1~n)}
=\Omega^{(2~3 \cdots n~1)}.  
\nonumber 
\end{eqnarray}
This implies that the $n$-vertex becomes symmetric or anti-symmetric 
under the cyclic permutation 
when $s$ preserves the orientation of ${\cal CM}_n^{\partial}$ or not.

\begin{figure}[t]
\begin{center}
\includegraphics[height=4cm]{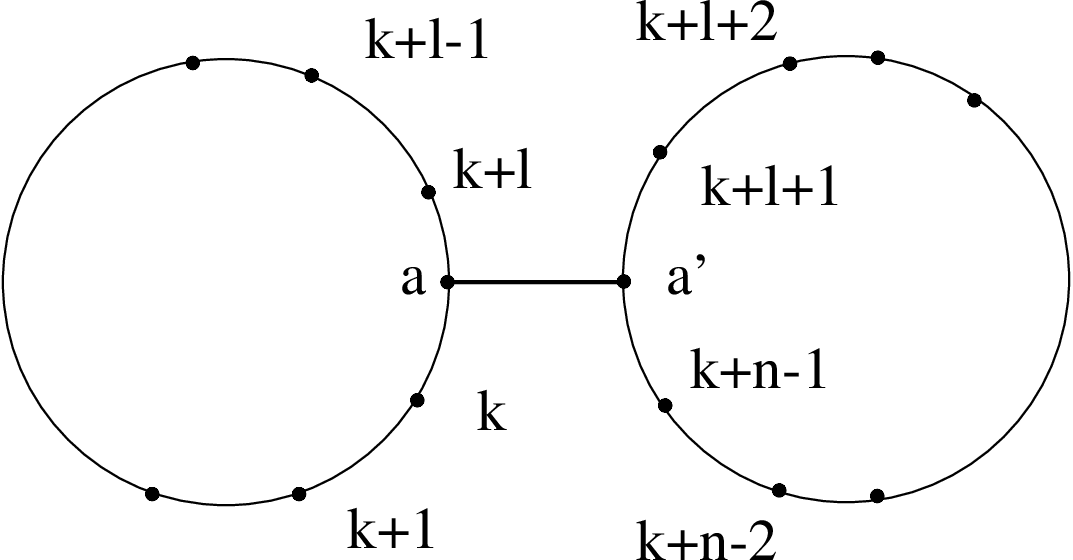}
\caption{{\small 
A typical configuration of $n$ points on 
$\partial D$ which belongs to 
${\cal V}_{(k~ \cdots~ k+l~ a)}(\zeta)\times 
{\cal V}_{(a'~k+l+1~ \cdots~ k+n-1)}(\zeta)$ 
($\subset \partial {\cal V}_n(\zeta))$. 
The solid line connecting two points $a$ and $a'$ 
represents open-string strip of length equal to 
$2\zeta$.}}
\label{boundaryVn-fig}
\end{center}
\end{figure}

The boundary $\partial {\cal V}_n(\zeta)$ is identified, as a set, 
with a suitable sum of 
${\cal V}_{l+2}(\zeta)\times {\cal V}_{n-l}(\zeta)$. 
${\cal V}_n(\zeta)$ is oriented by the standard orientation 
of $B_{n-3}$. 
Thereby $\partial {\cal V}_n(\zeta)$ is oriented. 
On the other hand,  
${\cal V}_{l+2}(\zeta)\times {\cal V}_{n-l}(\zeta)$ 
is also oriented by $B_{l-1}\times B_{n-l-3}$. 
Taking account of their orientations we have 
\begin{eqnarray}
\partial {\cal V}_n(\zeta)
=\frac{1}{2}
\sum_{k=1}^{n}\sum_{l=1}^{n-3}
(\pm)
{\cal V}_{(k~k+1~ \cdots ~k+l~a)}(\zeta)
\times 
{\cal V}_{(a'~k+l+1~ \cdots~ k+n-1)}(\zeta). 
\nonumber 
\end{eqnarray}
The factor $1/2$ is needed to avoid the double counting 
of the components. 
We write 
${\cal V}_{l+2}(\zeta)$ and ${\cal V}_{n-l}(\zeta)$ 
as 
${\cal V}_{(k~k+1~ \cdots~ k+l~a)}(\zeta)$ 
and 
${\cal V}_{(a'~k+l+1~ \cdots~ k+n-1)}(\zeta)$, 
in order to show that open-strings participating in the diagrams
are labeled in the clockwise order respectively by 
$(k,k+1,\cdots,k+l,a)$ and 
$(a',k+l+1,\cdots,k+n-1)$. 
(Figure \ref{boundaryVn-fig}).
%
%
The signature $(\pm)$ must be determined 
by a comparison of the orientations of 
$\partial {\cal V}_n(\zeta)$ 
and 
${\cal V}_{l+2}(\zeta)\times {\cal V}_{n-l}(\zeta)$.
As we have observed in the previous examples,  
action of the BRST charge on the open-string vertices  
is expected to provide a representation of 
the boundary operator $\partial$. 
If so, we will have 
\begin{eqnarray}
\Bll  1 \cdots n\!:\!\zeta \Br 
\Bigl(\sum_{i=1}^{n}Q^{(i)}\Bigr)
=
\frac{1}{2}\sum_{k=1}^{n}\sum_{l=1}^{n-3}(\pm)
\Bll 
\underbrace{k~k\!+\!1~  \cdots~ k\!+\!l~a}_{l+2}
\!:\!\zeta \Br 
\Bll  
\underbrace{a'~k\!+\!l\!+\!1~ \cdots~ k\!+\!n\!-\!1}_{n-l}
\!:\!\zeta \Br
\Bigl. S_{a'a}\Brr.  
\nonumber 
\end{eqnarray}

The $n$-vertex vanishes at $\zeta\!=\!0$ since 
${\cal V}_n(0)$ becomes a point.
The scale dependence of open-string vertices 
follows from the above speculation. 
The dependence comes from the ball ${\cal V}_n(\zeta)$ and 
the multiplier $\prod_ie^{-\zeta L_0^{(i)}}$. 
Variation of the $n$-vertex with respect to $\zeta$   
is a sum of the variations of these two. 
Therefore we have 
\begin{eqnarray}
\frac{d}{d\zeta}
\Bll  1~2 \cdots n\!:\!\zeta \Br  
&=&
\sum_{k=1}^{n}\sum_{l=1}^{n-3}
(\pm)
\Bll 
\underbrace{k~k\!+\!1~ \cdots~ k\!+\!l~a}_{l+2}
\!:\!\zeta \Br 
\Bll  
\underbrace{a'~ k\!+\!l\!+\!1~ \cdots~ k\!+\!n\!-\!1}_{n-l}
\!:\!\zeta \Br 
\left(b_0 \Bl S_{a'a}\Brr \right) 
\nonumber \\
&&-
\Bll  1~2~ \cdots~ n\!:\!\zeta \Br
\Bigl(\sum_{i=1}^{n}L_0^{(i)}\Bigr). 
\nonumber 
\end{eqnarray}

To carry out this program and examine   
the conjectural properties of open-string vertices,  
we need a systematic description of ${\cal CM}_n^{\partial}$. 
In particular, it is required 
so that the two possible orientations of each component 
appearing in $\partial {\cal V}_n(\zeta)$ 
must be compared with each other in a systematic way. 
Unfortunately we do not know such a description.  
We leave it as a future problem. 
Instead we would like to propose a possible solution 
in the following conjecture.

~

\noindent 
{\bf Conjecture}~
{\it 
Open-string $n$-vertex 
$\Bll 1 \cdots n\!:\!\zeta \Br 
\in ({\cal H}^{\otimes n})^*$  
($n \geq 3$) can be taken so that it enjoys  
the following action of the BRST charge,} 
\begin{eqnarray}
&&
\Bll 1 \cdots n\!:\!\zeta \Br  
\Bigl(\sum_{i=1}^{n}Q^{(i)}\Bigr)
\nonumber \\
&&~~~~=
-\frac{1}{2}
\sum_{k=1}^{n}\sum_{l=1}^{n-3}
(-)^{(n+1)(k+l+1)}
\Bll k~k\!+\!1~ \cdots~ k\!+\!l~a\!:\!\zeta \Br 
\Bll  a'~k\!+\!l\!+\!1~ \cdots~ k\!+\!n\!-\!1\!:\!\zeta \Br 
\Bigl. S_{a'a}\Brr,  
\nonumber \\ 
&&~~~~~
\label{conjecture 1}
\end{eqnarray}
{\it besides the following cyclic asymmetry with respect 
to the open-string indices,}  
\begin{eqnarray}
\Bll  1~2~\cdots~ n\!-\!1~n\!:\!\zeta \Br 
=(-)^{n+1}
\Bll  2~3~\cdots~ n~1\!:\!\zeta \Br.
\label{conjecture 2}
\end{eqnarray}
{\it Indices in the RHS of eq.(\ref{conjecture 1}), 
say $k\!+\!l$, are understood modulo $n$. }

~

\noindent
Open-string indices of the $n$-vertex are understood 
to label the Hilbert spaces ${\cal H}^{(i)}$  
attached to clockwise-ordered $n$ open-strings on $\partial D$. 
We regard the $n$-vertex $\Bll 1~ \cdots~ n:\zeta \Br$ 
as a vector of 
$({\cal H}^{(1)}\otimes \cdots \otimes {\cal H}^{(n)})^*$.  
It has the ghost number equal to $n$ 
and grassmannity $(-)^n$. 
We give a few comments on the above conjecture. 
First, the asymmetry in eq.(\ref{conjecture 2}) 
is consistent with the action of the BRST charge  
(\ref{conjecture 1}). 
Secondly, it can be checked that 
for the cases of $n\!=\!3,4$ and $5$,  
eqs.(\ref{conjecture 1}) and (\ref{conjecture 2}) 
reproduce correctly the corresponding results 
obtained in the previous subsection.

As we referred in the program at the beginning of 
this subsection, action of the BRST charge on the vertices 
is expected to provide a representation of the boundary operator 
$\partial$. For this to be realized, 
the conjectural action of the BRST charge must be nilpotent, at least.  
To show the nilpotency is a non-trivial test of the conjecture. 
If we accept eqs.(\ref{conjecture 1}) and (\ref{conjecture 2}),  
we have the following proposition.    
\begin{proposition}
\label{test for conjecture}
Action of the BRST charge given in (\ref{conjecture 1})
is nilpotent.
\begin{eqnarray}
\left\{
\Bll 1~2~ \cdots~ n \!:\!\zeta \Br
\Bigl(\sum_{i=1}^nQ^{(i)}\Bigr)
\right\} 
\Bigl(\sum_{i=1}^nQ^{(i)}\Bigr)=0.  
\label{nilpotency in conjecture}
\end{eqnarray}
\end{proposition}
To show this proposition, some complicated calculations  
are required. We provide the proof in Appendix. 
The cut-off scale dependence of the vertices 
can be read from the conjecture.  
\begin{proposition}
[Scale dependence of higher vertices]
\label{scale dependence of n-vertex}
Scale dependence of the open-string $n$-vertex 
($n\!\geq\! 3$) is described by 
\begin{eqnarray}
&&
\frac{d}{d \zeta}
\Bll  1~2 \cdots n\!:\!\zeta \Br 
\nonumber \\
&&~~~~=
-\sum_{k=1}^{n}\sum_{l=1}^{n-3}
(-)^{(n+1)(k+l+1)}
\Bll k~k\!+\!1~ \cdots~ k\!+\!l~a\!:\!\zeta \Br 
\Bll  a'~k\!+\!l\!+\!1~ \cdots~ k\!+\!n\!-\!1\!:\!\zeta \Br 
\Bigl(b_0 \Bl S_{a'a}\Brr \Bigr) 
\nonumber \\
&&~~~~~~
-
\Bll  1~2~ \cdots~ n\!:\!\zeta \Br 
\Bigl(\sum_{i=1}^{n}L_0^{(i)}\Bigr).
\label{cut-off scale dependence of n-vertex}
\end{eqnarray}
\end{proposition}
We note that 
eq.(\ref{cut-off scale dependence of n-vertex})
reproduces the previous results 
for the cases of $n\!=\!3,4$ and $5$.

~

\noindent 
We proceed on the rest of this paper by assuming the conjecture.


\section{$A_{\infty}$-Algebra In Low Energy Theory}

To develop low energy description of classical open-string 
field theory, we introduce the notion of homotopy associative 
algebra ($A_{\infty}$-algebra). $A_{\infty}$-algebra was introduced 
in \cite{Stasheff} and further investigated in \cite{Fukaya,FOOO} 
including deformation theory 
\footnote{We follow the convention used in \cite{Fukaya,FOOO}.}.
In string field theory, $A_{\infty}$ algebra was discussed in 
\cite{Witten CS, Zwiebach C-O}.

Let $C$ be a ${\bf Z}$-graded module. 
A graded module $\Pi C$ is defined by shifting the degree, 
$(\Pi C)^m \!\equiv\! C^{m+1} $. 
We write the shifted degree by $\epsilon$. 
We also define 
$B_k\Pi C\!\equiv\! (\Pi C)^{\otimes k }$ and 
$B\Pi C \!\equiv\! \bigoplus_k B_k\Pi C $. 
We consider a family of maps $m_k$ 
$: B_k\Pi C \rightarrow \Pi C$ of degree $\epsilon\!=\!1$. 
Each $m_k$ induces a map $d_k$ 
$: B\Pi C \rightarrow B\Pi C$ by 
\begin{eqnarray}
d_k
\Bigl(
x_1,\cdots, x_n
\Bigr)
\equiv 
\sum_{p=0}^{n-k}(-)^{\sum_{j=1}^{p}\epsilon(x_j)}
x_1\otimes \cdots \otimes x_p\otimes
m_k
\Bigl(
x_{p+1}\cdots x_{p+k}
\Bigr)
\otimes x_{p+k+1}\otimes 
\cdots \otimes x_n .
\end{eqnarray}
Then we put $\hat{d} \!\equiv\! \sum_k d_k$. 
\begin{definition}[$A_{\infty}$-algebra]
\label{def of A-algebra}
$\left(C,m_k \right)$ for $k\!=\!1,2,\cdots$ is said to be 
a $A_{\infty}$-algebra if $\hat{d}\hat{d}\!=\!0$. 
$\left(C,m_k \right)$ for $k\!=\!0,1,2,\cdots$ is said to be 
a weak $A_{\infty}$-algebra if $\hat{d}\hat{d}\!=\!0$.   
\end{definition}
Let $(C,m_k)$ be a $A_{\infty}$-algebra. The condition 
$\hat{d}\hat{d}\!=\!0$ becomes equivalent to the constraints,
$d_kd_1\!+\cdots+\!d_1d_k\!=\!0$ for $k \!\geq\! 1$.  
These can be written down in terms of $m_k$ as 
\begin{eqnarray}
\sum_{l=1}^k\sum_{p=0}^{k-l}
(-)^{\sum_{i=1}^p\epsilon(x_i)}
m_{k+1-l} \Bigl(x_1, \cdots ,x_p, 
           m_l  \bigl(x_{p+1},\cdots,x_{p+l}\bigr),
            x_{p+l+1},\cdots,x_k \Bigr)=0.
\label{A-algebra}
\end{eqnarray}
It may be instructive to comment on a first few series of 
these relations. 
\begin{eqnarray}
&&
m_1\Bigl(m_1\bigl(x\bigr)\Bigr)=0 , 
\label{k=1}
 \\
&& 
m_1\Bigl(m_2\bigl(x,y\bigr)\Bigr)\!+\!
m_2\Bigl(m_1\bigl(x\bigr),y\Bigr)
\!+\!(-)^{\epsilon(x)}
m_2\Bigl(x,m_1\bigl(y\bigr)\Bigr)=0 ,
\label{k=2} 
\\ 
&&
m_2\Bigl(m_2\bigl(x,y\bigr),z\Bigr)
\!+\!(-)^{\epsilon(x)}
m_2\Bigl(x,m_2\bigl(y,z\bigr)\Bigr)
\!+\!
m_1\Bigl(m_3\bigl(x,y,z\bigr)\Bigr)
\label{k=3} \\
&&~~~
\!+\!
m_3\Bigl(m_1\bigl(x\bigr),y,z\Bigr)
\!+\!(-)^{\epsilon(x)}
m_3\Bigl(x,m_1\bigl(y\bigr),z\Bigr)
\!+\!(-)^{\epsilon(x)+\epsilon(y)}
m_3\Bigl(x,y,m_1\bigl(z\bigr)\Bigr)
=0 .
\nonumber
\end{eqnarray}
The first equation implies that 
$m_1$ is a boundary operator and the second equation 
shows that $m_1$ is a derivation with respect to $m_2$. 
The third equation is related with an associativity relation. 
When $m_3$ vanishes, $m_2$ defines an associative 
algebra on $C$ by putting 
$x \cdot y \equiv (-)^{\epsilon(x)}m_2\Bigl(x,y\Bigr)$.  
When $m_3$ does not vanish,  
the algebra ``$\cdot$'' is not associative. 
But it induces an associative algebra on the cohomology 
$H^*(\Pi C\!:\!m_1)\equiv \mbox{Ker}~m_1/\mbox{Im}~m_1$.

Apart from general theory of $A_{\infty}$-algebra, 
we concentrate on classical open-string field theory. 
The open-string Hilbert space ${\cal H}$ is naturally 
${\bf Z}$-graded by the ghost number $G$. 
We introduce a $(-1)$-shifted ghost number operator $\epsilon$ by 
$\epsilon(A)\equiv G(A)\!-\!1$.  
By using the low energy open-string vertices 
we define a family of maps $m_k$ : 
$B_k\Pi {\cal H} \rightarrow \Pi {\cal H}$  
$(k \geq 1)$ as follows. 
\begin{definition}[$m_k$ at the cut-off scale $\zeta$]
\label{def of OSFT Ainfty}
For the case of $k\!=\!1$ let $m_1(A)$ be $ Q|A\rangle$.
For the case of $k \!\geq \!2$ we set
\footnote{$[x]$ is the maximum integer not greater than $x$.} 
\begin{eqnarray}
m_{k}
\Bigl(
A_1,A_2,\cdots ,A_{k}\!:\!\zeta
\Bigr) =
(-)^{\sum_{i=1}^{[k/2]}\epsilon(A_{k+1-2i})} 
\Bll
a'~1~2 \cdots k\!:\! \zeta 
\Bigr.
\Bigl| S_{a'a} \Brr 
\bl A_1 \brr_1 \bl A_2 \brr_2 \cdots \bl A_{k}\brr_{k}.
\label{def of OSFT mk}
\end{eqnarray}
\end{definition}
Our notation in eq.(\ref{def of OSFT mk}) emphasizes that  
the maps $m_k$ for $k\! \geq \!2$ depend on the cut-off scale 
$\zeta$ through the open-string vertices. 
Each map has the $(-1)$-shifted ghost 
number equal to one. This can be seen as follows. 
Since the ghost numbers of the $(k\!+\!1)$-vertex 
and inverse reflector are respectively equal to 
 $k\!+\!1$ and three, the ghost number of 
$m_k\Bigl(A_1,\cdots,A_k\!:\!\zeta \Bigr)$ 
becomes 
$2+\sum_{i=1}^k\epsilon (A_i)$ 
$\Bigl( 
=(k\!+\!1)\!+\!3\!
+\!\sum_{i=1}^kG(A_i)\!-\!2(k\!+\!1)
\Bigr)$. 
Hence 
$\epsilon \Bigl(m_k (A_1,\cdots,A_k\!:\!\zeta )\Bigr)
\!=\!1+\sum_{i=1}^k\epsilon(A_i)$. 
This means $\epsilon(m_k)\!=\!1$. 
The following is the main result of this paper.  
\begin{theorem}
[$A_{\infty}$-algebra in classical open-string field theory]
\label{A-algebra of OSFT}
$({\cal H},m_k)$ is a $A_{\infty}$-algebra.  
Namely the maps $m_k$ $(k \!\geq\! 1)$ 
given by definition \ref{def of OSFT Ainfty} 
satisfy the infinite set of algebraic relations 
(\ref{A-algebra}).
\end{theorem}
This theorem tells that the open-string vertices have 
the structure of $A_{\infty}$-algebra which is 
independent of the cut-off scale. 
In the microscopic description at $\zeta \!=\!0$, 
all the maps $m_k$ for $k\!\geq\!3$ vanish. 
The corresponding $A_{\infty}$-algebra is the open-string 
gauge algebra $(Q,\star )$. This is a non-commutative 
associative algebra.  On the other hand, in the macroscopic description 
the higher maps do not vanish and the gauge algebra becomes non-associative. 
The scale dependence of the $A_{\infty}$-algebra 
$({\cal H},m_k)$ becomes as follows. 
\begin{proposition}[Scale dependence of the $A_{\infty}$-algebra]
\label{scale dependence of OSFT Ainfty}
Scale dependence of the maps $m_k$ (\ref{def of OSFT mk}) 
for $k \!\geq\!2$ are described by 
\begin{eqnarray}
&&
\frac{\partial m_k}{\partial \zeta} 
\Bigl(
A_1,\cdots,A_k\!:\!\zeta  
\Bigr)
\nonumber \\
&&~~=
-2\sum_{l=2}^{k-1}\sum_{p=0}^{k-l}
  m_{k+1-l}\Bigl(
         A_1,\cdots,A_p, 
               b_0m_l\bigl(
                   A_{p+1},\cdots,A_{p+l}
                      \!:\!\zeta \bigr),
         A_{p+l+1},\cdots,A_k
            \!:\!\zeta \Bigr)
\nonumber \\
&&~~~~~-
   L_0m_k \Bigl(
         A_1,\cdots,A_k
      \!:\!\zeta \Bigr)
       -
   \sum_{p=0}^{k-1}
      m_k \Bigl(
         A_1,\cdots,A_p,
           L_0A_{p+1},
         A_{p+2},\cdots,A_k
      \!:\!\zeta \Bigr).  
\label{scale dependence of OSFT mk}
\end{eqnarray}
\end{proposition}
In the deformation theory  of 
$A_{\infty}$-algebra there exists a concept of 
equivalence relation  between two $A_{\infty}$-algebras. 
This is called homotopy-equivalence \cite{Fukaya,FOOO}. 
It is interesting to see whether the $A_{\infty}$-algebra 
at $\zeta \!\neq \! 0$ is homotopy-equivalent 
to the gauge algebra $(Q,\star )$ or not. 
If it is affirmative,  
integral of eqs.(\ref{scale dependence of OSFT mk}) 
will be the $A_{\infty}$-map 
\cite{Fukaya,FOOO} which realizes the homotopy-equivalence.  
To answer this question is important on the physical ground 
since eq.(\ref{scale dependence of OSFT mk}) 
plays an important role in our description of 
renormalization group of open-string field theory. 
Resolution of the question may shed light on geometry of the space 
of boundary field theories in two-dimensions.

~

\noindent
\underline{{\it Proof of Theorem \ref{A-algebra of OSFT}}}~: 
Let us abbreviate the vector 
$\Bll a'~1~2 \cdots k\!:\!\zeta \Bigr. 
\Bl S_{a'a} \Brr 
\bl A_1 \brr_1 \bl A_2 \brr_2 \cdots \bl A_{k} \brr_{k}$ 
by 
$\widetilde{m}_{k}
\Bigl(A_1,\!A_2,\!\cdots \!,A_{k}\!:\!\zeta \Bigr)$. 
We first rewrite $Q~\widetilde{m}_{k}$ $(k\!\geq \!2)$ as follows. 
\begin{eqnarray} 
&&
Q~ 
\widetilde{m}_{k}
\Bigl(A_1,A_2,\cdots ,A_{k}\!:\!\zeta \Bigr) 
\nonumber \\
&&=(-)^{k+1}
\Bll a'~1~2 \cdots k\!:\! \zeta \Br
\left(Q^{(a)} \Bl S_{a'a} \Brr \right) 
\bl A_1 \brr_1 \bl A_2 \brr_2 \cdots \bl A_{k} \brr_{k}
\nonumber \\
&&=(-)^k 
\Bll a'~1~2 \cdots k\!:\! \zeta \Bigr. 
\Bl S_{a'a} \Brr   
\left\{ \Bigl(\sum_{j=1}^{k}Q^{(j)}\Bigr) 
\bl A_1 \brr_1 \bl A_2 \brr_2 \cdots \bl A_{k} \brr_{k} \right\}
\label{proof Ainfty of OSFT 1} 
\\
&&~~
+(-)^k
\left\{ 
\Bll a'~1~2 \cdots k\!:\! \zeta \Br
\Bigl( Q^{(a')}+ \sum_{j=1}^{k}Q^{(j)} \Bigr) \right\}
\Bl S_{a'a} \Brr  
\bl A_1 \brr_1 \bl A_2 \brr_2 \cdots \bl A_{k} \brr_{k}
\label{proof Ainfty of OSFT 2},
\end{eqnarray}
where the BRST invariance of the inverse reflector 
besides its odd-grassmannity are used to show the second equality. 
We treat two terms 
(\ref{proof Ainfty of OSFT 1}) and (\ref{proof Ainfty of OSFT 2}) 
separately. 
The first term is written down easily to a form expressed in terms 
of $\widetilde{m}_k$.   
\begin{eqnarray}
&&
Eq.(\ref{proof Ainfty of OSFT 1}) 
\nonumber \\
&&=
(-)^k\sum_{p=0}^{k-1}(-)^{\sum_{i=1}^{p}G(A_i)}
\Bll a'~1 \cdots k \!:\! \zeta \Bigr. 
\Bl S_{a'a} \Brr   
\bl A_1 \brr_1 \cdots \bl A_p \brr_p 
\left(Q^{(p+1)}\bl A_{p+1} \brr_{p+1} \right)
\bl A_{p+2} \brr_{p+2} \cdots \bl A_{k} \brr_{k}
\nonumber \\
&&=
(-)^k\sum_{p=0}^{k-1}(-)^{p+\sum_{i=1}^{p}\epsilon(A_i)}
\widetilde{m}_k
\Bigl(
A_1,\cdots,A_p,QA_{p+1},A_{p+2},\cdots,A_k
\Bigr). 
\label{proof scale dependence of mk 1}
\end{eqnarray}
On the other hand we need some care to evaluate the second term. 
We compute the second term using the conjecture. 
By eq.(\ref{conjecture 1}) the second term becomes   
\begin{eqnarray}
&&
Eq.(\ref{proof Ainfty of OSFT 2}) 
\nonumber \\
&&=
(-)^k
\left\{ 
\Bll 0~1~2 \cdots k\!:\! \zeta \Br
\Bigl(\sum_{j=0}^{k}Q^{(j)}\Bigr)
\right\}
\Bl S_{0 a} \Brr  
\bl A_1 \brr_1 \bl A_2 \brr_2 \cdots \bl A_{k}\brr_{k}
\nonumber \\
&&=
(-)^{k+1}
\left\{
\sum_{l=2}^{k-1}\sum_{p=0}^{k-l}(-)^{kp}
\Bll 
\underbrace
{p\!+\!l\!+\!1~ \cdots~ k~0~1~\cdots~ p~b}_{k+2-l}
:\!\zeta \Br 
\Bll 
\underbrace
{b'~p\!+\!1~\cdots~p\!+\!l}_{l+1}
:\!\zeta \Br
\Bigl. S_{b'b}\Brr 
\right\}  
\nonumber \\ 
&&~~~~~~
\times \Bl S_{0a}\Brr \bl A_1\brr_1 \cdots \bl A_k \brr_k .
\label{proof scale dependence mk 1.2}
\end{eqnarray}
We then permute open-string indices of the first vertices 
according to eq.(\ref{conjecture 2}) as follows.
\begin{eqnarray}
&&
Eq.(\ref{proof scale dependence mk 1.2})
\nonumber \\ 
&&=
(-)^{k+1}
\left\{
\sum_{l=2}^{k-1}\sum_{p=0}^{k-l}
(-)^{kp+(k-l+1)p}
\Bll 
\underbrace{0~1~\cdots~p~b~p\!+\!l\!+\!1~\cdots~k}_{k+2-l}
:\!\zeta \Br
\Bll 
\underbrace{b'~p\!+\!1~\cdots~p\!+\!l}_{l+1}
:\!\zeta \Br
\Bigl. S_{b'b}\Brr 
\right\}
\nonumber \\
&&~~~~~~
\times 
\Bl S_{0a} \Brr \bl A_1\brr_1 \cdots \bl A_k\brr_k. 
\label{proof scale dependence of mk 1.5}
\end{eqnarray}
Finally we arrange eq.(\ref{proof scale dependence of mk 1.5}), 
taking account of the grassmannities, so that it is  
expressed in terms of $\widetilde{m}_k$. 
\begin{eqnarray} 
&&
Eq.(\ref{proof scale dependence of mk 1.5})
\nonumber \\
&&=
(-)^{k+1}
\sum_{l=2}^{k-1}\sum_{p=0}^{k-l}
(-)^{p+l\left(1+\sum_{i=1}^p \epsilon(A_i)\right)}
\Bll 
\underbrace{0~1~\cdots~p~b~p\!+\!l\!+\!1~\cdots~k}_{k+2-l}
:\!\zeta \Br
\Bigl. S_{0a}\Brr  
\bl A_1\brr_1 \cdots \bl A_{p}\brr_p 
\nonumber \\
&&~~~~~~
\times 
\Bll 
\underbrace{b'~p\!+\!1~\cdots~p\!+\!l}_{l+1}
:\!\zeta \Bigr. 
\Bl S_{b'b}\Brr 
\bl A_{p+1}\brr_{p+1} \cdots \bl A_{p+l} \brr_{p+l} 
\nonumber \\
&&~~~~~~~~~~~~
\times 
\bl A_{p+l+1}\brr_{p+l+1}\cdots \bl A_k \brr_k
\nonumber \\
&&=
(-)^{k+1}\sum_{l=2}^{k-1}\sum_{p=0}^{k-l}
(-)^{p+l\left(1+\sum_{i=1}^p \epsilon(A_i)\right)} 
\widetilde{m}_{k+1-l}
\Bigl(A_1,\cdots,A_p,
\widetilde{m}_l
\bigl(A_{p+1},\cdots,A_{p+l} \bigr),
A_{p+l+1},\cdots,A_k \Bigr). 
\nonumber \\
\label{proof scale dependence of mk 2}
\end{eqnarray}

The sum of eqs.(\ref{proof Ainfty of OSFT 1}) 
and (\ref{proof Ainfty of OSFT 2}) is 
$Q\widetilde{m}_k\Bigl(A_1,\cdots,A_k\Bigr)$.   
Using the expressions  
(\ref{proof scale dependence of mk 1}) and 
(\ref{proof scale dependence of mk 2}) 
we obtain the following relation of $\widetilde{m}_k$.  
\begin{eqnarray}
&&
Q~\widetilde{m}_{k}
\Bigl(
A_1,\!A_2,\!\cdots \!,A_{k}
\Bigr) 
\!+\!
\sum_{p=0}^{k-1}(-)^{k+p+1+\sum_{i=1}^{p}\epsilon(A_i)}
\widetilde{m}_k
\Bigl(
A_1,\cdots,QA_{p+1},\cdots,A_k 
\Bigr) 
\nonumber \\
&&
+\!
\sum_{l=2}^{k-1}\sum_{p=0}^{k-l}
(-)^{k+p+l\left(1+\sum_{i=1}^p \epsilon(A_i)\right)} 
\widetilde{m}_{k+1-l}
\Bigl(
A_1,\cdots,A_p,
  \widetilde{m}_l
      \bigl(A_{p+1},\cdots,A_{p+l} \bigr),
A_{p+l+1},\cdots,A_k 
\Bigr)
\nonumber \\
&&
=0. 
\label{A-algebra tilde}
\end{eqnarray}
This relation can be rewritten 
in terms of $m_k$ if one recalls the correspondence,  
\begin{eqnarray} 
\widetilde{m}_{k}
\Bigl(
A_1,\!A_2,\!\cdots \!,A_{k}\!:\!\zeta
\Bigr)
=(-)^{\sum_{i=1}^{[k/2]}\epsilon(A_{k+1-2i})}
{m}_{k}
\Bigl(
A_1,\!A_2,\!\cdots \!,A_{k}\!:\!\zeta  
\Bigr).
\nonumber 
\end{eqnarray} 
After straightforward calculations  
we finally find out that eq.(\ref{A-algebra tilde}) is 
rephrased to  
\begin{eqnarray}
\sum_{l=1}^{k}\sum_{p=0}^{k-l}
(-)^{\sum_{i=1}^{p}\epsilon(A_i)}
m_{k+1-l}
\Bigl(A_1, \cdots ,A_p, 
m_l
\bigl(
A_{p+1},\cdots ,A_{p+l}\!:\!\zeta  
\bigr),
A_{p+l+1},\cdots ,A_{k}\!:\!\zeta  
\Bigr)
=0.
\nonumber  
\end{eqnarray} 
This is nothing but the relation (\ref{A-algebra}). 
We omit the last part of the proof since the calculations, 
though they are straightforward, are needed to 
be done case-by-case and require some spaces.\P

~

\noindent
{\it \underline{Proof of Proposition \ref{scale dependence of OSFT Ainfty}}}~:
(We use the same notation as in the proof of 
Theorem \ref{A-algebra of OSFT}.) 
We first compute 
$\partial \widetilde{m}_k/\partial \zeta$ 
by using Proposition \ref{scale dependence of n-vertex}. 
\begin{eqnarray}
&&
\frac{\partial \widetilde{m}_k}{\partial \zeta}
\Bigl(
A_1,A_2,\cdots,A_{k}\!:\Lambda
\Bigr)
\nonumber \\
&&=
\left(
\frac{d}{d \zeta}
\Bll 0~1~2~\cdots~k\!:\!\zeta \Br
\right)\Bl S_{0 a} \Brr  
\bl A_1\brr_1 \bl A_2\brr_2 \cdots \bl A_{k}\brr_{k}
\nonumber \\
&&=-2 
\left(
\sum_{l=2}^{k-1}\sum_{p=0}^{k-l}(-)^{kp}
\Bll 
\underbrace{p\!+\!l\!+\!1~\cdots~k~0~1~\cdots~p~b}_{k+2-l}
\!:\!\zeta \Br
\Bll 
\underbrace{b'~p\!+\!1~\cdots~p\!+\!l}_{l+1}
\!:\!\zeta \Br
\Bigl( 
b_0 \Bl S_{b'b}\Brr 
\Bigr) 
\right) 
\label{proof dependence of mk 1} \\ 
&&~~~~~~
\times \Bl S_{0a}\Brr 
\bl A_1\brr_1 \cdots \bl A_k\brr_k  
\nonumber \\
&&~~-
\Bll 0~1 \cdots k\!:\!\zeta \Br
\left(
\Bigl(\sum_{i=0}^kL_0^{(i)} \Bigr)
\Bl S_{0a}\Brr 
\bl A_1\brr_1 \cdots \bl A_k\brr_k 
\right).  
\label{proof dependence of mk 2}
\end{eqnarray}
We arrange the first term 
(\ref{proof dependence of mk 1}) 
so that it is expressed in terms of $\widetilde{m}_k$. 
For this purpose we permute open-string indices of the first 
vertices by using the asymmetry (\ref{conjecture 2}) as follows.
\begin{eqnarray}
&&
Eq.(\ref{proof dependence of mk 1})
\nonumber \\
&&=-2 
\left(
\sum_{l=2}^{k-1}\sum_{p=0}^{k-l}
(-)^{kp+(k-l+1)p}
\Bll 
\underbrace{0~1~\cdots~p~b~p\!+\!l\!+\!1~\cdots~k}_{k+2-l}
\!:\!\zeta \Br
\Bll 
\underbrace{b'~p\!+\!1~\cdots~p\!+\!l}_{l+1}
\!:\!\zeta \Br 
\Bigl(
b_0 \Bl S_{b'b} \Brr 
\Bigr) 
\right)
\nonumber \\
&&~~~~~~~~~
\times \Bl S_{0a}\Brr 
\bl A_1\brr_1 \cdots \bl A_k\brr_k. 
\label{proof dependence of mk 2.5}
\end{eqnarray}
Then we rewrite eq.(\ref{proof dependence of mk 2.5}), 
taking account of the grassmannities of $|A_i\rangle$, 
into the following form. 
\begin{eqnarray} 
&&
Eq.(\ref{proof dependence of mk 2.5})
\nonumber \\ 
&&=-2
\sum_{l=2}^{k-1}\sum_{p=0}^{k-l}
(-)^{(l+1) \sum_{i=1}^p\epsilon(A_i)}
\Bll 
\underbrace{0~1~\cdots~p~b~p\!+\!l\!+\!1~\cdots~k}_{k+2-l}
\!:\!\zeta \Br
\Bigl. S_{0a}\Brr  
\bl A_1\brr_1 \cdots \bl A_{p}\brr_p 
\nonumber \\
&&~~~~~~~~~~~~~~~~~~
\times 
\Bigl(b_0^{(b)}
\Bll 
\underbrace{b'~p\!+\!1~\cdots~p\!+\!l}_{l+1}
\!:\!\zeta \Bigr.
\Bl S_{b'b}\Brr 
\bl A_{p+1}\brr_{p+1} \cdots \bl A_{p+l} \brr_{p+l}\Bigr) 
\nonumber \\
&&~~~~~~~~~~~~~~~~~~~~~~~~~~
\times 
\bl A_{p+l+1}\brr_{p+l+1}\cdots \bl A_k \brr_k
\nonumber \\
&&=-2 
\sum_{l=2}^{k-1}\sum_{p=0}^{k-l}
(-)^{(l+1)\sum_{i=1}^{p}\epsilon(A_i)} 
\widetilde{m}_{k+1-l}
\Bigl(A_1,\cdots,A_p,
b_0 \widetilde{m}_l
\bigl(A_{p+1},\cdots,A_{p+l} \bigr),
A_{p+l+1},\cdots,A_k \Bigr).
\nonumber \\
\label{proof dependence of mk 3}
\end{eqnarray}
While this, we can easily write down  
the second term (\ref{proof dependence of mk 2}) 
by means of $\widetilde{m}_k$ as follows. 
\begin{eqnarray}
Eq.(\ref{proof dependence of mk 2})
&=& 
-L_0
\widetilde{m}_{k}\Bigl(A_1,\!A_2,\!\cdots \!,A_{k}\Bigr)
-
\sum_{i=1}^k
\widetilde{m}_{k}\Bigl(A_1,\!\cdots,\!L_0A_i,\! 
\cdots, \!A_{k}\Bigr).  
\label{proof dependence of mk 4}
\end{eqnarray}

Eqs.(\ref{proof dependence of mk 3}) 
and 
(\ref{proof dependence of mk 4}) give  
the following expression of 
$\partial \widetilde{m}_k/\partial \zeta$ 
in terms of $\widetilde{m}_k$. 
\begin{eqnarray}
&&
\frac{\partial \widetilde{m}_k}{\partial \zeta}
\Bigl(A_1,\!A_2,\!\cdots \!,A_{k}\Bigr)
\nonumber \\
&&~~=-2
\sum_{l=2}^{k-1}\sum_{p=0}^{k-l}
(-)^{(l+1)\sum_{i=1}^{p}\epsilon(A_i)} 
\widetilde{m}_{k+1-l}
\Bigl(A_1,\cdots,A_p,
b_0 \widetilde{m}_l
\bigl(A_{p+1},\cdots,A_{p+l} \bigr),
A_{p+l+1},\cdots,A_k \Bigr)
\nonumber \\
&&~~~~~~
-L_0
\widetilde{m}_{k}
\Bigl(A_1,\!A_2,\!\cdots \!,A_{k}\Bigr)
-
\sum_{i=1}^k
\widetilde{m}_{k}
\Bigl(A_1,\!\cdots,\!L_0A_i,\! \cdots, \!A_{k}\Bigr)~~.  
\label{proof dependence of mk 5}
\end{eqnarray}
This describes the scale dependence of $\widetilde{m}_k$. 
By rewriting this equation in terms of $m_k$, 
we find out, after some calculations, 
that it becomes eq.(\ref{scale dependence of OSFT mk}).\P

~

The maps $m_k$ are introduced  
by the open-string vertices.   
Conversely the open-string vertices 
can be written in terms of $m_k$ as follows.   
\begin{proposition}
\label{open-string vertex by mk}
Let $k \!\geq\! 3$. We have the following identities. 
\begin{eqnarray}
\Bll 1 \cdots k\!:\!\zeta \Br
\bigl. A_1\brr_1 \cdots
\bl A_{k}\brr_{k}
&=&
(-)^{\sum_{i=1}^{[k/2]}\epsilon(A_{k+1-2i})}
\Bll \omega_{a b}\Br 
\bigl. A_1\brr_{a}
\Bl m_{k-1}\Bigl(A_2,\cdots,A_k\!:\!\zeta \Bigr)\Brr_b
\label{k-vertex by m(k-1) 1} \\
~~
\nonumber
\\
&=&
(-)^{\sum_{i=1}^{[k/2]}\epsilon(A_{k-2i})}
\Bll \omega_{a b} \Bl
\Bigl. m_{k-1}\Bigl(A_1,\cdots,A_{k-1}\!:\!\zeta \Bigr)\Brr_a 
\bl A_{k}\brr_b .
\label{k-vertex by m(k-1)}
\end{eqnarray}
\end{proposition}
\begin{remark}
This proposition becomes useful in the subsequent 
discussions. 
\end{remark}

~

\noindent
\underline{{\it Proof of Proposition \ref{open-string vertex by mk}}}~: 
We give a proof only for the case when $k$ is odd. 
The other case can be shown in the same manner.  
We rewrite 
$\Bll \omega_{a b}\Br \bigl.A_1\brr_{a}
\Bl m_{2p}\Bigl(A_2,\cdots,A_{2p+1}\Bigr)\Brr_b$ in terms of the 
open-string vertices as follows.  
\begin{eqnarray}
&& 
\Bll \omega_{a b}\Bl \bigl.A_1\brr_{a}
\Bl m_{2p}\Bigl(A_2,\cdots,A_{2p+1}\!:\!\zeta\Bigr)\Brr_b 
\nonumber \\
&&=
\Bll \omega_{ab}\Br \bigl.A_1\brr_1
\left( 
(-)^{\sum_{i=1}^p\epsilon(A_{2i})} 
\Bll 0~2~\cdots~2p\!+\!1\!:\!\zeta \Br
\Bigl. S_{0b}\Brr  
\bl A_2\brr_2\cdots \bl A_{2p+1}\brr_{2p+1}
\right)
\nonumber \\ 
&&=
(-)^{1+\sum_{i=1}^p\epsilon(A_{2i})}
\Bll 0~2~\cdots~2p\!+\!1\!:\!\zeta  \Br 
\Bll \omega_{ab}\Br \Bigl.S_{0b}\Brr  
\bl A_1\brr_a \bl A_2\brr_2\cdots
\bl A_{2p+1}\brr_{2p+1},
\label{proof v by m 1}
\end{eqnarray}
where we use odd grassmannities of the $(2p+1)$-vertex and 
the (inverse) reflector to show the second equality.  
Since we know 
$\Bll \omega_{12}\Br \Bigl.S_{23}\Brr \!=\!_3P_1 $, 
eq.(\ref{proof v by m 1}) can be evaluated into 
\begin{eqnarray} 
Eq.(\ref{proof v by m 1})
&=&
(-)^{1+\sum_{i=1}^p\epsilon(A_{2i})}
\Bll 0~2~\cdots~2p\!+\!1\!:\!\zeta \Br 
\Bigl(-_0P_a\Bigr)  
|A_1\rangle_a|A_2\rangle_2\cdots
|A_{2p+1}\rangle_{2p+1}
\nonumber \\
&=&
(-)^{\sum_{i=1}^p\epsilon(A_{2i})}
\Bll 1~2~\cdots~2p\!+\!1\!:\!\zeta \Br  
A_1\rangle_1|A_2\rangle_2\cdots
|A_{2p+1}\rangle_{2p+1}.
\label{proof v by m 2}
\end{eqnarray}
Hence we obtain eq.(\ref{k-vertex by m(k-1) 1}) with  $k\!=\!2p+1$.  
Similarly, by rewriting 
$\Bll \omega_{a b}\Bigr. 
\Bl m_{2p}\Bigl(A_1,\cdots,A_{2p}\Bigr)\Brr_a
|A_{2p+1}\rangle_{2p+1}$ 
in terms of the open-string vertices, 
we obtain eq.(\ref{k-vertex by m(k-1)}) 
with $k\!=\!2p+1$.\P    


\section{Low Energy Description Of Open-String Field Theory}
We use the Batalin-Vilkovisky (BV) formalism 
\cite{BV formalism}, \cite{BV formalism 2} 
for a low energy description of open-string field theory. 
This formalism was elegantly used in quantizing open-string 
field theory \cite{Thorn, Bochicchio}. 
Although the BV formalism was originally intended to quantize 
gauge invariant field theories which we already knew, 
it was used in closed-string field theory \cite{Zwiebach BV}
to find the unknown theory.

\subsection{Odd symplectic structure of open-string field theory}
In order to develop the BV formalism 
of open-string field theory we need to introduce 
an odd symplectic structure $\omega$ on 
the open-string Hilbert space ${\cal H}$.  
We define $\omega$ as follows. 
\begin{definition}[Odd symplectic structure of ${\cal H}$]
\label{odd symplectic structure of OSFT}
Let $\omega$ be an element of $({\cal H}^{\otimes 2})^*$ 
which is given by 
\begin{eqnarray}
\omega(A,B)=
(-)^{\epsilon(A)}\Bll \omega_{12} \Br
A \rangle_1 |B \rangle_2,
\label{odd symplectic structure}
\end{eqnarray}
where $\epsilon(A)$ is the $(-1)$-shifted ghost number of $A$. 
\end{definition}
The above bilinear form is 
non-degenerate on ${\cal H}$ 
because of the non-degeneracy of the BPZ pairing. 
The symmetry of the reflector implies 
$\omega(A,B)\!=\!-(-)^{\epsilon(A)\epsilon(B)}\omega(B,A)$. 
The selection rule of the BPZ pairing gives  
$\omega(A,B)\!\neq\! 0 \Rightarrow 
\epsilon(A)\!+\!\epsilon(B)\!=\!1$.

Let $\Bigl\{ \phi_a \Bigr\}_a$ be bases of ${\cal H}$. 
We take the conjugate bases  
$\Bigl\{\phi^a\Bigr\}_a$ 
such that they satisfy 
$\omega(\phi_a,\phi^b)\!=\!\delta_a^b$. 
The $(-1)$-shifted ghost number of $\phi^a$ is 
$\epsilon(\phi^a)\!=\!1\!-\!\epsilon(\phi_a)$. 
We put 
$\omega_{ab}\equiv \omega(\phi_a,\phi_b)$
and 
$\omega^{ab}\equiv \omega(\phi^a,\phi^b)$.
These matrix elements enjoy 
\begin{eqnarray}
\sum_c\omega^{ac}\omega_{cb}
=\sum_c\omega_{bc}\omega^{ca}=-\delta^a_b. 
\label{inverse of omega(ab)}
\end{eqnarray}
This may need a proof : 
It can be seen that these two bases are related with each other by 
$\phi_a\!=\!\sum_{b}\phi^b \omega_{ba}$ 
or equivalently 
$\phi^a\!=\!-\sum_b\phi_b\omega^{ba}$.
By using these relations we obtain 
$\phi_a\!=\!\sum_b\phi_b
\left(\sum_c(-)
\omega^{bc}\omega_{ca}
\right)$ 
and 
$\phi^a\!=\!
\sum_b\phi^b
\left(\sum_c(-)
\omega_{bc}\omega^{ca}
\right)$. 
These imply eq.(\ref{inverse of omega(ab)}).

Open-string field $\Phi$ is a vector of ${\cal H}$ with 
$\epsilon(\Phi)\!=\!0$. 
Hence $\Phi$ is restricted on a subspace of ${\cal H}$. 
In the BV formalism this restriction is removed 
\cite{Thorn, Bochicchio}
by introduction of fields and anti-fields.  
Let us expand open-string field by the bases $\phi_a$,  
\begin{eqnarray}
\Phi=\sum_at^a\phi_a.
\end{eqnarray}
Each coefficient $t^a$ is required to have the ghost number 
$G(t^a)\!=\!-\epsilon(\phi_a)$ 
and the grassmannity 
$(-)^{G(t^a)}$ 
so that $\Phi$ is still grassmann-odd with 
$\epsilon(\Phi)\!=\!0$. 
We can regard these coefficients $t^a$ 
as coordinates of a super-manifold 
(infinite-dimensional super-vector space).  
This manifold is endowed with 
the odd-symplectic structure $\omega$ if one 
identifies the open-string Hilbert space 
${\cal H}$ with its tangent space. 
Let $F(\Phi)$ be a functional of open-string field $\Phi$. 
We introduce hamiltonian vector 
$\Bigl|\partial F\!/\partial \Phi\Bigr\rangle$ 
$\in {\cal H}$ as follows.
\begin{definition}[Hamiltonian vector]
\label{hamilton vector}
Hamiltonian vector 
$\Bigl|\partial F\!/\partial \Phi \Bigr\rangle$ of a functional 
$F(\Phi)$ is a vector of ${\cal H}$.  
It is defined by the following first-order variation of $F$ 
with respect to $\Phi$. 
\begin{eqnarray}
\delta F(\Phi)
=
\omega \left( \delta \Phi, 
\left|\frac{\partial F}{\partial \Phi}\right \rangle \right),
\label{def of hamilton vector}
\end{eqnarray}
where $\delta \Phi$ is arbitrary vector 
with $\epsilon(\delta \Phi)\!=\!0$. 
\end{definition}
The ghost number of hamiltonian vector becomes 
$G\left(\Bl \partial F\!/\partial \Phi \Brr \right)\!=\!G(F)\!+\!2$. 
Expansion of the vector in terms of $\phi^a$ turns out to 
have the forms,  
\begin{eqnarray}
\left| 
\frac{\partial F}{\partial \Phi}
\right\rangle
=\sum_a|\phi^a\rangle  \partial_a^L F  
~~~~~~
\left(
=\sum_a F \partial_a^R |\phi^a \rangle
\right).
\label{hamilton vector in component}
\end{eqnarray}
We regard $F$ as a function of $t^a$ 
in the RHS of the above equations. 
We also introduce left- and right-differentials  
$\partial_a^{L} \!\equiv\! \partial^{L}/ \partial t^a$ 
and 
$\partial_a^{R}\!\equiv\! \partial^{R}/ \partial t^a$. 
$\partial^L_aF$ and $F\partial^R_a$ 
are defined by the first-order variation of $F$ with respect to $t^a$. 
\begin{eqnarray}
\delta F 
= 
\delta t^a \left(\partial^L_aF \right)  
~~~~~~~~
\left(
= 
\left(F\partial^R_a \right)\delta t^a 
\right).
\end{eqnarray}  
Let us derive eqs.(\ref{hamilton vector in component}).  
We put $\delta \Phi \!=\! \sum_a \delta t^a \phi_a$ 
and write down eq.(\ref{def of hamilton vector}) as follows.  
\begin{eqnarray}
\delta F = 
\sum_a \delta t^a 
\omega \left(\phi_a, 
       \left|\frac{\partial F}{\partial \Phi}\right \rangle 
        \right)
~~~~~~~~
\left(= 
-\sum_a \omega \left( 
  \left| \frac{\partial F}{\partial \Phi}\right \rangle, 
  \phi_a \right) \delta t^a 
\right). 
\end{eqnarray}
These expressions mean   
\begin{eqnarray} 
\partial_a^LF \!=\!
\omega \left( 
        \phi_a, 
   \left| \frac{\partial F}{\partial \Phi} \right  \rangle 
       \right ),~~~~~~~
F\partial^R_a\!=\! 
-\omega \left( 
   \left| \frac{\partial F}{\partial \Phi} \right \rangle 
   \phi_a 
        \right).
\label{super-derivative of F}
\end{eqnarray}
While this, any vector $A \in {\cal H}$ can be 
expanded in the following forms.
\begin{eqnarray}  
|A \rangle 
=
\sum_a | \phi^a \rangle \omega ( \phi_a ,A)
~~~~~~\left(=
-\sum_a \omega(A,\phi_a)|\phi^a \rangle 
\right).
\label{expansion of vector}
\end{eqnarray}
Let $A$ be 
$\Bigl| \partial F/ \partial \Phi \Bigr \rangle$ 
and compare eqs.(\ref{expansion of vector}) 
with eqs.(\ref{super-derivative of F}).   
Then we obtain eqs.(\ref{hamilton vector in component}). 
\begin{definition}[Anti-bracket $\left\{~,~\right\}$]
\label{anti-bracket}
Let $F$ and $K$ be functionals of open-string field $\Phi$. 
Their anti-bracket $\left\{F,K\right\}$ is defined by 
\begin{eqnarray}
\Bigl\{ F,K\Bigr\}= 
\omega \left(
  \left|\frac{\partial F}{\partial \Phi} \right \rangle,
  \left|\frac{\partial K}{\partial \Phi} \right \rangle 
        \right).
\label{def anti-bracket}
\end{eqnarray}
\end{definition}
\begin{remark}
By eqs.(\ref{hamilton vector in component}) 
the anti-bracket (\ref{def anti-bracket}) has the following 
expression.  
\begin{eqnarray}
\Bigl\{ F,K\Bigr\}=
\sum_{a,b}
F\partial^R_a\omega^{ab}\partial^L_bK.
\label{antibracket in component}
\end{eqnarray}
\end{remark}
\begin{proposition}
\label{properties of anti-bracket}
The anti-bracket $\left\{~,~\right\}$ 
defined by eq.(\ref{def anti-bracket}) 
enjoys the following properties.      
\begin{eqnarray}
&& 
G(\left\{F,K\right\})=G(F)+G(K)+1 ,
\label{G of anti-bracket}
\\
&& 
\bigl\{F,K \bigr\}=-(-)^{\epsilon(F)\epsilon(K)}
\bigl\{K,F \bigr\} ,
\label{symmetry of anti-bracket}
\\
&&
\Bigl\{F,\bigl\{K,M \bigr\}\Bigr\}
=\bigl\{F,K \bigr\}M+
(-)^{\epsilon(F)(\epsilon(K)+1)}K
\bigl\{F,M \bigr\},
\label{derivation of anti-bracket}
\\
&&
0=(-)^{\epsilon(F)\epsilon(M)}
\Bigl\{ \bigl\{F,K \bigr\},M \Bigr\}
+ \mbox{cyclic permutations}.
\label{Jacobi identity}
\end{eqnarray}
\end{proposition}
Eq.(\ref{derivation of anti-bracket}) means that 
the anti-bracket is a derivation with respect to itself. 
Eq.(\ref{Jacobi identity}) is the Jacobi identity.   
Thus the anti-bracket defines a super Lie algebra. 
This proposition can be shown by standard calculations and 
we omit the proof.

\subsection{Low energy action $S[\Phi\!:\!\zeta]$}
Low energy action $S_{eff}^{\zeta}[\Phi]$ 
of open-string field $\Phi$ can be obtained by 
integrating out all the contributions 
from length scale less than $\zeta$.  
It includes higher open-string interactions. 
These interactions are obtained from graphs which are 
one-particle irreducible with respect to the regularized 
propagator. We denote the classical part 
of $S_{eff}^{\zeta}[\Phi]$ by $S[\Phi\!:\!\zeta]$. 
This gives a low energy description of 
classical open string field theory. 
All the open-string vertices investigated previously 
become the elementary interactions 
at the cut-off scale $\zeta$ and contribute to $S[\Phi\!:\!\zeta]$. 
Thus we arrive at the following definition of $S[\Phi\!:\!\zeta]$. 
\begin{definition}
[Low energy action of classical open-string field theory]
Action of classical open-string field at the cut-off scale $\zeta$ 
is given by 
\begin{eqnarray}
S\Bigl[\Phi:\zeta \Bigr]=
\frac{1}{2}\Bll \omega_{12}\Br 
\Phi \rangle_1 
\left(Q^{(2)}|\Phi \rangle_2\right) 
+\sum_{k \geq 3}\frac{1}{k}
\Bll 1~2~\cdots~k:\zeta \Br 
\Phi \rangle_1|\Phi \rangle_2 \cdots |\Phi\rangle_k.
\label{def of low energy action}
\end{eqnarray}
\end{definition}
This action reduces to the microscopic action (\ref{microscopic action}) 
as $\zeta\!\rightarrow\!0$. 
In this subsection we examine the low energy description 
from the perspective of the BV formalism. 
For this purpose it is convenient to provide another form of the action  
by which the underlying BV structure becomes manifest. 
We rewrite the interactions in (\ref{def of low energy action}) 
by using Proposition \ref{open-string vertex by mk}.  
\begin{eqnarray}
\sum_{k \geq 3}\frac{1}{k}
\Bll 1~2 \cdots k\!:\!\zeta \Br 
\Phi \rangle_1|\Phi \rangle_2 \cdots |\Phi\rangle_k
&=& 
\sum_{k \geq 2}\frac{1}{k\!+\!1}
\Bll \omega_{12}\Br 
\Phi \rangle_1
\Bl  m_{k}\Bigl(\Phi^{k}\!:\!\zeta\Bigr)\Brr_2 
\nonumber \\
&=& 
\sum_{k \geq 2}\frac{1}{k\!+\!1}
\omega \left(\Phi,m_{k}\Bigl(\Phi^k\!:\!\zeta \Bigr)\right), 
\end{eqnarray}
where $m_k\Bigl(\Phi^k\!:\!\zeta \Bigr)$ is the abbreviation for 
$\textstyle{m_k\Bigl(\underbrace{\Phi,\cdots,\Phi}_{k}\!:\!\zeta \Bigr)}$. 
We will use similar abbreviation frequently. 
For instance,  
$m_{k_1+k_2+1}\Bigl(\Phi^{k_1},A,\Phi^{k_2}\!:\!\zeta \Bigr)$ 
stands for 
$m_{k_1+k_2+1}\Bigl(
\underbrace{\Phi,\cdots,\Phi}_{k_1},$
$A,\underbrace{\Phi,\cdots,\Phi}_{k_2}\!:\!\zeta \Bigr)$. 
The quadratic part of the action is just 
$\textstyle{\frac{1}{2}\omega \left(\Phi,m_1\Bigl(\Phi\Bigr)\right)}$. 
Thus we obtain another expression for $S[\Phi \!:\!\zeta]$, 
\begin{eqnarray}
S\Bigl[\Phi\!:\!\zeta \Bigr]=
\omega \left( \Phi,
\sum_{k \geq 1} 
\frac{1}{k\!+\!1}m_k 
      \Bigl(
         \Phi^k\!:\zeta 
      \Bigr)
      \right).
\label{low energy action}
\end{eqnarray}
\begin{proposition}[Equation of motion for $\Phi$]
\label{prop variation formula}
First-order variation of the action (\ref{def of low energy action}) 
with respect to $\Phi$ becomes as follows. 
\begin{eqnarray}
\delta S\Bigl[\Phi\!:\!\zeta \Bigr]=
\omega \left( \delta \Phi,~ 
   \sum_{k \geq 1}m_k\Bigl(\Phi^k\!:\!\zeta \Bigr) 
       \right).
\label{variation formula}
\end{eqnarray}
Therefore the equation of motion $\delta S\!=\!0$ is given by  
\begin{eqnarray}
\sum_{k \geq 1}m_k\Bigl(\Phi^k\!:\!\zeta \Bigr)=0.
\label{eq of motion}
\end{eqnarray}
\end{proposition}
We give a few comments on some relation with deformation theory 
\cite{Fukaya,FOOO} of $A_{\infty}$-algebra :  
Let $(C,m_k)$ be a $A_{\infty}$-algebra. 
Let $b \in  (\Pi C)^0$. 
Define a family of maps $m_k^b$ ($k\!=\!0,1,\cdots$) by     
\begin{eqnarray}
m_k^{b}\Bigl(x_1,\cdots,x_k \Bigr)=  
\displaystyle{\sum_{q_1,\cdots,q_{k+1}\geq 0}}
m_{k+q_1+\cdots+q_{k+1}}
\Bigl(
\underbrace{b,\!\cdots,\!b}_{q_1},
x_1,
\underbrace{b,\!\cdots,\!b}_{q_2},
\cdots,
\underbrace{b,\!\cdots,\!b}_{q_k},
x_k,
\underbrace{b,\!\cdots,\!b}_{q_{k+1}}
\Bigr).
\end{eqnarray}
It was shown in \cite{Fukaya,FOOO} that 
$(C,m_k^b)$ for $k=0,1,\cdots$ becomes 
a {\it weak} $A_{\infty}$-algebra. 
If $m_0^b(1)$ vanishes, it becomes a $A_{\infty}$-algebra.
So $m_0^b(1)$ is an obstruction in the deformation theory. 
Let us consider the particular case of open-string field theory. 
For any $\Phi \in (\Pi{\cal H})^0$ we obtain a 
{\it weak} $A_{\infty}$-algebra 
$({\cal H}, m_k^{\Phi})$. 
The obstruction is $m_0^{\Phi}\Bigl(1:\zeta \Bigr)$. 
Now we become aware that 
the equation of motion (\ref{eq of motion}) states  
vanishing the obstruction. 
For any classical solution $\Phi_{\sharp}$, 
even apart from the trivial one,  
we still have a $A_{\infty}$-algebra 
$({\cal H},m_k^{\Phi_{\sharp}})$.
In the BV formalism variational formula 
(\ref{variation formula}) determines 
hamiltonian vector of the action $S$. 
We also recognize that the hamiltonian vector 
is nothing but the obstruction of deformation. 
\begin{eqnarray}
\left|\frac{\partial S[\Phi\!:\!\zeta]}{\partial \Phi}
\right \rangle
=m_0^{\Phi}\Bigl(1\!:\!\zeta \Bigr).
\label{hamilton vector of S}
\end{eqnarray}
\begin{remark}
\label{shift of Phi in S}
The variational formula (\ref{variation formula}) is mod 
$O\left( (\delta\Phi)^2\right)$. 
By calculations similar to the proof of eq.(\ref{variation formula}) 
(which we give below), the exact result can be found out to be 
\begin{eqnarray}
S\Bigl[\Phi\!+\!\delta\Phi\!:\!\zeta\Bigr]=
S\Bigl[\Phi\!:\!\zeta\Bigr]+
\omega\left(\delta\Phi, 
\sum_{k \geq 0}
\frac{1}{k\!+\!1}m_k^{\Phi}
\Bigl(\delta\Phi\!:\!\zeta \Bigr) \right). 
\label{exact variation formula}
\end{eqnarray}
\end{remark}

~

\noindent
\underline{{\it Proof of Proposition \ref{prop variation formula}}}~: 
The variation can be evaluated as follows.  
\begin{eqnarray}
&&\delta S\Bigl[\Phi\!:\!\zeta\Bigr]
\nonumber \\
&&=
\frac{1}{2}
\Bll \omega_{12}\Br
\left\{
  |\delta \Phi \rangle_1 
    \left( Q^{(2)}|\Phi \rangle_2 \right)
  \!+\!  
  |\Phi \rangle_1 
    \left( Q^{(2)}|\delta \Phi \rangle_2 \right) 
\left\} 
+ 
\sum_{k \geq 3}\frac{1}{k}
\Bll 1 \cdots k\!:\!\zeta \Br 
\left( \sum_{i=1}^k 
|\Phi\rangle_1 \cdots |\delta \Phi \rangle_i \cdots |\Phi \rangle_k 
\right) \right. \right. 
\nonumber \\
&&=
\Bll \omega_{12}\Br 
\delta \Phi \rangle_1 
\left( Q^{(2)}|\Phi \rangle_2 \right)
+ 
\sum_{k \geq 3}\frac{1}{k} 
\left\{
\sum_{i=1}^k
\Bll i~i\!+\!1 \cdots i\!+\!k\!-\!1\!:\!\zeta \Br 
\delta \Phi\rangle_{i}
|\Phi \rangle_{i+1} 
\cdots 
|\Phi \rangle_{i+k-1} 
\right\}
\nonumber \\
&&=
\Bll \omega_{12}\Br
\delta \Phi \rangle_1 
\left( Q^{(2)}|\Phi \rangle_2 \right)
+ 
\sum_{k \geq 3}
\Bll 1~2 \cdots k\!:\!\zeta \Br 
\delta \Phi\rangle_{1}
|\Phi \rangle_{2} 
\cdots 
|\Phi \rangle_{k} 
\nonumber \\
&&=
\sum_{k \geq 1}
\Bll \omega_{12} \Br 
\delta \Phi \rangle_1 
\Bl m_k\Bigl(\Phi^k \!:\! \zeta\Bigr) \Brr_2 
\nonumber \\
&&=
\omega 
\left(
\delta \Phi, 
\sum_{k \geq 1}m_k
\Bigl(\Phi^k \!:\! \zeta \Bigr) 
\right).  
\end{eqnarray}
We use the asymmetry (\ref{conjecture 2}) of open-string vertices 
and the symmetry of the reflector to show the second equality 
in the above computation. 
We also use Proposition \ref{open-string vertex by mk} 
to show the fourth equality.\P 

~

In the previous section,  
in order to emphasize the perspective of renormalization group, 
open-string vertices are constructed by using the propagator 
in the Siegel gauge. 
These vertices are used in the definition of  
$S[\Phi\!:\!\zeta]$. Nevertheless it turns out that 
the action is a covariant classical action of open-string field. 
\begin{definition}[Gauge transformation of $\Phi$]
\label{def gauge transf}
For any vector $\rho \!\in\! {\cal H}$ with 
$\epsilon(\rho)\!=\!-1$ 
we define an infinitesimal transformation of $\Phi$ by
\footnote{We follow the convention of the deformation theory 
of $A_{\infty}$-algebra.} 
\begin{eqnarray}
\delta_{\rho}\Phi = m_1^{\Phi}\Bigl(\rho \!:\!\zeta \Bigr).
\label{gauge transf}
\end{eqnarray}
\end{definition}
When $\zeta \!=\!0$, this transformation reduces to 
$\delta_{\rho}\Phi
\!=\!
Q\rho\!+\!\Phi \star \rho 
\!-\! \rho \star \Phi$. 
This is precisely the gauge symmetry of the microscopic 
action (\ref{microscopic action}). 
In the case of $\zeta \!\neq\! 0$, 
it is unclear whether the transformation (\ref{gauge transf}) 
is still gauge symmetry of $S[\Phi\!:\!\zeta]$. 
We have the following proposition. 
\begin{proposition}[Gauge invariance of the action]
\label{gauge invariance}
The action (\ref{def of low energy action}) is invariant 
under the infinitesimal transformation (\ref{gauge transf}) 
of $\Phi$. 
\end{proposition}

~

\noindent 
\underline{{\it Proof of Proposition \ref{gauge invariance}}}~:
By using the variational formula (\ref{variation formula}) 
we rewrite $\delta_{\rho}S$ to the form,   
\begin{eqnarray}
\delta_{\rho}S\Bigl[\Phi\!:\!\zeta\Bigr]
&=& 
\omega 
\left(
\delta_{\rho}\Phi,~ m_0^{\Phi}\Bigl(1\!:\!\zeta \Bigr)
\right) 
\nonumber \\
&=&  
\omega 
\left( 
m_1^{\Phi}\Bigl(\rho\!:\!\zeta \Bigr),~
m_0^{\Phi}\Bigl(1\!:\!\zeta \Bigr)
\right).
\label{proof of gauge symmetry 1}
\end{eqnarray}
We further rewrite eq.(\ref{proof of gauge symmetry 1}) 
by using Proposition \ref{open-string vertex by mk} as follows.
\begin{eqnarray}
&&
Eq.(\ref{proof of gauge symmetry 1})
\nonumber \\
&&=
\sum_{q_1,q_2}
\omega
\left(
m_{q_1+q_2+1}
\Bigl(
\Phi^{q_1}, \rho, \Phi^{q_2}
\Bigr),~ 
m_0^{\Phi}(1)
\right)
\nonumber \\
&&=
\sum_{q_1,q_2}
\Bll \omega_{ab} \Br~ 
\Bigl. m_{q_1+q_2+1}\Bigl(\Phi^{q_1},\rho,\Phi^{q_2}\Bigr)\Brr_a  
\Bl~ m_0^{\Phi}(1)\Brr_b 
\nonumber \\
&&=
\sum_{q_1,q_2}(-)^{q_2}
\Bll 1  \cdots  q_1\!+\!q_2\!+\!2:\!\zeta \Br
\underbrace{\Phi \rangle_1 \cdots |\Phi\rangle_{q_1}}_{q_1} 
|\rho \rangle_{q_1+1} 
\underbrace{|\Phi\rangle_{q_1+2} \cdots |\Phi\rangle_{q_1+q_2+1}}_{q_2}
\Bl m_0^{\Phi}(1)\Brr_{q_1+q_2+2}.
\nonumber \\ 
\label{proof of gauge symmetry 2}
\end{eqnarray}
We replace the open-string fields  
by taking account of the asymmetry (\ref{conjecture 2}) 
of the open-string vertices 
and then rewrite eq.(\ref{proof of gauge symmetry 2}), 
again by using Proposition \ref{open-string vertex by mk} 
as follows. 
\begin{eqnarray}
&&
Eq.(\ref{proof of gauge symmetry 2})
\nonumber \\
&&=-
\sum_{q_1,q_2}
\Bll 1  \cdots  q_1\!+\!q_2\!+\!2:\!\zeta \Br
\underbrace{\Phi \rangle_1 \cdots |\Phi\rangle_{q_2}}_{q_2} 
\Bl m_0^{\Phi}(1) \Brr_{q_2+1} 
\underbrace{|\Phi\rangle_{q_2+2}\cdots |\Phi\rangle_{q_1+q_2+1}}_{q_1}
|\rho\rangle_{q_1+q_2+2}
\nonumber \\
&&=-
\sum_{q_1,q_2}
\omega
\left(
m_{q_1+q_2+1}
\Bigl(\Phi^{q_2},m_0^{\Phi}(1),\Phi^{q_1} \Bigr),~ 
\rho
\right)
\nonumber \\
&&=-
\omega 
\left( 
m_1^{\Phi}\Bigl(m_0^{\Phi}(1)\Bigr),~
\rho
\right). 
\label{m01 and omega}
\end{eqnarray}

Therefore $\delta_{\rho}S$ turns out to be 
\begin{eqnarray} 
\delta_{\rho}S\Bigl[\Phi\!:\!\zeta\Bigr]=
\omega \left(
\rho,
m_1^{\Phi}\Bigl(m_0^{\Phi}(1\!:\!\zeta):\!\zeta \Bigr)
\right).
\label{proof of gauge symmetry 3}
\end{eqnarray} 
Recall $({\cal H},m_k^{\Phi})$ is a {\it weak} $A_{\infty}$-algebra. 
Especially it follows from Definition 
\ref{def of A-algebra} that 
$m_1^{\Phi}\Bigl(m_0^{\Phi}(1)\Bigr)\!=\!0$. 
Thus $\delta_{\rho}S\!=\!0$.\P

~

In the BV formalism \cite{BV formalism}, \cite{BV formalism 2}
quantum master equation is 
the criterion for consistency of a quantum gauge theory. 
Its classical limit is called classical master equation. 
It is simply given by $\Bigl\{S,S\Bigr\}=0$, 
where $S$ is the classical BV action of the gauge theory.   
Classical master equation itself is not sufficient to ensure 
consistency of the quantum theory, but it must be satisfied  
at the classical level in order to obtain a consistent quantum 
gauge theory.
\begin{proposition}[Classical master equation]
\label{classical master equation}
The action (\ref{def of low energy action}) satisfies 
classical master equation,   
\begin{eqnarray}
\left\{ S\Bigl[\Phi\!:\!\zeta\Bigr],~
S\Bigl[\Phi\!:\!\zeta\Bigr] \right\}=0.
\label{classical BV eq}
\end{eqnarray}
\end{proposition}
The classical master equation (\ref{classical BV eq}) 
eventually reduces to the $A_{\infty}$-algebra $({\cal H},m_k)$
as we will find in the proof of this proposition. 
Physical significance of classical master equation is easy 
to see if we take another form of this equation. 
The anti-bracket in eq.(\ref{classical BV eq}) is equal to  
$\omega \left( m_0^{\Phi}\Bigl(1\!:\!\zeta \Bigr),
m_0^{\Phi}\Bigl(1\!:\!\zeta \Bigr)\right)$, 
where we identify the hamiltonian vector with the obstruction 
of the deformation theory.  
\begin{definition}[BRST transformation of $\Phi$]
\label{def of BRST transf}
BRST transformation ${\bf \delta_{BRS}}$
of open-string field $\Phi$ is defined   
by the hamiltonian vector (\ref{hamilton vector of S}). 
\begin{eqnarray}
{\bf \delta_{BRS}}\Phi =
m_0^{\Phi}\Bigl(1\!:\!\zeta \Bigr).
\label{BRST trans of Phi}
\end{eqnarray}
\end{definition}
${\bf \delta_{BRS}}\Phi$ has the ghost number two. 
Hence we can attach the ghost number one 
to ${\bf \delta_{BRS}}$. 
We can also see ${\bf \delta_{BRS}}$ is grassmann-odd. 
Physical significance of classical master equation 
may be observed in the following propositions. 
\begin{proposition}[Nilpotency of BRST transformation]
\label{prop nilpotency of BRST trans}
\begin{eqnarray}
{\bf \delta_{BRS}}\cdot 
{\bf \delta_{BRS}}=0.
\label{nilpotency of BRST trans}
\end{eqnarray}
\end{proposition}
\begin{proposition}[BRST invariance of the action]
\label{BRS invariance of the action} 
The variational formula (\ref{variation formula}) can be written as 
\begin{eqnarray}  
\delta S\Bigl[\Phi\!:\!\zeta\Bigr]=
\omega \Bigl(
\delta \Phi, 
{\bf \delta_{BRS}} \Phi 
\Bigr).
\label{variation formula 2}
\end{eqnarray}
In particular, if one takes ${\bf \delta_{BRS}}\Phi$ 
as $\delta \Phi$, we obtain    
\begin{eqnarray}
{\bf \delta_{BRS}}S\Bigl[\Phi\!:\!\zeta\Bigr]=0.
\label{BRST inv of S}
\end{eqnarray}
\end{proposition} 
Eq.(\ref{variation formula 2}) is a simple reinterpretation 
of eq.(\ref{variation formula}). The BRST invariance of the action 
follows from Proposition \ref{classical master equation} by using 
the formula (\ref{variation formula 2}).

~

\noindent 
\underline{{\it Proof of Proposition \ref{classical master equation}}}~:  
We apply the strategy taken in \cite{Zwiebach BV} 
for a proof of the quantum master equation of closed-string field theory. 
We first write down the anti-bracket 
$\omega \left( m_0^{\Phi}(1), m_0^{\Phi}(1)\right)$ 
as follows.  
\begin{eqnarray}
&&
\omega
\left(
m_0^{\Phi}(1),m_0^{\Phi}(1)
\right)
\nonumber \\
&&~~~~= 
\sum_{k \geq 2}
\left\{
\sum_{l=1}^{k-1}
\omega 
\left(
m_{k-l}\Bigl(\Phi^{k-l}\Bigr),
m_l\Bigl(\Phi^l\Bigr)
\right)
\right\}
\nonumber \\
&&~~~~=
\sum_{k \geq 2}
\left\{
\sum_{l=1}^{k-1}
\frac{k\!-\!l}{k} \omega 
\left(
m_{k-l}\Bigl(\Phi^{k-l}\Bigr),
m_l\Bigl(\Phi^l\Bigr)
\right)
\right\}
\!+\!
\sum_{k \geq 2}
\left\{
\sum_{l=1}^{k-1}
\frac{l}{k} \omega 
\left(
m_{k-l}\Bigl(\Phi^{k-l}\Bigr),
m_l\Bigl(\Phi^l \Bigr)
\right)
\right\}
\nonumber \\
&&~~~~=
\sum_{k \geq 2}
\frac{2}{k}
\left\{
\sum_{l=1}^{k-1}
(k\!-\!l) \times 
 \omega 
\left(
m_{k-l}\Bigl(\Phi^{k-l}\Bigr),
m_l\Bigl(\Phi^l \Bigr)
\right) 
\right\}. 
\label{antibracket 1}
\end{eqnarray}

To proceed on the computation 
we need a suitable identity :  
We rewrite 
$\omega 
\left(
m_{k-l}\Bigl(\Phi^{k-l}\Bigr),
m_l\Bigl(\Phi^l \Bigr)
\right)$,  
which appears in eq.(\ref{antibracket 1}),  
into the following form 
by using Proposition \ref{open-string vertex by mk}. 
\begin{eqnarray}
\omega 
\left( 
   m_{k-l}\Bigl(\Phi^{k-l}\Bigr),~
   m_l\Bigl(\Phi^l \Bigr) 
\right)
&=&-
\Bll \omega_{ab}\Br~ 
\Bigl.m_{k-l}\Bigl(\Phi^{k-l}\Bigr)\Brr_a 
\Bl~m_l\Bigl(\Phi^l\Bigr) \Brr_b
\nonumber \\
&=&-
\Bll 1 ~\cdots ~k\!-\!l\!+\!1\!:\!\zeta  \Br
\underbrace{\Phi \rangle_1 \cdots |\Phi \rangle_{k-l}}_{k-l} 
\Bl~m_l(\Phi^l) \Brr_{k-l+1}.
\label{antibracket 1.5}
\end{eqnarray}
We permute $(k-l)$ open-string fields in this equation  
by taking account of the asymmetry of the vertex and then 
rewrite the equation in terms of $m_k$ 
by using Proposition \ref{open-string vertex by mk}
as follows. 
\begin{eqnarray}
&&
Eq.(\ref{antibracket 1.5})
\nonumber \\
&&=
(-)^{i}
\Bll i \cdots k\!-\!l\!+\!1~~1 \cdots i\!-\!1\!:\!\zeta \Br
\underbrace{\Phi \rangle_i \cdots |\Phi\rangle_{k-l}}_{k-l-i+1}
\Bl~m_l(\Phi^l)\Brr_{k-l+1} 
\underbrace{|\Phi\rangle_{1} \cdots |\Phi \rangle_{i-1}}_{i-1}
\nonumber \\
&&=
\Bll \omega_{ab}\Br~
m_{k-l}\Bigl(\Phi^{k-l-i+1},m_l(\Phi^l),\Phi^{i-2}\Bigr)\Brr_a 
|\Phi \rangle_b 
\nonumber \\
&&=
\omega \left( 
m_{k-l}
\Bigl(\Phi^{k-l-i+1},m_l\bigl(\Phi^l \bigl),
\Phi^{i-2} \Bigr),~ 
\Phi 
\right).
\label{antibracket 2}
\end{eqnarray} 
Hence we obtain the following identity independent of $i$. 
\begin{eqnarray}
\omega 
\left(  
m_{k-l}
\Bigl(\Phi^{k-l-i} ,m_l\bigl(\Phi^l\bigr), \Phi^{i-1}\Bigr),
\Phi  
\right)
=
\omega \left( 
   m_{k-l}\Bigl(\Phi^{k-l}\Bigr),
   m_l\Bigl(\Phi^l\Bigr) 
\right).
\label{antibracket 3}
\end{eqnarray}

Now we proceed on the computation of eq.(\ref{antibracket 1}).    
By using the above identity it becomes as follows.
\begin{eqnarray}
Eq.(\ref{antibracket 1})&=& 
\sum_{k \geq 2}
\frac{2}{k}
\sum_{l=1}^{k-1}
\left\{
\sum_{i=1}^{k-l}
\omega 
\left( 
m_{k-l}
\Bigl(\Phi^{k-l-i},m_l\bigl(\Phi^l\bigr),\Phi^{i-1}\Bigr),~
\Phi 
\right)
\right\} 
\nonumber \\
&=& 
\sum_{k \geq 1}
\frac{2}{k\!+\!1}
\omega 
\left( 
\sum_{l=1}^k \sum_{i=0}^{k-l}
m_{k+1-l}
\Bigl(
  \Phi^{i},m_l\bigl(\Phi^l\bigr),\Phi^{k-l-i}
\Bigr),~
\Phi
\right). 
\label{antibracket 4}
\end{eqnarray} 
This vanishes identically 
due to the relation (\ref{A-algebra}) 
of the $A_{\infty}$-algebra $({\cal H},m_k)$. 
Thus we obtain 
$\omega \left( m_0^{\Phi}(1), 
m_0^{\Phi}(1)\right)\!=\!0$.\P 

~

\noindent
\underline{{\it Proof of Proposition \ref{prop nilpotency of BRST trans}}}~: 
We show ${\bf \delta_{BRS}}\cdot {\bf \delta_{BRS}}\Phi=0$. 
We first rewrite ${\bf \delta_{BRS}}\cdot {\bf \delta_{BRS}}\Phi$ 
as follows.   
\begin{eqnarray}
{\bf \delta_{BRS}}\cdot {\bf \delta_{BRS}}\Phi 
&=&
{\bf \delta_{BRS}}
\Bigl[m_0^{\Phi}(1)\Bigr]
\nonumber \\
&=&
\sum_{k \geq 1}{\bf \delta_{BRS}}
\Bigl[m_k\bigl(\Phi^k\bigr)\Bigr]
\nonumber \\
&=&
{\bf \delta_{BRS}}
\Bigl[m_1\bigl(\Phi \bigr)\Bigr]
+
\sum_{k \geq 2}{\bf \delta_{BRS}}
\Bigl[
\Bll 0~1 \cdots ~k\!:\!\zeta \Br
\Bigl.S_{0a}\Brr 
\bl \Phi\brr_1 \cdots \bl\Phi\brr_k 
\Bigr]
\nonumber \\
&=&-
m_1\Bigl({\bf \delta_{BRS}}\Phi \Bigr) 
+
\sum_{k \geq 2}(-)^{k}
\Bll 0~1 \cdots ~k:\zeta \Br
\Bigl.S_{0a}\Brr 
{\bf \delta_{BRS}} 
\Bigl[
\bl\Phi\brr_1 \cdots \bl\Phi\brr_k 
\Bigr].
\label{proof nilpotency of BRS 1}
\end{eqnarray}
The first term of eq.(\ref{proof nilpotency of BRS 1}) 
is equal to 
$-\!\sum_{k \geq 1}m_1\Bigl(m_k\bigl(\Phi^k\bigr)\Bigr)$. 
As for the second term we further compute by using 
Proposition \ref{open-string vertex by mk} as follows. 
\begin{eqnarray}
&&
\sum_{k \geq 2}(-)^{k}
\Bll 0~1 \cdots ~k:\zeta \Br
\Bigl.S_{0a}\Brr 
{\bf \delta_{BRS}} 
\Bigl[
\bl\Phi\brr_1 \cdots \bl\Phi\brr_k 
\Bigr]
\nonumber \\
&&~~~~~~=
\sum_{k \geq 2}
\left\{
\sum_{i=0}^{k-1}
(-)^{k+i}
\Bll 0~1 \cdots ~k:\zeta \Br
\Bigl.S_{0a}\Brr  
\bl\Phi\brr_1 \cdots \bl\Phi \brr_i 
\Bl {\bf \delta_{BRS}}\Phi \Brr_{i+1} 
\bl\Phi \brr_{i+2}\cdots \bl\Phi\brr_k 
\right\}
\nonumber \\
&&~~~~~~= 
\sum_{l \geq 1}\sum_{k \geq 2}
\left\{
\sum_{i=0}^{k-1}
(-)^{k+i}
\Bll 0~1 \cdots ~k:\zeta \Br
\Bigl. S_{0a}\Brr   
\bl \Phi\brr_1 \cdots \bl \Phi \brr_i
\Bl m_l(\Phi^l)\Brr_{i+1} 
\bl \Phi \brr_{i+2} \cdots \bl \Phi\brr_k 
\right\} 
\nonumber \\
&&~~~~~~=-
\sum_{l \geq 1}  
\sum_{k \geq 2}
\sum_{i=1}^{k}
m_{k}\Bigl(\Phi^{i},m_l\bigl(\Phi^l\bigr),\Phi^{k-1-i}\Bigr)
\nonumber \\
&&~~~~~~=-
\sum_{k \geq 1}
\left\{
\sum_{l=1}^{k-1}\sum_{i=0}^{k-l}
m_{k+1-l}\Bigl(\Phi^i, m_l\bigl(\Phi^l\bigr), 
\Phi^{k-l-i}\Bigr)
\right\}.
\label{proof nilpotency of BRS 2}
\end{eqnarray}

By putting these two terms together,  
we eventually obtain the following expression of 
${\bf \delta_{BRS}}\cdot {\bf \delta_{BRS}}\Phi$.   
\begin{eqnarray}
{\bf \delta_{BRS}}\cdot {\bf \delta_{BRS}}\Phi 
&=&
-\sum_{k \geq 1}m_1\Bigl(m_k\bigl(\Phi^k\bigr)\Bigr)
-\sum_{k \geq 1}
\left\{
\sum_{l=1}^{k-1}\sum_{i=0}^{k-l}
m_{k+1-l}\Bigl(\Phi^i, m_l\bigl(\Phi^l\bigr), 
\Phi^{k-l-i}\Bigr)
\right\}
\nonumber \\
&=& 
-\sum_{k \geq 1}
\left\{
\sum_{l=1}^{k}\sum_{i=0}^{k-l}
m_{k+1-l}\Bigl(\Phi^i, m_l\bigl(\Phi^l\bigr), 
\Phi^{k-l-i}\Bigr) 
\right\}. 
\label{proof nilpotency of BRS 3}
\end{eqnarray}
This actually vanishes by the $A_{\infty}$-relation.\P


\section{Renormalization Group Of Open-String Field Theory}
So far, 
open-string field $\Phi$ is simply a vector of 
the open-string Hilbert space ${\cal H}$. 
In the presence of the short-distance cut-off scale parameter 
$\zeta$, classical dynamics of open-string field 
is governed by the action $S[\Phi\!:\!\zeta]$. 
As concerns the classical dynamics,  
any consideration on {\it off-shell} open-string field is irrelevant. 
In particular, we can not find any relation between 
{\it off-shell} open-string fields at different scales. 
On the other hand, 
if one takes the renormalization group perspective, 
namely if one regards $S[\Phi\!:\!\zeta]$ 
as the classical part of $S_{eff}^{\zeta}[\Phi]$ 
which could be obtained integrated out all the contributions 
from length scale less than $\zeta$, 
{\it off-shell} open-string fields at different scales 
should be related with one another 
by renormalization group flow. 
In this section we study renormalization group 
of classical open-string field theory. 
Our formulation of renormalization group 
equation (RG equation) is based on Proposition 
\ref{scale dependence of OSFT Ainfty}.

Now open-string field $\Phi$ is allowed  
to depend on the cut-off scale parameter $\zeta$.
Equivalently the super-coordinates  $t^a$ which appear 
in the expansion by the bases $\phi_a$, depend on $\zeta$. 
\begin{eqnarray}
\Phi(\zeta)=\sum_at^a(\zeta)\phi_a.
\end{eqnarray}
RG equation determines the scale dependence 
of open-string field $\Phi$ 
or equivalently a flow $t^a(\zeta)$ on the super-manifold. 
\begin{definition}[RG equation of classical open-string field]
\label{def RG equation for OSFT}
Let $S[\Phi\!:\!\zeta]$ be the action (\ref{def of low energy action}). 
RG equation of open-string field $\Phi$ is defined by  
\begin{eqnarray}
\frac{d}{d\zeta}S
\Bigl[\Phi(\zeta)\!:\!\zeta \Bigr]=0. 
\label{def RG equation}
\end{eqnarray}
\end{definition}
The total derivation in the RG equation (\ref{def RG equation})
can be factorized into a sum of two terms. 
One is a derivation through the coordinates $t^a(\zeta)$. 
Using the variational formula (\ref{variation formula}) it is 
given by 
$\omega \Bigl(d\Phi /d\zeta ,
\Bl\partial S/ \partial \Phi \Brr\Bigr)$. 
The other is a derivation through the open-string vertices 
in the action. It is given by 
$\partial S/\partial \zeta$. 
The factorized form of the RG equation becomes as follows.   
\begin{eqnarray}
\omega \left(\frac{d\Phi}{d\zeta},
~\left|\frac{\partial S[\Phi\!:\!\zeta]}{\partial \Phi}
\right \rangle \right)=
-\frac{\partial S}{\partial \zeta}
\Bigl[\Phi\!:\!\zeta \Bigr]. 
\label{RG equation 1}
\end{eqnarray}
The RHS of this equation can be calculated using Proposition 
\ref{scale dependence of OSFT Ainfty}. 
Let $S_{int}[\Phi\!:\!\zeta]$ be the interaction part 
of the action (\ref{def of low energy action}). 
\begin{eqnarray}
S_{int}\Bigl[\Phi\!:\!\zeta \Bigr]\equiv 
\omega
\left(
\Phi,
\sum_{k \geq 2}\frac{1}{k\!+\!1}
m_k\Bigl(\Phi^k\!:\!\zeta\Bigr)
\right).  
\end{eqnarray}
The corresponding hamiltonian vector becomes 
\begin{eqnarray}
\left |\frac{\partial S_{int}[\Phi\!:\!\zeta]}
{\partial \Phi}\right \rangle 
= \sum_{k \geq 2}m_k\Bigl(\Phi^k\!:\!\zeta \Bigr).
\end{eqnarray} 
With this notation the scale dependence of the action 
is described as follows.  
\begin{proposition}
\label{prop scale dependence of S}
Derivation of the action (\ref{def of low energy action}) 
with respect to the cut-off scale parameter $\zeta$ 
has the following form.  
\begin{eqnarray}
\frac{\partial S}{\partial \zeta}
\Bigl[\Phi\!:\!\zeta \Bigr]
=
-\omega \left( 
b_0
\left|
\frac{\partial S_{int}[\Phi\!:\!\zeta]}{\partial \Phi}\right \rangle 
+L_0\Phi,~
\left |
\frac{\partial S_{int}[\Phi\!:\!\zeta]}{\partial \Phi}\right \rangle 
\right). 
\label{scale dependence of S}
\end{eqnarray}
\end{proposition}

~

Due to this proposition the RG equation (\ref{RG equation 1}) becomes  
\begin{eqnarray}
\omega \left(\frac{d\Phi}{d\zeta},
~\left|\frac{\partial S[\Phi\!:\!\zeta]}{\partial \Phi}
\right \rangle \right)=
\omega \left(  
b_0
\left|
\frac{\partial S_{int}[\Phi\!:\!\zeta]}{\partial \Phi}\right \rangle 
+L_0\Phi,~
\left |
\frac{\partial S_{int}[\Phi\!:\!\zeta]}{\partial \Phi}\right \rangle 
\right).
\label{RG equation 2}
\end{eqnarray}
Eq.(\ref{RG equation 2}) is  
a classical analogue of the Polchinski's RG equation 
\cite{Polchinski,Wilson-Kogut}. 
RG equation $a la$ Wilson or Polchinski is 
$dZ\Bigl(t(\zeta)\!:\!\zeta\Bigr)/d\zeta
\!=\!0$, where $Z$ is a partition function of a quantum field 
theory regularized by the short distance cut-off scale $\zeta$.   
Roughly speaking, correspondence 
with the RG equation (\ref{def RG equation}) can be seen 
if one puts $Z\!=\!e^{-S}$. 
The precise relation may be established 
by using the Legendre transformation as considered 
in \cite{Kubota}.
There appears a quantum correction in the original equation. 
It has the form  
$\sim \partial^2\!S_{int}/\partial \phi \partial \phi$. 
In our case quantum correction appears as the higher loop 
contribution of open-string.  
Description of RG equation of string-field  
along the line presented in (\ref{RG equation 2}) 
was first given for closed-string field theory \cite{Alwis}. 
Similarity with the BV master equation was emphasized there. 
Resemblance between the two was further investigated 
in \cite{Hata-Zwiebach}.

As we stated earlier, the RG equation (\ref{def RG equation}) 
is introduced to determine the $\zeta$-evolution of open-string 
field $\Phi$. Let us solve the RG equation (\ref{RG equation 2}) 
in such a form by which it becomes manifest. 
For this purpose 
we first rewrite the RHS of eq.(\ref{RG equation 2}) as follows. 
\begin{eqnarray}
&&
\omega 
\left(
b_0
\left|
\frac{\partial S_{int}}{\partial \Phi}
\right \rangle 
+
L_0\Phi
,~ 
\left |
\frac{\partial S_{int}}{\partial \Phi}
\right \rangle 
\right)
\nonumber \\
&&~~~~~~~~~~~~=
\omega 
\left(
b_0
\left|
\frac{\partial S}{\partial \Phi}
\right \rangle 
+
Qb_0\Phi
,~ 
\left |
\frac{\partial S_{int}}{\partial \Phi}
\right \rangle 
\right)
\nonumber \\
&&~~~~~~~~~~~~=
\omega 
\left(
b_0
\left|
\frac{\partial S}{\partial \Phi}
\right \rangle 
,~ 
\left |
\frac{\partial S_{int}}{\partial \Phi}
\right \rangle 
\right)
+ 
\omega 
\left(
Qb_0\Phi
,~ 
\left |
\frac{\partial S}{\partial \Phi}
\right \rangle 
\right) 
\nonumber \\
&&~~~~~~~~~~~~=
\omega 
\left(
b_0
\left|
\frac{\partial S_{int}}{\partial \Phi}
\right \rangle 
,~ 
\left |
\frac{\partial S}{\partial \Phi}
\right \rangle 
\right)
+
\omega 
\left(
Qb_0\Phi
,~ 
\left |
\frac{\partial S}{\partial \Phi}
\right \rangle 
\right) 
\nonumber \\
&&~~~~~~~~~~~~=
\omega 
\left(
b_0
\left|
\frac{\partial S_{int}}{\partial \Phi}
\right \rangle 
+
Qb_0\Phi
,~ 
\left |
\frac{\partial S}{\partial \Phi}
\right \rangle 
\right),
\label{proof RG equation 3-1}
\end{eqnarray}
where the anti-commutation relation, 
$\left\{Q,b_0\right\}\!=\!L_0$, is used to show the first 
equality. The second equality follows from $Q^2\!=\!0$ and 
the BRST invariance of the reflector. 
To show the third equality we use the asymmetry of $\omega$ 
besides eq.(\ref{T and reflector}). 
The RG equation (\ref{RG equation 2}) 
acquires the following expression 
by eq.(\ref{proof RG equation 3-1}).  
\begin{eqnarray}
\omega \left(
\frac{d\Phi}{d\zeta}
-
\left\{
b_0
\left|
\frac{\partial S_{int}[\Phi\!:\!\zeta]}{\partial \Phi}
\right \rangle 
+
Qb_0\Phi 
\right\}
,
~\left|\frac{\partial S[\Phi\!:\!\zeta]}{\partial \Phi}
\right \rangle 
\right)=0. 
\label{RG equation 3}
\end{eqnarray}
Thus we obtain 
\begin{proposition}
\label{prop RG equation of OSFT}
Open-string field $\Phi(\zeta)$ obeying the following 
first-order differential equation is a solution of the 
RG equation (\ref{def RG equation}).
\begin{eqnarray}
\frac{d \Phi}{d \zeta}=
b_0 \left|
\frac{\partial S_{int}[\Phi\!:\!\zeta]}{\partial \Phi}   
\right \rangle 
+Qb_0\Phi. 
\label{RG equation of OSFT}
\end{eqnarray}
\end{proposition}
We do not know whether eq.(\ref{RG equation of OSFT}) 
is equivalent to the RG equation (\ref{def RG equation}) or not 
since equivalence between eqs.(\ref{RG equation 3}) 
and (\ref{RG equation of OSFT}) is obscure. 
But this does not cause any problem.  
Skeptical reader can adopt eq.(\ref{RG equation of OSFT}) 
as a definition of RG equation of open-string field 
and keep eq.(\ref{def RG equation}) as a proposition.  
Eq.(\ref{RG equation of OSFT}) will be also called 
RG equation of open-string field.

~

\noindent
\underline{{\it Proof of Proposition \ref{prop scale dependence of S}}}~:
We first write down the partial derivation 
$\partial S[\Phi\!:\!\zeta]/\partial \zeta$ 
by using Proposition 
\ref{scale dependence of OSFT Ainfty}. 
\begin{eqnarray}
\frac{\partial S}{\partial \zeta}
\Bigl[\Phi\!:\!\zeta \Bigr]
&=&
\sum_{k \geq 2}\frac{1}{k\!+\!1}
\omega
\left(
\Phi,~
\frac{\partial m_k}{\partial \zeta}
\Bigl(\Phi^k\!:\!\zeta \Bigr)
\right)
\nonumber \\
&=&
-\sum_{k \geq 2}\frac{1}{k\!+\!1} 
\omega
\left(
\Phi,~ 
L_0m_k \Bigl(\Phi^k \Bigr) 
+
\sum_{p=0}^{k-1}
m_{k}
\Bigl(
\Phi^p, L_0\Phi, \Phi^{k-1-p}
\Bigr)
\right)
\label{eq 1} \\
&&
-2
\sum_{k \geq 2}\frac{1}{k\!+\!1}
\omega
\left(
\Phi,~ 
\sum_{l=2}^{k-1}\sum_{p=0}^{k-l}
m_{k+1-l}
\Bigl(
\Phi^p, 
b_0m_l\bigl(\Phi^l \bigr), \Phi^{k-l-p}
\Bigr)
\right).
\label{eq 2}
\end{eqnarray}

We treat two terms (\ref{eq 1}) and (\ref{eq 2}) separately.  
We rewrite the first term 
by using Proposition \ref{open-string vertex by mk} 
as follows. 
\begin{eqnarray}
&&
\omega
\left(
\Phi,~ 
L_0m_k\Bigl(\Phi^k \Bigr) 
+
\sum_{p=0}^{k-1}
m_{k}
\Bigl(
\Phi^p, L_0\Phi, \Phi^{k-1-p}
\Bigr)
\right)
\nonumber \\
&&=
\Bll \omega_{ab}\Br \Phi\rangle_a 
\left(L_0^{(b)}\Bl m_k\Bigl(\Phi^k\Bigr)\Brr_b \right) 
+
\sum_{p=0}^{k-1}
\Bll \omega_{ab}\Br 
\bigl. \Phi \brr_a 
\Bl m_k\Bigl(\Phi^p,L_0\Phi,\Phi^{k-1-p}\Bigr)\Brr_b
\nonumber \\
&&=
\Bll \omega_{ab}\Br 
\bigl.\Phi \brr_a
\left(
L_0^{(b)}
\Bl m_k\Bigl(\Phi\Bigr)\Brr_b 
\right) 
\nonumber \\
&&~~+\sum_{p=0}^{k-1}
\Bll 1~ \cdots~ k\!+\!1\!:\!\zeta \Br
\underbrace{\bigl.\Phi \brr_1 \cdots \bl\Phi \brr_{p+1}}_{p+1} 
\left( L_0^{(p+2)}\bl\Phi\brr_{p+2}\right)
\underbrace{\bl\Phi\brr_{p+3} \cdots \bl\Phi \brr_{k+1}}_{k-p-1}.
\label{proof dependence S 0}
\end{eqnarray}
Due to the property (\ref{T and reflector}) 
of the reflector we have 
$\Bll \omega_{12}\Br$ 
$\left(L_0^{(1)}|A \rangle_1\right)|B\rangle_2$. 
$=$
$\Bll \omega_{12}\Br A \rangle_1$
$\left(L_0^{(2)}|B\rangle_2\right)$.
We permute open-string fields in the second term 
of eq.(\ref{proof dependence S 0}) by taking account of   
the asymmetry (\ref{conjecture 2}) of the vertices, and 
then rewrite eq.(\ref{proof dependence S 0}) 
again in terms of $m_k$ 
by using Proposition \ref{open-string vertex by mk} 
as follows.
\begin{eqnarray}
Eq.(\ref{proof dependence S 0})
&=&  
\Bll \omega_{ab}\Br 
\left(L_0^{(a)}\bl\Phi \brr_a\right)
\Bl m_k\Bigl(\Phi^k\Bigr)\Brr_b 
\nonumber \\
&&
+\sum_{p=0}^{k-1}
\Bll p\!+\!1~p\!+\!2~ \cdots~ p\!:\!\zeta \Br 
\left( L_0^{(p+1)}\bl\Phi\brr_{p+1}\right)
\underbrace{\bl \Phi\brr_{p+2} \cdots \bl\Phi \brr_p}_{k}
\nonumber \\
&=& 
\Bll \omega_{ab}\Br
\left(L_0^{(a)}\bl\Phi \brr_a\right)
\Bl m_k\Bigl(\Phi^k\Bigr)\Brr_b 
\nonumber \\
&&+k \times 
\Bll 1~2~ \cdots~ k\!+\!1\!:\!\zeta \Br 
\left( L_0^{(1)}\bl\Phi\brr_1\right)
\underbrace{\bl\Phi\brr_{2} \cdots \bl\Phi \brr_{k+1}}_{k} 
\nonumber \\
&=& 
(k\!+\!1) \times 
\omega 
\left( 
L_0\Phi, m_k\Bigl(\Phi^k\Bigr) 
\right).
\label{proof dependence of S 1}
\end{eqnarray}
As the result we obtain the following expression of 
eq.(\ref{eq 1}).   
\begin{eqnarray}
Eq.(\ref{eq 1}) 
&=& 
-
\sum_{k \geq 2}
\omega 
\left( 
L_0\Phi,~m_k\Bigl(\Phi^k\Bigr) 
\right) 
\nonumber \\
&=&
-
\omega 
\left( 
L_0\Phi,~ 
\left|
\frac{\partial S_{int}}{\partial \Phi} 
\right \rangle 
\right).
\label{proof dependence of S 2}
\end{eqnarray}

Nextly we examine the second term. By using Proposition 
\ref{open-string vertex by mk} we rewrite eq.(\ref{eq 2}) 
as follows. 
\begin{eqnarray}
&&
\omega
\left(
\Phi,~ 
\sum_{l=2}^{k-1}\sum_{p=0}^{k-l}
m_{k+1-l}
\Bigl(
\Phi^p, 
b_0m_l\bigl(\Phi^l\bigr), \Phi^{k-l-p}
\Bigr)
\right)
\nonumber \\
&&~~~~~=
\sum_{l=2}^{k-1}\sum_{p=0}^{k-l}
\Bll \omega_{ab} \Br 
\bigl. \Phi \brr_a 
\Bl~ m_{k+1-l}\Bigl(\Phi^p, b_0m_l(\Phi^l),\Phi^{k-l-p}\Bigr) \Brr_b 
\nonumber \\
&&~~~~~= 
\sum_{l=2}^{k-1}\sum_{p=0}^{k-l}
\Bll 1 \cdots k\!+\!2\!-\!l\!:\!\zeta \Br
\underbrace{\bigl.\Phi \brr_1 \cdots \bl\Phi \brr_{p+1}}_{p+1} 
\left( b_0 \Bl m_l\Bigl(\Phi^l\Bigr) \Brr_{p+2}\right) 
\underbrace{\bl\Phi \brr_{p+3} \cdots \bl\Phi \brr_{k+2-l}}_{k-l-p} 
\nonumber \\
&&~~~~~=
\sum_{l=2}^{k-1}
(k\!+\!1\!-\!l)
\times 
\Bll 1 \cdots k\!+\!2\!-\!l\!:\! \zeta \Br
\left( b_0\Bl m_l\Bigl(\Phi^l\Bigr)\Brr_1 \right) 
\underbrace{\bl\Phi \brr_2 \cdots \bl\Phi \brr_{k+2-l}}_{k+1-l} 
\nonumber \\
&&~~~~~= 
\sum_{l=2}^{k-1}
(k\!+\!1\!-\!l)\times 
\omega 
\left( 
b_0m_l\Bigl(\Phi^l\Bigr),~
m_{k+1-l}\Bigl(\Phi^{k+1-l}\Bigr) 
\right).  
\label{proof dependence of S 3}
\end{eqnarray}
In this computation, 
in addition to the use of Proposition \ref{open-string vertex by mk},  
particularly to show the third equality 
we use the asymmetry (\ref{conjecture 2}) of the vertex.  
We can rewrite eq.(\ref{proof dependence of S 3}) 
in a convenient form. We notice the equality, 
\begin{eqnarray}
\omega 
\left( 
b_0m_l\Bigl(\Phi^l\Bigr),~
m_{k+1-l}\Bigl(\Phi^{k+1-l}\Bigr) 
\right)
=
\omega 
\left( 
b_0m_{k+1-l}\Bigl(\Phi^{k+1-l}\Bigr),~
m_l\Bigl(\Phi^l\Bigr) 
\right).
\end{eqnarray} 
By using this equality we write down 
eq.(\ref{proof dependence of S 3}) to the following form.  
\begin{eqnarray}
&&
\sum_{l=2}^{k-1}
(k\!+\!1\!-\!l)\times 
\omega 
\left( 
b_0m_l\Bigl(\Phi^l\Bigr),~
m_{k+1-l}\Bigl(\Phi^{k+1-l}\Bigr) 
\right)
\nonumber \\
&&~~~~= 
\sum_{l=2}^{k-1}
\frac{k\!+\!1\!-\!l}{2}
\omega 
\left( 
b_0m_l\Bigl(\Phi^l\Bigr),~
m_{k+1-l}\Bigl(\Phi^{k+1-l}\Bigr) 
\right) 
+
\sum_{l=2}^{k-1}
\frac{l}{2}
\omega 
\left( 
b_0m_{k+1-l}\Bigl(\Phi^{k+1-l}\Bigr),~
m_l\Bigl(\Phi^l\Bigr)
\right)    
\nonumber \\
&&~~~~= 
\sum_{l=2}^{k-1}
\frac{(k\!+\!1\!-\!l)\!+\!l}{2}
\omega 
\left( 
b_0m_l\Bigl(\Phi^l\Bigr),~
m_{k+1-l}\Bigl(\Phi^{k+1-l}\Bigr) 
\right) 
\nonumber \\
&&~~~~= 
\frac{k\!+\!1}{2}
\sum_{l=2}^{k-1}
\omega 
\left( 
b_0m_l\Bigl(\Phi^l\Bigr),~
m_{k+1-l}\Bigl(\Phi^{k+1-l}\Bigr) 
\right)
\label{proof dependence of S 4}
\end{eqnarray}
Hence the second term is found out to be 
\begin{eqnarray}
Eq.(\ref{eq 2})
&=& 
-
\sum_{k \geq 2}\sum_{l=2}^{k-1}
\omega 
\left( 
b_0m_l\Bigl(\Phi^l\Bigr),~
m_{k+1-l}\Bigl(\Phi^{k+1-l}\Bigr) 
\right)
\nonumber \\
&=& 
- 
\omega 
\left(b_0
\left|
\frac{\partial S_{int}}{\partial \Phi} 
\right \rangle,~
\left| 
\frac{\partial S_{int}}{\partial \Phi} 
\right \rangle 
\right).
\label{proof dependence of S 5}
\end{eqnarray}

Sum of (\ref{eq 1}) and (\ref{eq 2}) is equal to 
$\partial S/\partial \zeta$.  
Expressions (\ref{proof dependence of S 2}) 
and (\ref{proof dependence of S 5}) show that it is 
precisely eq.(\ref{scale dependence of S}).\P


\section{Analysis Of RG Flow In Siegel Gauge}
If we impose the Siegel gauge condition 
$b_0\Phi\!=\!0$ on open-string field,  
the RG equation (\ref{RG equation of OSFT}) 
obtains a simple form,    
\begin{eqnarray}
\frac{d \Phi}{d \zeta}=
b_0 \left|
\frac{\partial S_{int}[\Phi\!:\!\zeta]}{\partial \Phi}   
\right \rangle.  
\label{RGE Siegel gauge}
\end{eqnarray}
Clearly the evolution of $\Phi$ 
governed by eq.(\ref{RGE Siegel gauge})  
preserves the gauge condition. 
In this section we investigate 
the RG equation (\ref{RGE Siegel gauge}).  
Throughout this section the Siegel gauge condition 
is imposed on open-string field.

~

\noindent
Let ${\cal H}^S$ be 
subspace of the open-string Hilbert space ${\cal H}$ 
consisting of states  
which satisfy the Siegel gauge condition. 
Any vector $\phi \in {\cal H}^S$ does not contain $c_0$.  
The odd symplectic structure $\omega$ becomes null 
on ${\cal H}^S$. Instead of $\omega$ 
we introduce another bilinear form $g$.
\begin{definition}[Bilinear form $g$ on ${\cal H}^S$] 
\label{def g}
For any vectors $A,B \in {\cal H}^S$, 
we define $g(A,B)$ by 
\begin{eqnarray}
g(A,B)=
\Bll \omega_{12} \Br 
c_0^{(1)}|A\rangle_1 |B\rangle_2. 
\label{g}
\end{eqnarray}
\end{definition}
The bilinear form $g$ is non-degenerate on ${\cal H}^S$. 
$\Bll \omega_{12} \Br\!=\! \Bll \omega_{21} \Br$ implies 
$g(A,B)\!=\!(-)^{\epsilon(A)\epsilon(B)+\epsilon(A)+\epsilon(B)}$
$g(B,A)$. 
Selection rule (\ref{selection rule of BPZ}) 
of the BPZ pairing gives 
$g(A,B)\!\neq\!0$ 
$\Rightarrow$ 
$\epsilon(A)+\epsilon(B)\!=\!0$.

Let $\Bigl\{\phi_i\Bigr\}_i$ be bases of ${\cal H}^S$. 
We can take the conjugate bases $\Bigl\{\phi^i\Bigr\}_i$ 
such that they satisfy 
$g(\phi^i,\phi_j)\!=\!\delta^i_j$. 
The $(-1)$-shifted ghost number of $\phi^i$ is 
$\epsilon(\phi^i)\!=\!-\epsilon(\phi_i)$. 
We put 
$g_{ij}\!\equiv\!g(\phi_i,\phi_j)$ 
and 
$g^{ij}\!\equiv\!(-)^{\epsilon(\phi_j)}g(\phi^i,\phi^j)$. 
It can be seen that 
two bases are related with each other by 
$\phi^i\!=\!\sum_jg^{ij}\phi_j$ 
or equivalently by  
$\phi_i\!=\!\sum_jg_{ij}\phi^j$. 
Hence $g^{ij}$ is the inverse of $g_{ij}$.  
\begin{eqnarray}
\sum_kg^{ik}g_{kj}=\sum_kg_{jk}g^{ki}=\delta_j^i.
\end{eqnarray}

In the Siegel gauge,  
open-string field $\Phi$ has an expansion of the form     
$\Phi \!=\!\sum_i t^i \phi_i$. 
Each coefficient $t^i$ has the ghost number 
$G(t^i)\!=\!-\epsilon(\phi_i)$ 
and the grassmannity $(-)^{G(t^i)}$. 
They are coordinates of the super-submanifold 
constrained by the gauge condition. 
We write the $\zeta$-evolution of $\Phi$ as  
\begin{eqnarray}
\Phi(\zeta)=\sum_i t^i(\zeta)\phi_i. 
\end{eqnarray} 
RG flow $t^i(\zeta)$ on the super-manifold 
is determined from eq.(\ref{RGE Siegel gauge}) 
by using this expansion. 
We choose $\phi_i$ to be the eigenvectors of $L_0$.    
\begin{eqnarray}
L_0|\phi_i\rangle = \Delta_i |\phi_i \rangle . 
\end{eqnarray}

\subsection{Classical solution and RG equation}
Classical solutions of open-string field theory are  
stationary configurations of classical open-string 
field action. 
In our description of open-string field theory,  
we have different classical actions depending 
on the cut-off scale one considers. 
The classical solutions or equivalently solutions of 
the equation of motion (\ref{eq of motion}) are expected  
to depend on the cut-off scale parameter $\zeta$.  
It becomes unclear whether there exists any relation 
between classical solutions at different scales. 
We start our discussion by resolving this question.

Let $\Phi_{\sharp}$ be a solution 
of the equation motion at a given scale $\zeta_0$, 
\begin{eqnarray}
\left. m_0^{\Phi}
\Bigl(1\!:\!\zeta_0 \Bigr)
\right|_{\Phi=\Phi_{\sharp}}=0.
\end{eqnarray}
We examine the RG flow of this classical solution. 
Let $\Phi_{\sharp}(\zeta)$ be the solution of the RG equation 
(\ref{RGE Siegel gauge}) which satisfies the initial value condition 
$\Phi_{\sharp}(\zeta_0)\!=\!\Phi_{\sharp}$.
We obtain the following proposition.
\begin{proposition}
\label{RGF of classical solution}
Suppose we have a classical solution $\Phi_{\sharp}$ 
at a given scale $\zeta_0$. 
The RG flow of $\Phi_{\sharp}$ generates 
a family of classical solutions 
$\Bigl\{ \Phi_{\sharp}(\zeta)\Bigr\}_{\zeta}$.
Namely, at any value of $\zeta$, $\Phi_{\sharp}(\zeta)$ 
satisfies the equation of motion,  
$m_0^{\Phi_{\sharp}(\zeta)}\Bigl(1\!:\!\zeta \Bigr)\!=\!0$. 
\end{proposition}
To explain an implication of this proposition, 
we rewrite the RG equation (\ref{RGE Siegel gauge}), 
by using the anti-commutation relation  
$\left\{Q,b_0\right\}\!=\!L_0$, 
to the form,  
\begin{eqnarray}
\frac{d\Phi(\zeta)}{d\zeta}=
-L_0\Phi(\zeta)
+b_0m_0^{\Phi(\zeta)}
\Bigl(1\!:\!\zeta \Bigr).
\label{RGE Siegel gauge 2}
\end{eqnarray}
Then we recognize that 
the RG flow of a classical solution is described 
by the simple equation, 
\begin{eqnarray}
\frac{d\Phi_{\sharp}(\zeta)}{d\zeta}
=-L_0\Phi_{\sharp}(\zeta), 
\label{RG flow of classical solution}
\end{eqnarray}
because $\Phi_{\sharp}(\zeta)$ satisfies the equation of motion 
as stated in the above proposition. 
This evolution equation is easily integrated 
and provides the following relation.    
\begin{eqnarray}
\Phi_{\sharp}(\zeta)=
e^{-\zeta L_0}\Phi_{\sharp}(0). 
\label{relation classical solutions}
\end{eqnarray}
This means that 
any classical solution in the microscopic description 
($\zeta\!=\!0$)
still continues to be a classical solution in the low 
energy description 
($\zeta\!>\!0$) 
modulo (\ref{relation classical solutions}).  
The classical action of $\Phi_{\sharp}(\zeta)$ 
is independent of $\zeta$. We have 
$S[\Phi_{\sharp}(\zeta)\!:\!\zeta]\!=\!
S^{cubic}[\Phi_{\sharp}(0)]$.

In terms of the super-coordinates $t^i$ 
we can write eq.(\ref{relation classical solutions}) as 
\begin{eqnarray}
t_{\sharp}^i(\zeta)e^{\zeta \Delta_i}
=t_{\sharp}^i(0).   
\label{relation classical solutions 2}
\end{eqnarray}
Hence any classical solution in the microscopic description 
gives rise to a trivial flow on the super-manifold.  
It may sound unpleasant since we hope to interpret it  
as a fixed point of the RG flow. 
We can save this situation by introducing 
rescaled coordinates instead of $t^i$ :
Let $\Phi$ be open-string field at the cut-off scale $\zeta$. 
We expand it in the form,  
$\Phi\!=\!\sum_i T^i e^{-\zeta \Delta_i}\phi_i$. 
The coefficients $T^i$ define the rescaled coordinates. 
The RG flow $\Phi(\zeta)$ determines a flow $T^i(\zeta)$.  
It is related with $t^i(\zeta)$ by 
\begin{eqnarray}
T^i(\zeta)
= t^i(\zeta)e^{\zeta \Delta_i}.
\label{Ti}
\end{eqnarray}
Eq.(\ref{relation classical solutions 2}) becomes      
$T^i_{\sharp}(\zeta)\!=\!T^i_{\sharp}(0)$.  
Classical solutions are fixed points of the flow of $T^i$.

~

Now let us show Proposition \ref{RGF of classical solution}.
The following lemma becomes useful in the proof. 
\begin{lemma}
\label{scale dependence of m0}
Let $\Phi$ be a open-string field. 
The scale dependence of $m_0^{\Phi}(1\!:\!\zeta)$ 
is described by  
\begin{eqnarray}
\frac{\partial m_0^{\Phi}}{\partial \zeta}
\Bigl(1\!:\!\zeta\Bigr)
=
m_1^{\Phi}
\Bigl(L_0\Phi\!:\!\zeta \Bigr)
-2
m_1^{\Phi}
\left( b_0m_0^{\Phi}\left(1\!:\!\zeta\right)
       \!:\!\zeta\right)
+
\bigl(L_0\!-\!2b_0Q \bigr)
m_0^{\Phi}
\Bigl(1\!:\!\zeta\Bigr).
\label{scale dependence of m0 eq}
\end{eqnarray}
\end{lemma}

~

\noindent
\underline{{\it Proof of Lemma \ref{scale dependence of m0}}}~: 
We first compute 
$\partial m_0^{\Phi}/\partial \zeta$ by using Proposition 
\ref{scale dependence of OSFT Ainfty}.
\begin{eqnarray}
\frac{\partial m_0^{\Phi}}{\partial \zeta}
\Bigl(1\!:\!\zeta\Bigr)
&=&
\sum_{k \geq 2}
\frac{\partial m_k}{\partial \zeta}
\left(
\Phi^k\!:\! \zeta 
\right)
\nonumber \\ 
&=&
-
\sum_{k \geq 2}
\left\{
L_0m_k
\left( 
\Phi^k\!:\!\zeta 
\right)
+
\sum_{p=0}^{k-1}
m_k
\left(
\Phi^p, 
L_0\Phi, 
\Phi^{k-1-p}\!:\!\zeta
\right) 
\right\}
\label{proof scale dependence m0 1}
\\
&&
-2
\sum_{k \geq 2}
\sum_{l=2}^{k-1}\sum_{p=0}^{k-l}
m_{k+1-l}
\Bigl(
\Phi^p,
b_0m_l
\bigl(
\Phi^l\!:\!\zeta
\bigr), 
\Phi^{k-l-p}\!:\!\zeta  
\Bigr).
\label{proof scale dependence m0 2}
\end{eqnarray}

We treat two terms 
(\ref{proof scale dependence m0 1}) 
and 
(\ref{proof scale dependence m0 2}) separately. 
We write down the first term into the following form. 
\begin{eqnarray}
Eq.(\ref{proof scale dependence m0 1})
&=& 
-\left\{ 
L_0
\left(
\sum_{k \geq 1}m_k
\Bigl(\Phi^k
\Bigr)
\!-\!
Q\Phi 
\right)
+
\left(
\sum_{k \geq 1}\sum_{p=0}^{k-1}
m_k
\Bigl( 
\Phi^p,L_0\Phi,\Phi^{k-p-1}
\Bigr)
\!-\!
QL_0\Phi
\right)
\right\}
\nonumber \\
&=&
-
L_0
\left(
m_0^{\Phi}
\bigl( 1 \bigr)-Q\Phi 
\right)
-
\left(
m_1^{\Phi}
\bigl(L_0\Phi \bigr)
-QL_0\Phi
\right)
\nonumber \\
&=&
-L_0m_0^{\Phi}\bigl(1\bigr)
-m_1^{\Phi}\bigl(L_0\Phi\bigr)
+2L_0Q\Phi. 
\label{proof scale dependence m0 3}
\end{eqnarray}
As for the second term, we arrange it as follows.
\begin{eqnarray}
Eq.(\ref{proof scale dependence m0 2})
&=&
-2\sum_{k \geq 2}\sum_{l=1}^{k-1}\sum_{p=0}^{k-l}
m_{k+1-l}
\Bigl(
\Phi^p,
b_0m_l \bigl(\Phi^l \bigr),\Phi^{k-l-p}
\Bigr)
+2
\sum_{k\geq 2}\sum_{p=0}^{k-1}
m_k
\Bigl(
\Phi^p,b_0Q\Phi,\Phi^{k-1-p}
\Bigr)
\nonumber \\
&=&
-2
\left\{
m_1^{\Phi}\Bigl(b_0m_0^{\Phi}(1)\Bigr)
-
Qb_0m_0^{\Phi}(1)
\right\}
+2
\left\{
m_1^{\Phi}\Bigl(b_0Q\Phi\Bigr)
-
Qb_0Q\Phi
\right\}
\nonumber \\
&=&
-2
m_1^{\Phi}\Bigl(b_0m_0^{\Phi}(1)\Bigr)
+2
m_1^{\Phi}\Bigl(L_0\Phi\Bigr)
+2
Qb_0m_0^{\Phi}(1)
-2
L_0Q\Phi. 
\label{proof scale dependence m0 4}
\end{eqnarray}
Using expressions (\ref{proof scale dependence m0 3}) 
and (\ref{proof scale dependence m0 4}),  
sum of two terms 
(\ref{proof scale dependence m0 1}) and 
(\ref{proof scale dependence m0 2}) 
becomes the RHS of 
eq.(\ref{scale dependence of m0 eq}).\P

~

\noindent
\underline{{\it Proof of Proposition \ref{RGF of classical solution}}}~: 
Let $\Phi(\zeta)$ be a solution of the RG equation 
(\ref{RGE Siegel gauge}).
We compute $d m_0^{\Phi(\zeta)}/d \zeta$ as follows. 
\begin{eqnarray}
\frac{d}{d \zeta}
m_0^{\Phi(\zeta)}
\Bigl(1\!:\!\zeta \Bigr)
&=& 
\sum_{k \geq 1}
\frac{d}{d \zeta}
m_k 
\Bigl(\Phi(\zeta)^k\!:\! \zeta \Bigr)
\nonumber \\
&=& 
\sum_{k \geq 1}
m_k 
\Bigl(
d \Phi^k/d \zeta \!:\!\zeta
\Bigr)
+
\frac{\partial m_0^{\Phi}}{\partial \zeta}
\Bigl(1\!:\! \zeta  
\Bigr)
\nonumber \\ 
&=&
m_1^{\Phi(\zeta)}
\Bigl(
d \Phi/d \zeta \!:\!\zeta 
\Bigr)
+
\frac{\partial m_0^{\Phi}}{\partial \zeta}
\Bigl(
1\!:\! \zeta
\Bigr). 
\label{proof RGF classical solution 1}
\end{eqnarray}
We further evaluate eq.(\ref{proof RGF classical solution 1}) 
by using Lemma \ref{scale dependence of m0} 
as follows.  
\begin{eqnarray}
Eq.(\ref{proof RGF classical solution 1})
&=&
m_1^{\Phi(\zeta)}
\Bigl( 
d\Phi/d \zeta \!:\!\zeta 
\Bigr)
+
m_1^{\Phi(\zeta)}
\Bigl(L_0\Phi\!:\!\zeta \Bigr)
\nonumber \\
&&~~~~~~~
-2
m_1^{\Phi(\zeta)}
\Bigl( b_0m_0^{\Phi(\zeta)}
\bigl(1\!:\!\zeta \bigr)
       \!:\!\zeta
\Bigr)
+
(L_0\!-\!2b_0Q)
m_0^{\Phi(\zeta)}
\Bigl(1\!:\!\zeta \Bigr) 
\nonumber \\
&=&
(L_0\!-\!2b_0Q)
m_0^{\Phi(\zeta)}
\Bigl(1\!:\!\zeta \Bigr)
-
m_1^{\Phi(\zeta)}
\Bigl(b_0m_0^{\Phi(\zeta)}
\bigl(1\!:\!\zeta \bigr)\!:\!\zeta  
\Bigr),
\label{proof RGF classical solution 2}
\end{eqnarray}
where we use the RG equation (\ref{RGE Siegel gauge 2}) to
show the last equality. 
Thus we find out 
\begin{eqnarray}
\frac{d}{d \zeta}
m_0^{\Phi(\zeta)}
\Bigl(1\!:\!\zeta \Bigr)
=
(L_0\!-\!2b_0Q)
m_0^{\Phi(\zeta)}
\Bigl(1\!:\!\zeta \Bigr)
-
m_1^{\Phi(\zeta)}
\Bigl(b_0m_0^{\Phi(\zeta)}
\bigl(1\!:\!\zeta \bigr)\!:\!\zeta 
\Bigr). 
\label{proof RGF classical solution 3}
\end{eqnarray}

Now we consider the particular case of 
$\Phi_{\sharp}(\zeta)$. 
At $\zeta_0$ it satisfies 
$m_0^{\Phi_{\sharp}(\zeta_0)}(1\!:\!\zeta_0)\!=\!0$. 
The RHS of eq.(\ref{proof RGF classical solution 3}) 
vanishes there. 
This means  
$dm_0^{\Phi(\zeta)}(1\!:\!\zeta)/d\zeta
\left|_{\zeta=\zeta_0}
\!=\!0 \right.$. 
Thus, for any $\zeta$ sufficiently close to $\zeta_0$,  
we have $m_0^{\Phi(\zeta)}(1\!:\!\zeta)\!=\!0$.
For such $\zeta$, 
again by eq.(\ref{proof RGF classical solution 3}), 
$m_0^{\Phi(\zeta)}(1\!:\!\zeta)$ vanishes. 
Repeating this argument, we finally obtain 
$m_0^{\Phi(\zeta)}(1\!:\!\zeta)\!=\!0$ for 
arbitrary $\zeta$.\P

\subsection{Beta functions}
Let us express the RG equation (\ref{RGE Siegel gauge}) 
as a flow equation of the super-coordinates $T^i$.
We first introduce the free energy $F$ as 
a function of $T^i$ and $\zeta$. 
\begin{definition}[Free energy]
\begin{eqnarray}
F(T,\zeta)\equiv
S[\Phi\!:\!\zeta]
\left|_{\Phi=\sum_iT^ie^{-\zeta \Delta_i}\phi_i}
\right.
\label{F-function}
\end{eqnarray}
\end{definition}
\begin{proposition}[Beta function]
\label{def beta function} 
The RG equation (\ref{RGE Siegel gauge}) is equivalent 
to the following evolution equations of $T^i$. 
\begin{eqnarray}
\frac{d T^i}{d \zeta}
=\beta^i(T,\zeta),
\label{RGE by beta}
\end{eqnarray}
where beta functions $\beta^i$ are given by 
\begin{eqnarray}
\beta^i(T,\zeta) 
=
\sum_j 
g_{ren}^{ij}(\zeta)
\frac{\partial^L F(T,\zeta)}{\partial T^j}.
\label{def of beta}
\end{eqnarray}
Here $g_{ren}^{ij}(\zeta)$ in eqs.(\ref{def of beta}) 
is the inverse of the renormalized bilinear form,   
$g^{ren}_{ij}(\zeta)
\!\equiv\!
e^{-\zeta(\Delta_i+\Delta_j)}
g_{ij}$. 
\end{proposition}
Fixed points of the evolution equations (\ref{RGE by beta}) are 
zeros of the beta functions. 
According to eqs.(\ref{def of beta}) these zeros are solutions of 
$\partial^LF/\partial T^i\!=\!0$ since $g^{ij}$ is 
non-degenerate on ${\cal H}^S$. 
The free energy $F$ is given by eq.(\ref{F-function}). 
Therefore zeros of the beta functions are nothing but 
classical solutions of open-string field theory.

Perturbative expansions 
(expansions at $T\!=\!0$) of the beta functions 
can be read from the perturbative expansion of 
$F(T,\zeta)$. They have the following forms 
(cf. proof of proposition \ref{def beta function}).
\begin{eqnarray}
\beta^i(T,\zeta)=
\Delta_iT^i
+\sum_{j,k,l}g_{ren}^{ij}(\zeta)C_{jkl}^{ren}(\zeta)T^kT^l
+O(T^3).  
\end{eqnarray}
Here $C_{ijk}^{ren}(\zeta)$ 
are the renormalized three points functions, 
$C_{ijk}^{ren}(\zeta)\!\equiv\!
e^{-2\zeta(\Delta_i+\Delta_j+\Delta_k)}C_{ijk}$, 
where we put 
$C_{ijk}\!=\! 
(-)^{G(T^i)+G(T^j)+G(T^j)G(T^k)}
\Bll 1~2~3 \Br 
\phi_i \brr_1
\bl \phi_j \brr_2 \bl \phi_k \brr_3$.
Contrary to our naive expectation, the beta functions given by  
eqs.(\ref{def of beta}) depend on the cut-off scale parameter 
$\zeta$ {\it explicitly}. 
In a quantum field theory,  
a standard argument to show that 
beta functions of the theory 
have no explicit dependence on the cut-off scale 
is based on a simple dimensional analysis.  
In this argument it is assumed 
that there is at most only one dimensionful parameter 
whatever regularization scheme one chooses.  
The regularization we choose for open-string field theory does 
not satisfy this property. 
We formulate open-string field theory from the perspective 
of {\it two-dimensional} chiral CFT. 
But the regularization we choose is simply to put a restriction 
on length of open-string evolution. It is a regularization of
{\it one-dimension}. Actually we have two length scales. The missing 
scale is length of open-string itself. If we restore string length scale 
$l_s$ in the argument, there appear two dimensionful 
parameters $\zeta l_s$ and $l_s$ in our regularization. 
\begin{remark}
Dependence of the beta functions on the cut-off scale parameter 
can be seen as follows : 
We put $\Phi(T,\zeta)\!\equiv\!
\sum_iT^ie^{-\zeta \Delta_i}\phi_i$. 
By using the RG equation (\ref{RGE Siegel gauge}) we obtain   
the following expression for the beta functions. 
\begin{eqnarray}
\sum_i\beta^i\phi_i=
e^{\zeta L_0}b_0
m_0^{\Phi(T,\zeta)}
\Bigl(1\!:\!\zeta \Bigr).
\label{beta by m0}
\end{eqnarray}
We partial-differentiate eq.(\ref{beta by m0}) with respect to $\zeta$. 
By using Lemma \ref{scale dependence of m0} it turns out to be 
\begin{eqnarray}
\sum_i
\frac{\partial \beta^i}{\partial \zeta}\phi^i
=
-2
e^{\zeta L_0}b_0
\sum_{k_1+k_2\geq 1}
m_{k_1+k_2+1}
\Bigl( 
\Phi^{k_1},
e^{-\zeta L_0}\sum_j\beta^j\phi_j, 
\Phi^{k_2}
\!:\!\zeta  
\Bigr).
\label{d beta by m1}
\end{eqnarray}
\end{remark}

This remark or proposition \ref{RGF of classical solution} 
shows that zeros of the beta functions (\ref{def of beta}) 
are independent of $\zeta$.  
Let us suppose that $0$ and 
$T_c\!=\!(T_c^i)_i$ $(\neq 0)$ be zeros of the beta 
functions at $\zeta\!=\!0$. Namely 
$\beta^i(T,\zeta\!=\!0)\Bl_{T=0,T_c}=0$. 
Since they are the zeros even at $\zeta\!>\!0$, 
the beta functions must have the following forms.  
\begin{eqnarray}
\beta^i(T,\zeta)=
\sum_{N,M}\beta^i_{N,M}(\zeta)T^N(T-T_c)^M, 
\end{eqnarray}
where 
$N\!=\!(n_i)_i$ and $M\!=\!(m_i)_i$ are multi-indices 
taking values in ${\bf Z}_{\geq 1}$.  
This means that $\zeta$-dependence of the beta functions 
is absorbed into the coefficients $\beta^i_{N,M}$.

The RG equation of open-string field 
is originally introduced by eq.(\ref{def RG equation}) 
in Definition \ref{def RG equation for OSFT}. 
In the Siegel gauge we can write down this equation 
as the $\mbox{Gell-Mann}\!-\!\mbox{Low}$ equation for the free energy.
\begin{eqnarray}
\frac{\partial F(T,\zeta)}{\partial \zeta}
+
\sum_i \beta^i
\frac{\partial^LF(T,\zeta)}{\partial T^i}
=0.
\label{Callan-Symanzik 1}
\end{eqnarray}
This is an equation which 
determines the evolution of $T^i$. 
The RG flow $T(\zeta)$ is a solution of this equation.
By using eqs.(\ref{def of beta})
we can rewrite the $\mbox{Gell-Mann}\!-\!\mbox{Low}$ 
equation (\ref{Callan-Symanzik 1}) as follows.
\begin{eqnarray}
\frac{\partial F(T,\zeta)}{\partial \zeta}
=-\sum_{i,j}
\beta^i g_{ij}^{ren} \beta^j. 
\label{Callan-Symanzik 2}
\end{eqnarray}
By a slight look of this expression 
besides (\ref{def of beta}) of the beta functions,  
one may feel a similarity between 
the free energy $F$ (\ref{F-function}) and 
Zamolodochikov's $c$-function.  
But it is merely a formal resemblance
\footnote{Nevertheless, 
possibility of taking classical open-string field 
action as an analogue of $c$-function, which one calls 
$g$-function, was discussed in \cite{Kutasov-Marino-Moore}. 
The argument there was based on the assumption that the free energy 
does not depend on $\zeta$ explicitly. 
A discrepancy between the two is also discussed recently 
in \cite{Fujii-Itoyama}.}.

In a conventional approach to open-string field theory, 
presumably assuming a suitable decoupling of the ghost sector,  
one uses mostly the matter Hilbert space ${\cal H}_{matter}$ 
rather than the open-string Hilbert space. 
We identify ${\cal H}_{matter}$ with a subspace of ${\cal H}^S$ 
by the map 
$i$ 
$:~{\cal H}_{matter} \rightarrow {\cal H}^S$, 
where $i(O)\equiv c_1O$.
To present a physical application of the proposition,  
we follow the conventional approach 
although the restriction on ${\cal H}_{matter}$ 
might cause a problem on the RG equation. 
Practically we assume that the RG flow or the $\zeta$-evolution 
of $\Phi$ is closed on ${\cal H}_{matter}$\footnote{This might be 
shown by a detailed analysis of the symmetric $3$-vertex.}.

The bilinear form $g$, 
when restricted on ${\cal H}_{matter}$,  
turns out to have the form. 
\begin{eqnarray}
g(c_1O,c_1O')=
\Bigl\langle 
I[\varphi_{O}](\infty)\varphi_{O'}(0) 
\Bigr\rangle, 
\label{BPZ metric of matter}
\end{eqnarray}
where the RHS is the standard BPZ pairing of the matter CFT. 
Eq.(\ref{BPZ metric of matter}) can be shown 
if one computes the LHS by using the oscillator representation 
(\ref{reflector by modes}) of the reflector.  
Let $\Bigl\{O_I\Bigr\}_I$ be bases of ${\cal H}_{matter}$. 
$O_I$ are chosen as the eigenvectors of $L_0^{matter}$ :
$L_0^{matter}O_I\!=\!\Delta_{I}O_I$.
We put $g_{IJ}\!\equiv\!g(c_1O_I,c_1O_J)$. 
These define a positive-definite symmetric bilinear form on 
${\cal H}_{matter}$. 
Open-string field is now supposed to be 
a vector of ${\cal H}_{matter}$. 
We put $\Phi\!=\!
\sum_IT^Ie^{-\zeta (\Delta_I-1)}c_1O_I$. 
If one takes the $\sigma$-model viewpoint of the matter theory, 
$\Phi$ describes a generic perturbation $\sum_IT^I\varphi_{O_I}$ 
of the $\sigma$-model and $\beta^I$ 
are the beta functions associated with this perturbation.
Under this circumstance the free energy $F$ can be thought as 
a generating function of all correlation functions of the matter theory.   
Eqs.(\ref{def of beta}) can be written down as follows.   
\begin{eqnarray}
\frac{\partial F(T,\zeta)}{\partial T^I}
&=&
\sum_I g_{IJ}^{ren} \beta^I. 
\label{matter theory description} 
\end{eqnarray}
Closed-string version of eq.(\ref{matter theory description}) 
can be found in \cite{Polyakov's black book}, where it was stated 
as a conjecture.

~

\noindent
\underline{{\it Proof of Proposition \ref{def beta function}}}~: 
Let us put 
$\Phi\!=\!
\sum_iT^ie^{-\zeta \Delta_i}
\phi_i$. 
We can express the coefficients $T^i$ by 
$T^i\!=\!
e^{\zeta \Delta_i}
g(\phi^i,\Phi)$. 
Then $dT^i/d\zeta$ acquires the following form. 
\begin{eqnarray}
\frac{dT^i}{d \zeta}= 
e^{\zeta \Delta_i}
g
\left(
\phi^i, \frac{d \Phi}{d \zeta}
\right)
+
\Delta_i T^i. 
\label{proof beta 0}
\end{eqnarray}
We evaluate the RHS of eq.(\ref{proof beta 0}) 
by using the RG equation (\ref{RGE Siegel gauge}) 
as follows.
\begin{eqnarray}
\frac{dT^i}{d \zeta}= 
e^{\zeta \Delta_i}
g
\left(
\phi^i, 
b_0
\left| \frac{\partial S_{int}}{\partial \Phi}
\right \rangle
\right)
+
\Delta_i T^i.
\label{proof beta 1}
\end{eqnarray}

We treat two terms of eq.(\ref{proof beta 1}) separately. 
The bilinear form in the first term can be expressed 
as a pairing by $\omega$ as follows.     
\begin{eqnarray}
g
\left(
\phi^i,
b_0 
\left| \frac{\partial S_{int}}{\partial \Phi}
\right \rangle
\right)
&=&
\Bll  \omega_{12} \Br
\Bigl(
c_0^{(1)}
\bl \phi^i \brr_1
\Bigr)
\left(
b_0^{(2)}
\left|
\frac{\partial S_{int}}{\partial \Phi}
\right \rangle_2
\right)
\nonumber \\
&=&
(-)^{\epsilon(\phi^i)}
\Bll \omega_{12} \Br
\left(
b_0^{(1)}c_0^{(1)}
\bl \phi^i \brr_1
\right)
\left|
\frac{\partial S_{int}}{\partial \Phi}
\right \rangle_2
\nonumber \\
&=&
\omega
\left(
\phi^i,
\left|
\frac{\partial S_{int}}{\partial \Phi}
\right \rangle
\right), 
\label{proof beta 2}
\end{eqnarray}
where the last equality follows from the gauge condition,  
$b_0\phi^i\!=\!0$ besides the anti-commutation relation, 
$\left\{c_0,b_0\right\}\!=\!1$. 
We remark that $\phi^i$ is expressed in terms of $\Phi$ as 
follow.
\begin{eqnarray} 
\phi^i
=\sum_j
e^{\zeta \Delta_j}g^{ij} 
\frac{\partial^L\Phi}{\partial T^j}.
\end{eqnarray}  
By using this expression for $\phi^i$ in eq.(\ref{proof beta 2}), 
it follows from Definition \ref{hamilton vector} that  
the first term of eq.(\ref{proof beta 1}) becomes   
\begin{eqnarray}
e^{\zeta \Delta_i}
g
\left(
\phi^i, 
b_0
\left| \frac{\partial S_{int}}{\partial \Phi}
\right \rangle
\right)
=
\sum_j
e^{\zeta (\Delta_i+\Delta_j)}
g^{ij}
\frac{\partial^L}{\partial T^j}
\left\{
S_{int}\Bigl[\Phi\!:\!\Lambda\Bigr]
\left|_{\Phi\!=\!
\sum_kT^ke^{-\zeta \Delta_k}\phi_k}
\right. 
\right\}.
\label{proof beta 3}
\end{eqnarray}

Nextly we examine the second term of eq.(\ref{proof beta 1}). 
In the Siegel gauge, the BRST charge $Q$ contributes to 
$\omega(\Phi,Q\Phi)$ as an operator $c_0L_0$. Thus we have 
the following identity. 
\begin{eqnarray}
\omega
\Bigl(
\Phi,Q\Phi
\Bigr)
=
g
\Bigl(
\Phi,L_0\Phi
\Bigr). 
\label{proof beta 4}
\end{eqnarray}
This means that the quadratic term of 
the open-string field action 
has the following expression in terms of $T^i$. 
\begin{eqnarray}
\omega
\left(
\Phi,
\frac{1}{2}Q\Phi
\right)
\left|_{\Phi\!=\!
\sum_kT^ke^{-\zeta \Delta_k}\phi_k}
\right.
=
\frac{1}{2}
\sum_{i,j}
T^ie^{-\zeta (\Delta_i+\Delta_j)}
g_{ij}\Delta_jT^j.
\label{proof beta 5}
\end{eqnarray}
Then we write down the second term 
of eq.(\ref{proof beta 1}) as a differentiation of 
$\omega(\Phi,Q\Phi)$ by $T$. 
\begin{eqnarray}
\Delta_iT^i=
\sum_j
e^{\zeta (\Delta_i+\Delta_j)}
g^{ij}
\frac{\partial^L}{\partial T^j}
\left\{
\omega
\left(
\Phi,
\frac{1}{2}Q\Phi
\right)
\left|_{\Phi\!=\!
\sum_kT^ie^{-\zeta \Delta_k}\phi_k}
\right. 
\right\}.
\label{proof beta 6}
\end{eqnarray}

By using the expressions 
(\ref{proof beta 3}) and (\ref{proof beta 6}) 
we proceed on the computation of eq.(\ref{proof beta 1}) 
as follows.  
\begin{eqnarray}
\frac{dT^i}{d\zeta}
&=&
\sum_j
e^{\zeta (\Delta_i+\Delta_j)}
g^{ij}
\frac{\partial^L}{\partial T^j}
\left[
\left\{
S_{int}\Bigl[\Phi\!:\!\zeta \Bigr]
+
\omega
\left(
\Phi,
\frac{1}{2}Q\Phi
\right)
\right\}
\left|_{\Phi\!=\!
\sum_kT^ie^{-\zeta \Delta_k}\phi_k}
\right. 
\right]
\nonumber \\
&=& 
\sum_j
e^{\zeta (\Delta_i+\Delta_j)}
g^{ij}
\frac{\partial^L}{\partial T^j}
\left[
S\Bigl[\Phi\!:\!\zeta \Bigr]
\left|_{\Phi\!=\!
\sum_kT^ie^{-\zeta \Delta_k}\phi_k}
\right. 
\right]
\nonumber \\
&=&
\sum_j
e^{\zeta (\Delta_i+\Delta_j)}
g^{ij}
\frac{\partial^LF(T,\zeta)}{\partial T^j}.
\end{eqnarray}
Thus we obtain eq.(\ref{def of beta}).\P

\subsection{Perspective of world-sheet boundary theory}

To close this section we hint an interpretation 
of the previous results in terms of 
world-sheet boundary theories. 
In particular we clarify the role of the cut-off 
scale parameter $\zeta$. 
In the next section 
we further develop this perspective.

~

(Tree) open-string diagrams can be mapped holomorphically 
to the upper half-plane $\mbox{Im}z>0$. 
These maps are called Mandelstam maps. 
External open-strings in $n$-diagram 
are mapped to $n$ different points on the real line. 
The inverse of Mandelstam maps are given by using quadratic 
differentials on ${\bf CP}_1$.   
There exists one-to-one correspondence \cite{Strebel} between 
the sets of open-string tree diagrams and quadratic 
differentials on ${\bf CP}_1$ which satisfy suitable 
conditions.

\begin{figure}
\psfrag{0}{$0$}
\psfrag{i}{$i$}
\psfrag{sqrt3}{$\sqrt{3}$}
\psfrag{-sqrt3}{$-\sqrt{3}$}
\psfrag{z=x+iy}{$z=x+iy$}
\begin{center}
\includegraphics[height=7cm]{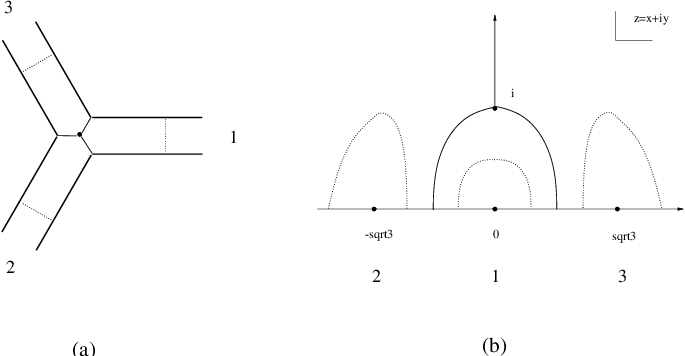}
\caption{
{\small 
(a) Trivalent open-string diagram. 
(b) The image of trivalent diagram on the upper half-plane. }}
\label{mandelstam0}
\end{center}
\end{figure}

Let us explain the use of quadratic differentials 
by a simple example. 
Trivalent open-string diagram (Figure \ref{mandelstam0} (a)) 
can be mapped to the upper half-plane as depicted in 
Figure \ref{mandelstam0} (b). 
Adding the mirror image we obtain Figure \ref{fiz}. 
There ${\bf CP}_1$ is exactly covered by $f_i(V_i)$ 
$(i\!=\!1,2,3)$.  
Recall $V_i$ is the unit disk $|v_i| \leq 1$ and 
$f_i$ is the holomorphic map (\ref{maps of 3 vertex in z}). 
The quadratic differential associated with 
Figure \ref{mandelstam0} (b) ( or Figure \ref{fiz}) is 
\begin{eqnarray}
\varphi=
\frac{9(1+z^2)}{z^2(z^2-3)^2}dz^{\otimes 2}. 
\label{qad 3 vertex}
\end{eqnarray}
It has second-order poles at $0$ and $\pm \sqrt{3}$ 
and first-order zeros at $\pm i$. Also it has the 
following expansion in the neighborhood of each pole. 
\begin{eqnarray}
\varphi =
\left\{ \frac{1}{(z-z_P)^2}+ \cdots \right\}
dz^{\otimes 2},  
\label{quad near pole}
\end{eqnarray}
where $P$ denotes the pole.
Let $\sqrt{\varphi}^{(i)}$ be a differential on 
$f_i(V_i)$ which satisfies 
$(\sqrt{\varphi}^{(i)})^{\otimes 2}=\varphi$. 
Particularly we can choose $\sqrt{\varphi}^{(i)}$ such that 
\begin{eqnarray}
f_i^*\sqrt{\varphi}^{(i)}=\frac{dv_i}{v_i}. 
\label{quad 3 vertex v}
\end{eqnarray}
We then introduce an analytic function $\rho_i$ on 
$f_i(V_i)$ by an integral of $\sqrt{\varphi}^{(i)}$ 
in the following manner. 
\begin{eqnarray}
\rho_1(z)-\rho_1(0)=
\int_0^z \sqrt{\varphi}^{(1)}. 
\end{eqnarray} 
Due to the property (\ref{quad 3 vertex v}) 
we can regard $e^{\rho_i}$ as a holomorphic map 
from $f_i(V_i)$ to $V_i$. 
It is the inverse of $f_i$. 
\begin{eqnarray}
e^{\rho_i}\circ f_i= 1. 
\end{eqnarray} 
The $i$-th strip in Figure \ref{mandelstam0} (a) 
is parametrized by $\rho_i=\tau_i+i\sigma_i$.

~

\begin{figure}
\psfrag{2 zeta}{$2 \zeta$}
\psfrag{a}{$a$}
\psfrag{-a}{$-a$}
\psfrag{1/a}{$1/a$}
\psfrag{-1/a}{$-1/a$}
\psfrag{ib}{$ib$}
\psfrag{i/b}{$i/b$}
\psfrag{i}{$i$}
\psfrag{0}{$0$}
\psfrag{z=x+iy}{$z=x+iy$}
\begin{center}
\includegraphics[height=7cm]{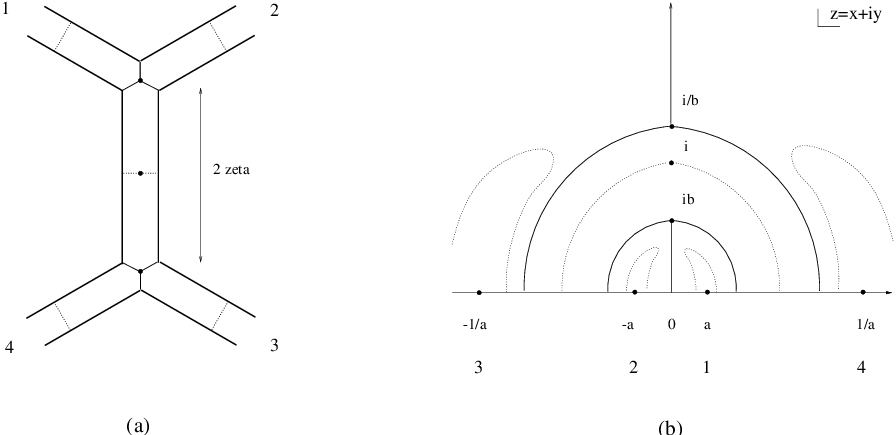}
\caption{
{\small 
(a) The $s$-channel diagram which appears in 
$\partial {\cal V}_4(\zeta)$. 
(b) The $s$-channel diagram is mapped holomorphically 
to the upper half-plane. The mid-points of the trivalent 
vertices are mapped to $ib$ and $i/b$. The external 
open-strings $1,2,3$ and $4$ are mapped respectively to 
$\pm a$ and $\pm 1/a$ on the real line. }}
\label{mandelstam}
\end{center}
\end{figure}

Consider the $s$-channel diagram which appears in 
$\partial {\cal V}_4(\zeta)$. 
See Figure \ref{mandelstam} (a). 
It can be mapped to the upper half-plane. 
The image is depicted in Figure \ref{mandelstam} (b). 
The four open-strings are mapped to 
$\pm a$ and $\pm 1/a$ on the real line. 
The mid-points of the trivalent vertices are 
mapped to $ib$ and $i/b$.  
($0<a,b<1$). 
$a$ and $b$ are suitable functions of $\zeta$. 
These functions are determined 
from a consideration on a quadratic differential 
associated with Figure \ref{mandelstam} (b).

The quadratic differential needs to have  
second-order poles at $\pm a$ and $\pm 1/a$ and 
first-order zeros at $\pm ib$ and $\pm i/b$. 
In the neighborhood of each pole it must have  
the expansion (\ref{quad near pole}).
These conditions restrict the quadratic differential 
to the following form.  
\begin{eqnarray}
\varphi =
\frac{4a^2(a^4-1)^2}{(a^2+b^2)(1+a^2b^2)}
\frac{(z^2+b^2)(1+b^2z^2)}{(z^2-a^2)^2(1-a^2z^2)^2} 
dz^{\otimes 2}. 
\label{quad 4vertex}
\end{eqnarray}
Let $\sqrt{\varphi}$ be a differential defined locally 
(in each image of the open-string strip and its mirror image). 
We further need to impose the condition which 
fixes width of the internal strip to $\pi$. 
\begin{eqnarray}
\int_{|z|=1}\sqrt{\varphi}=2\pi i. 
\label{varphi pi}
\end{eqnarray} 
This allows us to solve $b$ in terms of $a$. 
Thus obtained quadratic differentials describe open-string 
$4$-diagrams. Finally we need to relate $a$  with $\zeta$ 
so that $\varphi$ describes Figure \ref{mandelstam} (a). 
It can be achieved by imposing the condition,  
\begin{eqnarray}
\int_{ib}^i\sqrt{\varphi}=\zeta, 
\label{varphi zeta}
\end{eqnarray}
where the integral is taken along the imaginary axis.

The asymptotic forms of 
$a(\zeta)$ and $b(\zeta)$ 
can be written down explicitly. 
They are expected to approach to zero 
as $\zeta$ goes to $\infty$. When $a$ is nearly zero we can 
solve the condition (\ref{varphi pi}) in the following form. 
\begin{eqnarray}
b(a)=\sqrt{3}a+O(a^2). 
\end{eqnarray}
Therefore we approximate the quadratic differential 
(\ref{quad 4vertex}) by 
$\frac{z^2+3a^2}{(z^2-a^2)^2}dz^{\otimes 2}$. 
The condition (\ref{varphi zeta}) will be replaced by  
the following one. 
\begin{eqnarray}
\int_{i\sqrt{3}a}^i
\frac{\sqrt{z^2+3a^2}}{z^2-a^2}dz
=\zeta. 
\label{varphi zeta 2}
\end{eqnarray}    
This integral is easily evaluated and we obtain 
\begin{eqnarray}
a(\zeta)=\frac{2}{3\sqrt{3}}e^{-\zeta}
+O(e^{-3\zeta}). 
\label{a(zeta)}
\end{eqnarray}
The approximation used above becomes consistent 
when $a$ is sufficiently small. 
It is when $\zeta$ is sufficiently large.

~

Each strip of an open-string diagram has a k\"{a}hler metric 
$d\rho d\overline{\rho}$. ($\rho=\tau+i\sigma$ is a holomorphic 
coordinate of strip.) They induce a k\"{a}hler metric on the upper 
half-plane. We call it bulk metric $d^2s_{bulk}$. 
If one extends the bulk metric to the real line, 
it becomes singular at the points where the open-strings 
are inserted. From the perspective of world-sheet boundary 
theories we should rescale the bulk metric near the real line 
such that the rescaled metric provides a smooth metric 
on the real line. World-sheet boundary theory will be 
constructed on the real line  by using such a smooth metric 
$d^2s_{boundary}$.  
A convenient choice of a smooth metric 
on the boundary is 
\begin{eqnarray}
d^2s_{boundary}=dxdx. 
\label{boundary metric}
\end{eqnarray}
If we use the metric (\ref{boundary metric}) on the real line, 
the regularization employed so far in open-string field theory 
corresponds to a point-splitting regularization of 
short-distance behavior of a boundary theory. 
The estimation (\ref{a(zeta)}) shows that the point-splitting 
is prescribed by the boundary length scale $\delta l$ with the order  
\begin{eqnarray}
\delta l  \sim e^{-\zeta}.  
\label{delta l}
\end{eqnarray}

The low energy action $S[\Phi\!:\!\zeta]$ can be understood 
as an analogue of a generating function of all correlation 
functions of a world-sheet boundary theory regularized  
by the point-splitting (\ref{delta l}).  
$\zeta$ needs to be sufficiently large. In the next section we 
identify it with the so-called boundary open-string field 
theory.


\section{Boundary Open-String Field Theory}
In this section we discuss a relation between two different 
formulations of classical open-string field theory. 
One is the formulation given in the previous sections of this paper. 
It is based on the Wilson renormalization group of world-sheet theory 
in which the microscopic action is identified with the cubic action 
(\ref{microscopic action}) and the cut-off scale parameter is 
introduced by length of open-string strip (which appears in the Schwinger 
representation of the open-string propagator). 
The other is the so-called boundary open-string field theory.  
It is proposed in \cite{Witten BOSFT} intended for the background 
independent description of open-string field theory. 
To discuss their relation we need to introduce boundary states. 
These are states of the closed-string Hilbert space. 
String vertices which describe interactions of open-strings with a 
single closed-string are required. 
We start this section by a consideration of the moduli problem 
relevant to constructions of these vertices.  
Our discussion in this section is not rigorous although it is 
physically motivated. We believe that perspectives presented in this 
section become useful in our understanding of duality of open- and closed- 
strings or holography in string theory.

~

\noindent
Let $\widehat{{\cal M}}^1_n$ be the set of 
clockwise-ordered $n$ different points 
$(z_1,\cdots,z_n)$ on the boundary of 
{\it one-punctured} two-disk. 
Let $z_0$ be the puncture. 
\begin{eqnarray}
\widehat{{\cal M}}^1_n = 
\left\{
(z_0;z_1,\cdots,z_n)
\left|
\begin{array}{c}
z_i \in \partial D
~~(1\!\leq\!i\!\leq\!n),~~
z_0 \in D,
\\
z_i\!\neq\! z_j ~~\mbox{if}~~i\!\neq\!j.  
\end{array}
\right.\right\}.
\label{hatM(1,n)}
\end{eqnarray}
We also let $\widehat{{\cal M}}^0_n$ be the set of 
clockwise-ordered $n$ different points $(z_1,\cdots,z_n)$ 
on the boundary of two-disk. 
\begin{eqnarray}
\widehat{{\cal M}}^0_n =  
\left\{
(z_1,\cdots,z_n)
\left|
\begin{array}{c}
z_i \in \partial D
~~(1\!\leq\!i\!\leq\!n),
\\
z_i\!\neq\! z_j ~~\mbox{if}~~i\!\neq\!j.  
\end{array}
\right.\right\}.
\label{hatM(0,n)}
\end{eqnarray}
$SL_2({\bf R})$ acts on the both sets in the standard manner.
We put 
${\cal M}^1_n \equiv \widehat{{\cal M}}^1_n/SL_2({\bf R})$ 
for $n \geq 1$, and 
${\cal M}^0_n \equiv \widehat{{\cal M}}^0_n/SL_2({\bf R})$ 
for $n \geq 3$. 
In the previous sections we denote ${\cal M}^0_n$ by 
${\cal M}^{\partial}_n$. 
We can always put the puncture at the origin of two-disk by using  
the $SL_2({\bf R})$. Thereby ${\cal M}^1_n$ is identified with 
$\widehat{{\cal M}}^0_n/U(1)$, where $U(1)$ is the rotation 
group of $\partial D$.

\begin{figure}[t]
\begin{center}
\includegraphics[height=4.5cm]{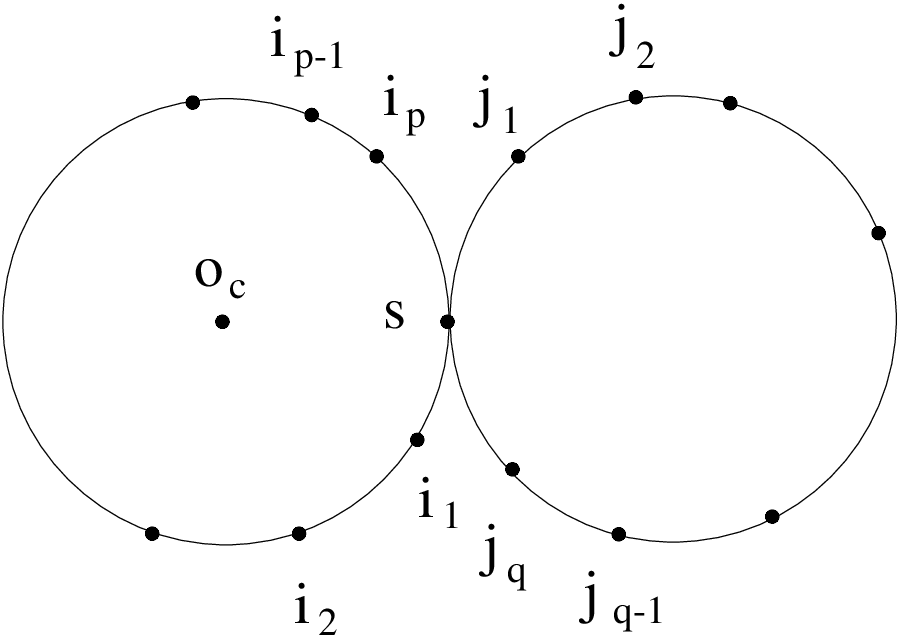}
\caption{
{\small 
A typical configuration at the infinities  
${\cal M}^1_{(I,s)}\times {\cal M}^0_{(s,J)}$ of 
${\cal M}^1_n$.}}
\label{boundary-closed fig}
\end{center}
\end{figure}

To describe infinities of ${\cal M}^1_n$, 
we introduce a pair of multi-indices $(I,J)$,
where the multi-indices 
$I\!=\!(i_1,\cdots,i_p)$ $(p\!\geq\!0)$ 
and $J\!=\!(j_1,\cdots,j_q)$ $(q\!\geq\!2)$
satisfy the conditions : 
$I \cap J\!=\!\emptyset$ and 
$(i_1,\cdots,i_p,j_1,\cdots,j_q)
=(1,\cdots,n)$ mod cyclic permutations. 
At infinities of ${\cal M}^1_n$ there appear 
${\cal M}^1_{(I,s)}\times {\cal M}^0_{(s,J)}$ 
for all the pairs $(I,J)$. ``$s$'' denotes the singular 
point in the configuration at the infinities. 
See Figure \ref{boundary-closed fig}. 
%
%
%
Stable compactification of ${\cal M}^1_n$, which we call 
${\cal CM}^1_n$, is defined inductively by adding 
${\cal CM}^1_{(I,s)}
\times 
{\cal CM}^0_{(s,J)}$ 
to the infinities of ${\cal M}^1_n$. 
\footnote{${\cal CM}^0_{(s,J)}$ is 
${\cal CM}^{\partial}_{(s,J)}$ 
in the previous notation.}
\begin{eqnarray}
{\cal CM}^1_n
=
{\cal M}^1_n 
\cup 
\left\{ \bigcup_{(I,J)}
{\cal CM}^1_{(I,s)}
\times 
{\cal CM}^0_{(s,J)}
\right\},
\label{CM(1,n)}
\end{eqnarray}
where ${\cal CM}^1_1 \equiv {\cal M}^1_1$ is a point.
Topologically ${\cal CM}_n^1$ becomes 
a $(n\!-\!1)$-dimensional ball $B_{n-1}$. 
This can be seen from the identification of 
${\cal M}^1_n$ with $\widehat{{\cal M}}^0_n/U(1)$. 
We fix orientation of ${\cal CM}^1_n$ by the standard orientation 
of $B_{n-1}$. Taking account of the orientations we obtain 
\begin{eqnarray}
\partial {\cal CM}^1_n
=
\sum_{(I,J)}
(\pm){\cal CM}^1_{(I,s)}
\times 
{\cal CM}^0_{(s,J)}. 
\label{boundary of CM(0,1)}
\end{eqnarray}
The signature $(\pm)$ must be determined by a comparison 
of the orientations of 
$\partial {\cal CM}^1_n$ and 
${\cal CM}^1_{(I,s)}\times {\cal CM}^0_{(s,J)}$. 
Later we will discuss about this problem.

In the previous sections  
we introduced ${\cal V}_n^0(\zeta)\equiv{\cal V}_n(\zeta)$, 
which is the subset of ${\cal CM}^0_n$,  
in order to obtain open-string vertices at the cut-off scale $\zeta$. 
Analogous consideration is also possible at the present situation : 
We identify ${\cal M}^1_n$ with $\widehat{{\cal M}}^0_n/U(1)$. 
Consider the configurations near the infinities. 
In these configurations some of $n$ points (on $\partial D$) are getting 
``close'' to one another. We can measure how close they are by length of 
open-string strips. Length of open-string strips provide a local coordinates 
at least near the boundary of ${\cal CM}^1_n$. 
If we put a restriction on their length by the scale parameter $\zeta$, 
the set of forbidden configurations provides  
a neighborhood of the infinities.  
Let us denote this neighborhood by ${\cal P}_n^1(\zeta)$. 
We put ${\cal V}_n^1(\zeta)\equiv
{\cal CM}^1_n\setminus {\cal P}^1_n(\zeta)$. 
This plays the same role as ${\cal V}^0_n(\zeta)$ in the construction 
of string vertex. ${\cal V}_n^1(\zeta)$ is topologically $B_{n-1}$. 
Thus obtained ${\cal V}_n^1(\zeta)$ for $\zeta >0$ satisfies 
\begin{eqnarray}
\partial {\cal V}^1_n(\zeta)
=
\sum_{(I,J)}
(\pm){\cal V}^1_{(I,s)}(\zeta)
\times 
{\cal V}^0_{(s,J)}(\zeta).
\label{boundary of V(0,1)(Lambda)}
\end{eqnarray}
This equation plays an important role of our understanding 
of boundary states. 

~

\noindent
Boundary states are introduced as states of closed-string. 
For their description we prepare some terminology 
of closed-string field theory. 
We follow the convention used in \cite{Zwiebach BV}. 
Let ${\cal H}^c_{matter}$ and ${\cal H}^c_{ghost}$ be respectively 
the matter and ghost Hilbert spaces of critical bosonic string. 
We put 
${\cal H}_{aux}\equiv {\cal H}^c_{matter}\otimes {\cal H}^c_{ghost}$. 
The $SL_2$-invariant vacuum $|0\rangle$ of 
the auxiliary Hilbert space ${\cal H}_{aux}$ is set to have no ghost number. 
The $SL_2$-invariant vacuum $\langle 0|$ of the dual Hilbert space 
${\cal H}_{aux}^*$ is also set to have no ghost number. 
Both states are grassmann-even. 
Dual pairing between ${\cal H}_{aux}$ and ${\cal H}^*_{aux}$ is prescribed 
based on $\langle0|c_{-1}\bar{c}_{-1}
c_0\bar{c}_0c_1\bar{c}_1|0\rangle \!=\!1$, 
where $c_{\pm 1,0}$ and $\bar{c}_{\pm 1,0}$ are respectively 
the ghost zero modes of chiral and anti-chiral parts on ${\bf CP}_1$. 
Closed-string Hilbert space ${\cal H}_c$ 
consists of vectors $\psi$ of ${\cal H}_{aux}$ 
which satisfy the conditions, 
$b_0^-\psi\!=\!L_0^-\psi\!=\!0$
\footnote{We put 
$L_0^{\pm}\!\equiv\! L_0\pm\bar{L}_0$, 
$b_0^{\pm}\!\equiv\! b_0\pm\bar{b}_0$ 
and 
$c_0^{\pm}\!\equiv\! (c_0\pm\bar{c}_0)/2$.}. 
Closed-string field $\Psi$ is a grassmann-even vector of ${\cal H}_c$ 
and has the ghost number $G(\Psi)\!=\!2$.

Closed-string Hilbert space has an odd symplectic structure 
$\omega^c$. We write it in the form, 
$\omega^c(A,B)=
\langle \omega^c_{12}|A\rangle_1|B\rangle_2$ 
for $A,B \in {\cal H}_c$. 
$\langle \omega^c_{12}|$ is an element of 
$({\cal H}_c^{\otimes 2})^*$. 
It is defined by the restriction of $\langle R_{12}^c|c_0^{-(2)}$ 
on ${\cal H}_c$. 
Here $\langle R_{12}^c|$ is the reflector of ${\cal H}_{aux}$.  
It is given by using the BPZ pairing as follows.  
$\langle R_{12}^c|A\rangle_1|B\rangle_2\equiv\langle A^T|B\rangle$ 
for any $A,B \in {\cal H}_{aux}$. 
The reflector is a grassmann-even vector and 
has the ghost number equal to six. 
Hence the odd symplectic form $\langle \omega^c_{12}|$ is a grassmann-odd 
vector and has the ghost number equal to seven. 
The selection rule of $\omega^c$ becomes :
$\omega^c(A,B)\!\neq\!0 \Longrightarrow G(A)+G(B)\!=\!5$. 
The inverse of the odd symplectic form is called sewing ket. 
We denote it by $|S^c_{12}\rangle$. It is an element of 
${\cal H}_c^{\otimes 2}$ and satisfies 
$\langle \omega^c_{12}|S^c_{23}\rangle \!=\!_3P_1$. 
The sewing ket is grassmann-odd and has the ghost number 
equal to five.

We want to gain ultimately a field theory which describes 
interactions of open- and closed-strings. 
If such a field theory consistently exists, 
we need to have all the string vertices for these interactions. 
Recent attempts in this direction can be found in 
\cite{Zwiebach C-O, Kugo et al}. 
Nevertheless the construction is still beyond completion. 
One of the reasons is that we still do not have a suitable physical 
perspective of closed-strings by open-strings 
or open-strings by closed-strings. 
We also wish to find such a physical perspective 
in the following discussion. 
Our discussion is based on several assumptions. 
They are physically reasonable and acceptable if a field 
theory of interacting open- and closed-strings exists 
consistently.

We assume an existence of string vertices 
$\Bll{\bf 0_c};1 ~\cdots~ n\!:\!\zeta \Br$ for $n\!\geq\!1$. 
Here the index ${\bf 0_c}$ represents a closed-string on $D$  
and the indices $i$ $(1\!\leq\!i\!\leq\!n)$ label clockwise-ordered 
$n$ open-strings on $\partial D$. 
The vertices describe interactions between a single closed-string 
and $n$ open-strings at the cut-off scale $\zeta$. 
If one takes the closed-string picture, an incoming closed-string 
is reflected at $\partial D$ and at the same time decaying into 
open-strings. 
The interaction term   
$\Bll {\bf 0_c};1~\cdots~n:\zeta\Br
\bigl.\Psi\brr_{{\bf 0_c}}
\bl\Phi\brr_1 \cdots \bl\Phi\brr_n$ should not vanish. 
Recalling the ghost numbers $G(\Psi)\!=\!2$ and $G(\Phi)\!=\!1$, 
this requires that the ghost number of the vertex 
$\Bll {\bf 0_c};1~ \cdots~ n:\zeta\Br$ is equal to 
$n\!+\!4$. Also for the non-vanishing, 
odd-grassmannity of $\Phi$ requires the following 
cyclic asymmetry with respect to the open-string indices. 
\begin{eqnarray}
\Bll {\bf 0_c};1~2~\cdots~ n\!-\!1~n:\zeta\Br
=
(-)^{n+1}
\Bll {\bf 0_c};2~3~\cdots~ n~1:\zeta \Br.
\label{asymmetry of (1,n)-vertex}
\end{eqnarray}

Explicit construction of the vertices 
was examined in \cite{Hata-Nojiri} 
for the cases of $n\!=\!1,2$ by a slightly different formulation. 
In our formulation it will be generalized based on a conjectural 
map (or a section of the Hilbert bundle on ${\cal CM}^1_n$),  
\begin{eqnarray}
\begin{array}{ccc}
{\cal CM}^1_n & 
-\!\!\!-\!\!\!-\!\!\!\longrightarrow & 
({\cal H}_c\times {\cal H}_o^{\otimes n})^* \\
\Sigma &
\mapsto &
\langle \Sigma |
\end{array}~~~~~.
\label{state of CM(1,n)} 
\end{eqnarray}
This map gives rise to  
a $({\cal H}_c \times {\cal H}_o^{\otimes n})^*$-valued 
$(n\!-\!1)$-form $\langle \Omega |$ on ${\cal CM}^1_n$. 
The vertex will be defined as an integration of thus obtained 
$\langle \Omega|$ on ${\cal V}^1_n(\zeta)$. 
\begin{eqnarray}
\Bll {\bf 0_c};1~\cdots~n:\zeta \Br
=
\int_{{\cal V}^1_n(\zeta)}
\langle \Omega|^{(1~ \cdots~ n)}
\prod_{i=1}^ne^{-\zeta L_0^{(i)}}. 
\label{(1,n)-vertex}
\end{eqnarray}
As is the case of open-string field theory, 
action of the BRST charge on the vertices 
is expected to be a representation of the boundary operator $\partial$. 
Since it is an interacting theory of closed- and open-strings, 
the BRST charge in question should be a sum of the closed-string 
BRST charge $Q_c$ and the open-string BRST charge $Q_o$.  
Hence we arrive at the following 
conjectural action of the BRS charge. 
\begin{eqnarray}
\Bll {\bf 0_c};1~ \cdots~ n:\zeta \Br
\Bigl(Q_c^{({\bf 0_c})}\!+\!\sum_{i=1}^n Q_o^{(i)}\Bigr)
=
\sum_{(I,J)}
(\pm)
\Bll {\bf  0_c};I,a:\zeta \Br
\Bll a',J:\zeta \Bigr.
\Bl S_{a'a}^{o}\Brr, 
\label{Q-action on (1,n)-vertex}
\end{eqnarray} 
where we denote the open-string inverse reflector by 
$\Bl S_{a'a}^o \Brr$.

\subsubsection*{Boundary states}
So far, our consideration of the string vertices 
in this section excludes the case of $n\!=\!0$.  
In this particular case we have the state 
$\Bll {\bf 0_c}\Br \in {\cal H}_c^*$. 
(It is different from the $SL_2$-invariant vacuum 
$\langle 0 |$.) 
This is closed-string vertex which describes 
a simple reflection of a single closed-string at $\partial D$. 
It is a BRST invariant state and has the ghost number four. 
If we take the dual by using the sewing ket of closed-string, 
it acquires a familiar form \cite{Hata-Hashimoto}. 
\begin{eqnarray}
\Bl B \Brr \equiv 
\Bll {\bf 0_c} \Br 
\Bigl. S_{{\bf 0_c *_c}}^{c}\Brr .
\label{Ishibashi state}
\end{eqnarray}
This is a boundary state of closed-string 
and is called the Ishibashi state. It satisfies 
\begin{eqnarray}
Q_c \Bl B \Brr =0,~~~~~~~G\Bigl( \Bl B \Brr \Bigr)=3.  
\end{eqnarray}
Due to the correct ghost number  
we have the non-vanishing coupling 
$\omega^c( \Psi, |B \rangle)$ with closed-string field.  
Strictly speaking, the Ishibashi state  
is not a state of ${\cal H}_c$ because its norm turns out 
to be divergent. 
Without any contradiction there is no local field operator 
in $2D$ CFT which corresponds to this state. 
For a suitable class of $2D$ CFT,  
physically acceptable bases of solutions of 
$Q_c|B \rangle=0$ are found in \cite{Cardy} and 
called the Cardy bases. 
Roughly speaking, they are characterized 
by the reflection conditions  
of closed-string at $\partial D$. Of course 
it can be rephrased as the boundary conditions 
of open-strings when they interact with 
the closed-string via the vertex (\ref{(1,n)-vertex}). 
For simplicity we impose the Neumann condition 
on the reflection of closed-string.

Let us consider the following boundary state 
in the presence of an open-string field $\Phi$.  
\begin{eqnarray}
\Bigl|B[\Phi\!:\!\zeta]\Bigr\rangle 
\equiv 
e^{-\zeta L_0^{+(*_c)}}
\left\{
\Bll {\bf 0_c}\Br \Bigl. S_{{\bf 0_c *_c}}^{c} \Brr
+\sum_{k \geq 1}\frac{1}{k}
\Bll {\bf 0_c};1 \cdots k:\zeta \Br 
\Bigl. S^{c}_{{\bf 0_c *_c}}\Brr 
\bl \Phi\brr_1 \cdots \bl \Phi\brr_k 
\right\}. 
\label{boundary state bphi}
\end{eqnarray}
The ghost number of this state becomes three. This means that 
we have a non-vanishing coupling with closed-string field 
$\Psi$ of the form, 
$\omega^c\Bigl( 
\Psi,\Bigl|B[\Phi\!:\!\zeta]\Bigr\rangle
\Bigr)$. 
If one takes the closed-string picture, 
it describes an interaction of a closed-string 
with a reservoir of open-strings. 
Due to this interaction or a possible decay to open-strings 
the boundary state (\ref{boundary state bphi}) is not invariant 
under the action of the closed-string BRST charge $Q_c$. 
Our first investigation is about an interpretation of this action 
from the open-string viewpoint.

\subsubsection*{Role of $Q_c$ in open-string field theory} 
Let us examine the boundary state 
(\ref{boundary state bphi}) from the open-string picture. 
We regard the boundary state 
as a ${\cal H}_c$-valued function of $\Phi$. 
Then the corresponding hamiltonian vector 
$\Bigl|
\partial B[\Phi]/\partial \Phi
\Bigr \rangle$ 
is an element of ${\cal H}_c \times {\cal H}_o$. 
It is given by the variational formula 
(\ref{def of hamilton vector}), 
\begin{eqnarray}
\delta \Bigl| B[\Phi\!:\!\zeta] \Bigr\rangle 
=
\omega^o  
\left( \delta \Phi,~ 
\left|\frac{\partial B[\Phi\!:\!\zeta]}{\partial \Phi}
\right\rangle 
\right), 
\label{def of h-vector of state bphi}
\end{eqnarray}
where we denote the open-string odd symplectic structure 
by $\omega^o$. After a little calculation we find out the 
following expression of the hamiltonian vector. 
\begin{eqnarray}
\left |
\frac{\partial B[\Phi\!:\!\zeta]}{\partial \Phi}
\right \rangle 
=
e^{-\zeta L_0^{+(*_c)}}
\sum_{k \geq 1}
\Bll {\bf 0_c};1~ \cdots~ k\!-\!1~k:\zeta \Br
\Bigl. S^{c}_{{\bf 0_c*_c}}\Brr  
\bl \Phi\brr_1 \cdots \bl \Phi\brr_{k-1}
\Bl S^{o}_{k *_o}\Brr .  
\label{hamilton vector of boundaru state}
\end{eqnarray}

We now compute action of the closed-string BRST charge on the 
boundary state (\ref{boundary state bphi}). 
By a simple evaluation it can be rewritten in the following form. 
\begin{eqnarray}
Q_c
\Bigl|B[\Phi\!:\!\zeta]\Bigr\rangle 
&=&
\omega^o\! 
\left(
Q_o\Phi,~
\left| 
\frac{\partial B[\Phi\!:\!\zeta]}{\partial \Phi}
\right \rangle 
\right) 
\nonumber \\
&&+
e^{-\zeta L_0^{+({\bf *_c})}}
\sum_{k \geq 1}
\frac{(-)^{k+1}}{k}
\Bll {\bf 0_c}; 1~\cdots~ k:\zeta \Br 
\Bigl(Q_c^{({\bf 0_c})}\!+\!\sum_{i=1}^kQ_o^{(i)}\Bigr)
\Bl S^{c}_{{\bf 0_c *_c}} \Brr  
\bl \Phi \brr_1 \cdots \bl\Phi \brr_k. 
\nonumber \\
\label{action of Qc on B}
\end{eqnarray}
To proceed computation of the second term 
of eq.(\ref{action of Qc on B}) 
we need to use the conjectural form (\ref{Q-action on (1,n)-vertex}). 
The result is a suitable weighted sum of the quantities,  
\begin{eqnarray}
\omega^o
\left( 
m_p\Bigl(\Phi^p\!:\!\zeta \Bigr),~
e^{-\zeta L_0^{+({\bf *_c})}}
\Bll {\bf 0_c};1~\cdots~ q~0_o:\zeta \Br 
\Bigl. S^c_{{\bf 0_c *_c}}\Brr  
\bl \Phi\brr_1 \cdots \bl \Phi\brr_q 
\Bl S^o_{0_o *_o}\Brr  
\right), 
\label{component of Qc on bphi}
\end{eqnarray}
for $p \!\geq\! 2$ and $q \!\geq\! 0$. 
Their weights are determined by the signatures 
in eq.(\ref{Q-action on (1,n)-vertex}) 
or equivalently in eq.(\ref{boundary of V(0,1)(Lambda)}).   
So our computation may stop here. 
What we can learn from eq.(\ref{component of Qc on bphi})
is the appearance of the $A_{\infty}$-algebra. 
If one recalls that hamiltonian vector 
of the classical open-string field action  
(\ref{def of low energy action}) is the sum of $m_k$, 
one may infer that the weighted sum 
eventually gives rise to the following result. 
\begin{eqnarray}
Q_c
\Bigl |B[\Phi\!:\!\zeta]
\Bigr \rangle
=
\omega^o
\left( 
\left| 
\frac{\partial S[\Phi\!:\!\zeta]}{\partial \Phi}
\right \rangle,~
\left| \frac{\partial B[\Phi\!:\!\zeta]}{\partial \Phi}
\right \rangle
\right), 
\label{conjecture of Qc on bphi}
\end{eqnarray}
where 
$\Bigl|\partial S[\Phi\!:\!\zeta]/\partial \Phi \Bigr \rangle$ 
is the hamiltonian vector given by (\ref{hamilton vector of S}).

Eq.(\ref{conjecture of Qc on bphi}) must be tested.  
First of all, the above  $Q_c$-action needs to be nilpotent,  
$Q_c \left( Q_c \Bigl|B[\Phi]\Bigr\rangle \right)$
$=0$. 
Let us show this : 
By using the open-string anti-bracket 
we rewrite the RHS of eq.(\ref{conjecture of Qc on bphi}) as   
$\Bigl\{S[\Phi],|B[\Phi]\rangle \Bigr\}$. 
We then compute $Q_c\left(Q_c\Bigl|B[\Phi]\Bigr\rangle\right)$ 
as follows.
\begin{eqnarray}
Q_c\left(Q_c \Bigl|B[\Phi]\Bigr\rangle \right) 
&=& 
(-)Q_c
\left\{ 
\Bigl|B[\Phi]\Bigr\rangle,
S[\Phi] 
\right\} 
\nonumber \\
&=& 
(-)\left\{ 
Q_c\Bigl|B[\Phi]\Bigr\rangle,
S[\Phi]
\right\}
\nonumber \\
&=& 
\left\{
\Bigl\{ 
\Bigl|B[\Phi]\Bigr\rangle, 
S[\Phi] 
\Bigr\},
S[\Phi]
\right\}
\nonumber \\
&=& 
-\frac{1}{2}
\left\{ 
\Bigl\{S[\Phi],S[\Phi]\Bigr\},
\Bigl|B[\Phi]\Bigr\rangle 
\right\},   
\label{jacobi of SSB}
\end{eqnarray}
where we use the Jacobi identity (\ref{Jacobi identity}) 
to show the last equality. 
Since we have the classical master equation 
$\Bigl\{S,S\Bigr\}\!=\!0$ by Proposition 
\ref{classical master equation}, 
eq.(\ref{jacobi of SSB}) vanishes identically.  
Another consistency may be obtained if we can find the signatures 
in eq.(\ref{Q-action on (1,n)-vertex}) 
so that they provide a representation 
of the boundary operator $\partial$ and also 
give rise to eq.(\ref{conjecture of Qc on bphi}). 
As such a solution we finally find out the following ones. 
\begin{eqnarray}
&&
\Bll {\bf 0_c};1~ \cdots~ n:\zeta \Br 
\Bigl( Q_c^{({\bf 0_c})}\!+\!\sum_{i=1}^nQ_o^{(i)} \Bigr)
\nonumber \\
&&=
\sum_{l=0}^{n-2}\sum_{k=1}^{n}
(-)^{k(n+1)+(l+1)n}
\Bll {\bf  0_c};
\underbrace{k~ \cdots~ k\!+\!l\!-\!1~a}_{l+1}
:\zeta \Br  
\Bll 
\underbrace{a'~k\!+\!l~ \cdots~ k\!+\!n\!-\!1}_{n+1-l}
:\zeta \Br 
\Bigl. S_{a'a}^{o}\Brr . 
\end{eqnarray}
These are consistent with the asymmetry 
(\ref{asymmetry of (1,n)-vertex}).

Eq.(\ref{conjecture of Qc on bphi}) itself 
was alluded previously  
from a different perspective 
in \cite{Hata-Nojiri,Zwiebach C-O}. 
These authors intended to construct  
new symmetries of open-string field theory. 
Their argument is as follows.   
Let $\psi \in {\cal H}_c$ be a BRST-closed state 
($Q_c\psi\!=\!0$) with $G(\psi)\!=\!2$. 
Associated with such a state  
we can introduce a non-vanishing function  
$F_{\psi}(\Phi)\!\equiv\!
\omega^c\Bigl( \psi,|B(\Phi)\rangle \Bigr)$. 
Then, since $Q_c\psi$ vanishes,
eq.(\ref{conjecture of Qc on bphi}) implies  
$\Bigl\{S[\Phi],F_{\psi}(\Phi)\Bigr\}\!=\!0$.   
This means that $F_{\psi}$ generates a symmetry of 
the theory. As a trivial example we can consider 
the massless $U(1)$ gauge field $a_{\mu}$ 
(of open-string) in presence of the massless 
anti-symmetric tensor field $b_{\mu \nu}$ 
(of closed-string). 
In this simple case 
infinitesimal transformations, 
$\delta a_{\mu}\!=\!\eta_{\mu}$ and 
$\delta b_{\mu \nu}\!=\!
\partial_{\mu} \eta_{\nu}\!-\!\partial_{\nu}\eta_{\mu}$,  
correspond to a symmetry in question. 
$\eta_{\mu}$ is identified with a suitable component 
of $\psi\!=\!Q_c\eta$.

Significance of the $Q_c$-action (\ref{conjecture of Qc on bphi}) 
should be stressed. 
The BRST transformation 
${\bf \delta_{BRS}}$ 
of open-string field $\Phi$ is defined by 
eq.(\ref{BRST trans of Phi}).  
It is just the hamiltonian vector of $S[\Phi\!:\!\zeta]$. 
By using the variational formula (\ref{def of hamilton vector}) 
we can rewrite the $Q_c$-action into 
\begin{eqnarray}
Q_c \Bigl| B[\Phi\!:\!\zeta] \Bigr\rangle 
=
{\bf \delta_{BRS}}
\Bigl| B[\Phi\!:\!\zeta] \Bigr\rangle. 
\label{Qc=Qo}
\end{eqnarray}
Thus the closed-string BRST charge 
induces the BRST transformation 
of open-string field.  
Eq.(\ref{Qc=Qo}) becomes a key in the construction of 
boundary open-string field theory.

\subsubsection*{Boundary open-string field theory}
A framework for background independent 
open-string field theory was proposed 
in \cite{Witten BOSFT} based on the BV formalism. 
It is conventionally called boundary open-string field theory. 
This formulation was further investigated 
in \cite{Shatashivili,Witten BOSFT2,Li-Witten}. 
Construction based on the BV formalism requires a triple 
$\left( ({\cal X},\omega),V\right)$. 
${\cal X}$ is a super-manifold.  
$\omega$ is an odd symplectic structure of ${\cal X}$ 
and $V$ is a nilpotent fermionic vector field 
on ${\cal X}$. 
Having such a triple, 
the BV action, if it exists, 
is obtained as the hamiltonian function of $V$. 
The nilpotency of $V$ ensures the classical master equation. 
And the BRST transformation of the theory is given by $V$. 
In order to construct boundary open-string field theory, 
a hypothetical ``space of all open-string world-sheet theories'' 
is taken \cite{Witten BOSFT} as ${\cal X}$. 
$\omega$ is determined by  
correlation functions of world-sheet theories. 
$V$ is practically identified 
with the {\it closed}-string BRST charge $Q_c$.

We want to interpret the macroscopic open-string field theory 
presented in this paper as a boundary open-string field theory.

Our formulation is based on the Wilson renormalization group. 
We start at the trivial classical solution $\Phi\!=\!0$. 
This solution describes a flat Minkowski space ${\bf R}^{(1,25)}$   
and is a conformal point of the above ${\cal X}$.  
We regards it as a UV fixed point. 
The open-string Hilbert space ${\cal H}$ used in this paper 
is the tangent of ${\cal X}$ at $\Phi\!=\!0$. 
A local coordinate patch of ${\cal X}$ centered 
at this point will be given by the tangent space 
via the exponential map. 
Our naive expectation is that,  
excluding the critical points,  
theory at $\Phi\!\neq\!0$ on this patch is described 
by the classical action $S[\Phi\!:\!\zeta]$ 
(or the fluctuation given in Remark \ref{shift of Phi in S}). 
To pursue this naive picture we need to interpret the scale 
parameter $\zeta$ correctly.

The regularization employed in this paper corresponds to 
a point-splitting regularization of short-distance on the 
boundary when $\zeta$ is sufficiently large. The point-splitting 
is prescribed by the boundary length scale 
$a \sim e^{-\zeta}$ measured by the boundary metric $dxdx$. 
Consider a space of all open-string world-sheet cut-off theories 
which cut-off length scale is $a$. 
The tangent space at $\Phi\!=\!0$, which we denote by 
${\cal Y}(a)$, is identified with the open-string 
Hilbert space ${\cal H}$. 
The RG flow in the previous sections defines a map 
\begin{eqnarray}
{\cal R}_{\zeta'}^{\zeta}~~~~~
{\cal Y}(e^{-\zeta}) 
\longrightarrow 
{\cal Y}(e^{-\zeta'}), 
\end{eqnarray}
where $\zeta > \zeta'$. 
The previous construction of the RG flow 
is the reverse of the above map. 
This is due to the different interpretation of the scale 
parameter $\zeta$. On the boundary $e^{-\zeta}$ 
plays the role of the short distance cut-off. 
Both spaces ${\cal Y}(e^{-\zeta})$ and ${\cal Y}(e^{-\zeta'})$
are identified with ${\cal H}$. 
The RG flow becomes the flow on ${\cal H}$ generated by 
\begin{eqnarray}
\frac{d\Phi}{d \zeta}=
-\left\{ 
b_0 \left| 
\frac{\partial S_{int}[\Phi\!:\!\zeta]}{\partial \Phi}
\right \rangle
+Qb_0\Phi 
\right\}. 
\end{eqnarray}

The classical action $S[\Phi\!:\!\zeta]$ defines  
a boundary open-string field theory constructed on 
${\cal Y}(e^{-\zeta})$. 
By Proposition \ref{BRS invariance of the action}, 
the action $S[\Phi\!:\!\zeta]$ satisfies 
$\delta S[\Phi\!:\!\zeta]\!=\!\omega
\Bigl(\delta\Phi,{\bf \delta_{BRS}}\Phi \Bigr)$. 
Thus it is the hamiltonian function of 
${\bf \delta_{BRS}}$. 
We now regards ${\bf \delta_{BRS}}$ 
as a nilpotent fermionic vector field 
on ${\cal Y}(e^{-\zeta})$.  
As we find in eq.(\ref{Qc=Qo}), it can be identified 
with the {\it closed}-string BRST charge $Q_c$. 
Therefore geometrical ingredients of both theories become same.

In order to define 
a ``space of all open-string world-sheet theories'', 
we first need to take a continuum limit 
by letting $\zeta \rightarrow \infty$. 
The existence of infinitely many irrelevant operators 
\footnote{States $\phi_i$ of ${\cal H}^S$ which 
satisfy $\Delta_i >0$.}
causes serious problems. 
The hypothetical space ${\cal X}$
is supposed to be as follows in the original description 
\cite{Witten BOSFT}.
Let $I_0$ be a two-dimensional Lagrangian (on $D$) which describes 
a fixed bulk closed-string background. 
One considers two-dimensional Lagrangians of the form, 
\begin{eqnarray}
I=I_0+I_{\partial D},
\end{eqnarray}
where $I_{\partial D}$ is a suitable boundary term 
describing the coupling to external open-strings. 
\begin{eqnarray}
I_{\partial D}=
\int_{\partial D}\!d\theta~ 
{\cal V}(X,b,c), 
\end{eqnarray}
where $d\theta$ is the standard line element of $S^1$. 
The space ${\cal X}$ is roughly introduced as the space 
of Lagrangians $I$ with $I_0$ fixed and $I_{\partial D}$ 
allowed to vary. In order to obtain ${\cal X}$ as such,  
one needs to define $I_{\partial D}$ correctly by a suitable 
regularization and then take a continuum limit. 
In \cite{Li-Witten} \cite{K-V} such a prescription was partially 
examined. It is amusing to reconsider the results of \cite{Li-Witten}
\cite{K-V} from the perspective of our formulation.

\appendix
\newpage
\section{Appendix}
In this appendix we give a proof of Proposition 
\ref{test for conjecture} presented in the text. 
We first compute the LHS of eq.(\ref{nilpotency in conjecture}) 
by using eq.(\ref{conjecture 1}) as follows. 
\begin{eqnarray}
&& 
\left\{
\Bll 1~2~ \cdots~ n \!:\!\zeta \Br
\Bigl(\sum_{i=1}^nQ^{(i)}\Bigr)
\right\} 
\Bigl(\sum_{i=1}^nQ^{(i)}\Bigr)
\nonumber \\
&&=
-\frac{1}{2}
\sum_{k=1}^{n}\sum_{l=1}^{n-3}
(-)^{(n+1)(k+l+1)}
\Bll 
\underbrace{k~ \cdots~ k\!+\!l~a}_{l+2} \!:\!\zeta \Br  
\Bll 
\underbrace{a'~k\!+\!l\!+\!1~ \cdots~ k\!+\!n\!-\!1}_{n-l}
\!:\! \zeta \Br 
\Bigl. S_{a'a} \Brr 
\Bigl( \sum_{j=1}^{n}Q^{(j)} \Bigr)
\nonumber \\
&&=
\frac{1}{2}
\sum_{k=1}^{n}\sum_{l=1}^{n-3}
(-)^{(n+1)(k+l+1)+n-l}
\left\{
\Bll 
k~ \cdots~ k\!+\!l~a \!:\!\zeta
\Br  
\Bigl(Q^{(a)}+\sum_{j=k}^{k+l}Q^{(j)} \Bigr)
\right\}
\nonumber \\ 
&&~~~~~~~~~~~~~~~~~~~~~~~~~~~~~~~~~~~~~~~~~~~~~~~~~~~~
\times 
\Bll 
a'~k\!+\!l\!+\!1~ \cdots~ k\!+\!n\!-\!1 
\!:\! \zeta
\Br 
\Bigl. S_{a'a} \Brr 
\label{append 1}
\\
&&~~+
\frac{1}{2}
\sum_{k=1}^{n}\sum_{l=1}^{n-3}
(-)^{(n+1)(k+l+1)}
\Bll 
k~ \cdots~ k\!+\!l~a \!:\!\zeta \Br 
\nonumber \\
&&~~~~~~~~~~~~~~~~~~~~~~~
\times  
\left\{
\Bll 
a'~k\!+\!l\!+\!1~ \cdots~ k\!+\!n\!-\!1 
\!:\! \zeta \Br 
\Bigl(Q^{(a')}+\sum_{j=k+l+1}^{k+n-1}Q^{(j)} \Bigr)
\right\}
\Bl S_{a'a} \Brr , 
\label{append 2}
\end{eqnarray}
where we use the BRST invariance 
(\ref{BRST invariance of inverse reflector}) 
of the inverse reflector to show the last equality. 
We rewrite (\ref{append 1}) by replacing two vertices in the equation.  
Taking account of the asymmetries (\ref{conjecture 2}) 
and the grassmannities, it becomes as follows. 
\begin{eqnarray}
(\ref{append 1})
&=& 
\frac{1}{2}\sum_{k=1}^n\sum_{l=1}^{n-3}
(-)^{(n+1)k}
\Bll 
\underbrace{k\!+\!l\!+\!1~ \cdots~ k\!+\!n\!-\!1}_{n-l}
\!:\!\zeta
\Br 
\nonumber \\
&&~~~~
\times 
\left\{ 
\Bll 
\underbrace{a'~k~ \cdots~ k\!+\!l}_{l+2}
\!:\! \zeta
\Br 
\Bigl(
Q^{(a')}\!+\!\sum_{j=k}^{k+l}Q^{(j)}
\Bigr)
\right\} 
\Bl S_{a'a} \Brr.  
\label{append 3}
\end{eqnarray} 
After a slight change of the open-string indices  
eq.(\ref{append 3}) turns out to be the second term (\ref{append 2}).  
Thus we obtain the following expression 
for the LHS of (\ref{nilpotency in conjecture}).
\begin{eqnarray}
&& 
\left\{
\Bll 1~2~ \cdots~ n \!:\!\zeta \Br
\Bigl(\sum_{i=1}^nQ^{(i)}\Bigr)
\right\} 
\Bigl(\sum_{i=1}^nQ^{(i)}\Bigr)
\nonumber \\
&&=
\sum_{k=1}^{n}\sum_{l=1}^{n-3}
(-)^{(n+1)(k+l+1)}
\Bll 
\underbrace{k~\cdots~ k\!+\!l~a}_{l+2} \!:\!\zeta \Br 
\nonumber \\
&&~~~~~~~~~~~~~~~~~~~~~~~
\times  
\left\{
\Bll 
\underbrace{a'~k\!+\!l\!+\!1~ \cdots~ k\!+\!n\!-\!1}_{n-l} 
\!:\! \zeta \Br 
\Bigl(Q^{(a')}\!+\!\sum_{j=k+l+1}^{k+n-1}Q^{(j)} \Bigr)
\right\}
\Bl S_{a'a} \Brr.
\label{append 4}
\end{eqnarray}

\begin{figure}[t]
\begin{center}
\includegraphics[height=5cm]{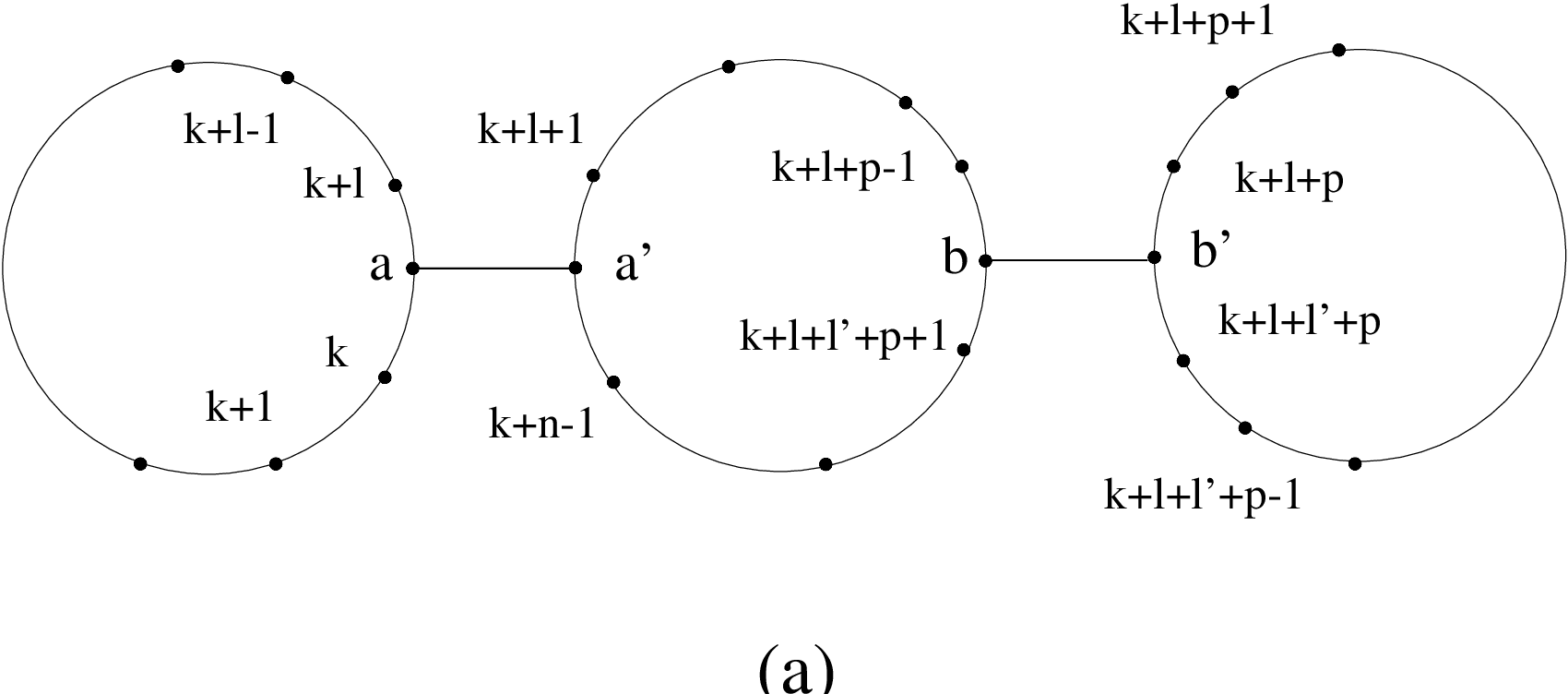} ~~~~~~~~~~~~~~~~~~
\includegraphics[height=5.2cm]{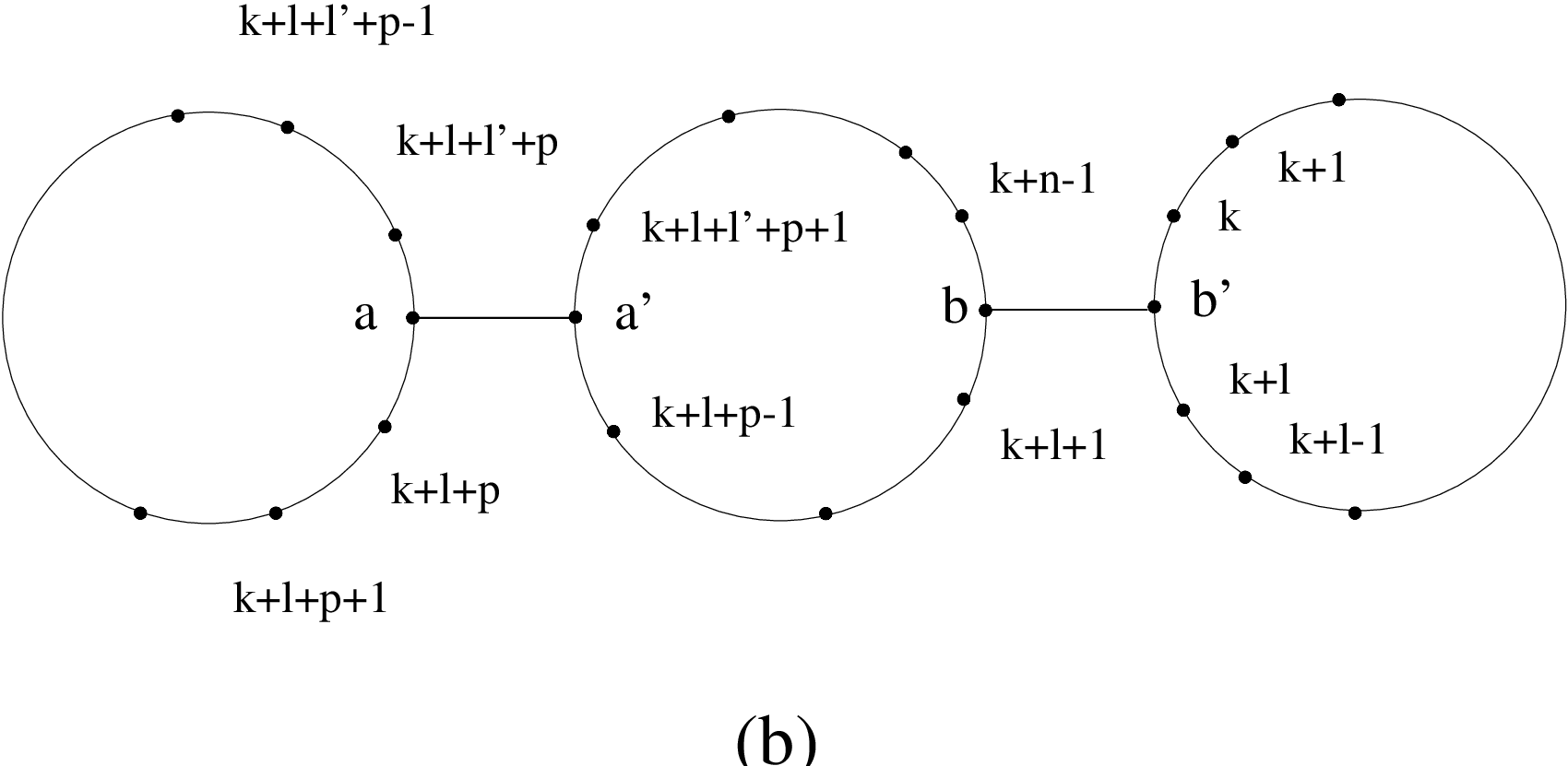} 
\caption{{\small 
(a) Open-string diagram which appears in eq.(\ref{append 5}).
(b) Open-string diagram which appears in eq.(\ref{append 6}). }}
\label{appendix fig}
\end{center}
\end{figure}

In order to compute the action of the BRST charge 
in the RHS of eq.(\ref{append 4}), the following lemma becomes useful. 
\begin{lemma}
\label{lemma appendix}
The action (\ref{conjecture 1}) of the BRST charge 
can be written as follows. 
\begin{eqnarray}
&&
\Bll 1~ \cdots~ n\!:\!\zeta \Br  
\Bigl(\sum_{i=1}^{n}Q^{(i)}\Bigr)
\nonumber \\
&&~~~~=
-
\sum_{l=1}^{n-3}~\sum_{k=1}^{n-l-1}
(-)^{(n+1)(k+1)}
\Bll 
\underbrace{k\!+\!l\!+\!2 ~\cdots~ n~1~2 \cdots k~a}_{n-l}
\!:\!\zeta
\Br 
\Bll  
\underbrace{a'~k\!+\!1~ \cdots~ k\!+\!l\!+\!1}_{l+2}
\!:\!\zeta
\Br 
\Bigl. S_{a'a}\Brr.   
\nonumber \\
\label{eq lemma appendix}
\end{eqnarray}
\end{lemma}
We omit the proof of this lemma. 
By using this lemma  our computation goes as follows. 
\begin{eqnarray}
&&
Eq.(\ref{append 4}) 
\nonumber \\ 
&&=
\sum_{k=1}^{n}\sum_{l=1}^{n-3}
(-)^{(n+1)(k+l+1)}
\Bll 
\underbrace{k~\cdots~ k\!+\!l~a}_{l+2} 
\!:\!\zeta
\Br 
\nonumber \\
&&~~~~
\times  
\left\{
\sum_{l'=1}^{n-l-3}~\sum_{p=1}^{n-l-l'-1}
(-)^{(n-l+1)(p+1)+1} 
\Bll 
\underbrace{
k\!+\!l\!+\!l'\!+\!p\!+\!1 \cdots a'~
\cdots~ k\!+\!l\!+\!p\!-\!1~b}_{n-l-l'}
\!:\! \zeta
\Br 
\right.
\nonumber \\
&&~~~~~~~~~~~~~~~~~~
\left.
\times 
\Bll 
\underbrace{
b'~k\!+\!l\!+\!p~ \cdots~  k\!+\!l\!+\!l'\!+\!p}_{l'+2}
\!:\! \zeta
\Br 
\Bigl. S_{b'b} \Brr 
\right\}
\Bl S_{a'a} \Brr
\nonumber \\
&&=
\sum_{k=1}^{n}\sum_{l=1}^{n-3}
~\sum_{l'=1}^{n-l-3}~\sum_{p=1}^{n-l-l'-1}
(-)^{(n+1)(k+p+l)+l(l+p)+1}
\Bll 
\underbrace{k~ \cdots~ k\!+\!l~a}_{l+2} 
\!:\!\zeta
\Br 
\nonumber \\
&&~~~~~~~~~~~~~~~~~~~~~~~~~~~~~~~~~~~
\times 
\Bll 
\underbrace{
k\!+\!l\!+\!l'\!+\!p\!+\!1~ \cdots~ a'~
\cdots~ k\!+\!l\!+\!p\!-\!1~b}_{n-l-l'}
\!:\! \zeta 
\Br 
\nonumber \\
&&~~~~~~~~~~~~~~~~~~~~~~~~~~~~~~~~~~~~~~~~~~~~
\times 
\Bll 
\underbrace{
b'~k\!+\!l\!+\!p~ \cdots~  k\!+\!l\!+\!l'\!+\!p}_{l'+2}
\!:\! \zeta
\Br 
\Bigl. S_{b'b} \Brr \Bl S_{a'a} \Brr. 
\label{append 5}
\end{eqnarray}
Terms appearing in eq.(\ref{append 5}) are conveniently depicted in 
Figure \ref{appendix fig}-(a).

We want to show that eq.(\ref{append 5}) vanishes identically. 
For this purpose we exchange the first and third vertices 
in eq.(\ref{append 5}) taking account of the grassmannities, 
and then rewrite eq.(\ref{append 5}) by using the asymmetries 
(\ref{conjecture 2}) of the vertices. 
We obtain the following expression for eq.(\ref{append 5}). 
\begin{eqnarray}
&&
Eq.(\ref{append 5}) 
\nonumber \\
&&=
\sum_{k=1}^{n}\sum_{l=1}^{n-3}~
\sum_{l'=1}^{n-l-3}~\sum_{p=1}^{n-l-l'-1}
(-)^{(n+1)(k+l')+l'(l+l'+p)}
\Bll 
\underbrace{k\!+\!p\!+\!l ~\cdots~ k\!+\!p\!+\!l\!+\!l'~a}_{l'+2} 
\!:\!\zeta
\Br 
\nonumber \\
&&~~~~~~~~~~~~~~~~~~~~~~~~~~~~~~~~~~~
\times 
\Bll 
\underbrace{
k\!+\!l\!+\!1~ \cdots~ a'~
~\cdots~ k\!+\!n\!-\!1~b}_{n-l-l'}
\!:\! \zeta 
\Br 
\nonumber \\
&&~~~~~~~~~~~~~~~~~~~~~~~~~~~~~~~~~~~~~~~~~~~~
\times 
\Bll 
\underbrace{
b'~k~ \cdots~  k\!+\!l}_{l+2}
\!:\! \zeta 
\Br 
\Bigl. S_{b'b} \Brr \Bl S_{a'a} \Brr. 
\label{append 6} 
\end{eqnarray}
Terms appearing in eq.(\ref{append 6}) are conveniently 
depicted in Figure \ref{appendix fig}-(b). 
%
%
%
%

To compare two expressions 
(\ref{append 5}) and (\ref{append 6}) of 
the LHS of eq.(\ref{nilpotency in conjecture}),  
we replace the open-string indices in eq.(\ref{append 6}) 
so that they match with those in eq.(\ref{append 5}). 
We first exchange $l \leftrightarrow l'$ 
and then put 
$k'\!\equiv\!k+p+l' $ and 
$p'\!\equiv\!n-p-l-l'$. 
With these replacements eq.(\ref{append 6}) becomes as follows. 
\begin{eqnarray}
&&
Eq.(\ref{append 6})
\nonumber \\
&&
=
\sum_{k=1}^{n}\sum_{l=1}^{n-3}~
\sum_{l'=1}^{n-l-3}~\sum_{p=1}^{n-l-l'-1}
(-)^{(n+1)(k+l)+l(l+l'+p)}
\Bll 
\underbrace{k\!+\!p\!+\!l'~ \cdots~ k\!+\!p\!+\!l\!+\!l'~a}_{l+2} 
\!:\!\zeta 
\Br 
\nonumber \\
&&~~~~~~~~~~~~~~~~~~~~~~~~~~~~~~~~~~~
\times 
\Bll 
\underbrace{
k\!+\!l'\!+\!1~ \cdots~ a'~
\cdots~ k\!+\!n\!-\!1~b}_{n-l-l'}
\!:\! \zeta 
\Br 
\nonumber \\
&&~~~~~~~~~~~~~~~~~~~~~~~~~~~~~~~~~~~~~~~~~~~~
\times 
\Bll 
\underbrace{
b'~k~ \cdots~  k\!+\!l'}_{l'+2}
\!:\! \zeta 
\Br 
\Bigl. S_{b'b} \Brr \Bl S_{a'a} \Brr 
\nonumber \\
&&
=
\sum_{k'=1}^{n}\sum_{l=1}^{n-3}
~\sum_{l'=1}^{n-l-3}~\sum_{p'=1}^{n-l-l'-1}
(-)^{(n+1)(k'+p'+l)+l(l+p')}
\Bll 
\underbrace{k'~ \cdots~ k'\!+\!l~a}_{l+2} 
\!:\!\zeta
\Br 
\nonumber \\
&&~~~~~~~~~~~~~~~~~~~~~~~~~~~~~~~~~~~
\times 
\Bll 
\underbrace{
k'\!+\!l\!+\!l'\!+\!p'\!+\!1~ \cdots~ a'~
\cdots~ k'\!+\!l\!+\!p'\!-\!1~b}_{n-l-l'}
\!:\! \zeta 
\Br 
\nonumber \\
&&~~~~~~~~~~~~~~~~~~~~~~~~~~~~~~~~~~~~~~~~~~~~
\times 
\Bll 
\underbrace{
b'~k'\!+\!l\!+\!p'~ \cdots~  k'\!+\!l\!+\!l'\!+\!p'}_{l'+2}
\!:\! \zeta
\Br 
\Bigl. S_{b'b} \Brr \Bl S_{a'a} \Brr. 
\nonumber \\ 
\label{append 7}
\end{eqnarray} 
If one compares this expression with eq.(\ref{append 5}), 
one finds that it is equal to $(-1)\times$eq.(\ref{append 5}). 
This means that eq.(\ref{append 5})$=(-1)\times$eq.(\ref{append 5}). 
Therefore eq.(\ref{append 5}) must vanish.    
We complete the proof of Proposition \ref{test for conjecture}.\P


\newpage
\section*{Acknowledgements} 
The interpretation as a boundary string field theory in the last 
section benefitted from discussions with A. Tsuchiya, 
A. Fujii, K. Higashijima and K. Murakami. 
The interpretation is modified to some extent 
including the addition of subsection 7.3 in the 
revised version. It benefitted from discussions 
with H. Kawai.  We also thank T. Kugo and T. Eguchi. 
We are grateful to T. Kubota and N. Yokoi 
for useful discussions at the early stage of this work, 
and to T. Kimura for a helpful instruction to draw figures 
in the text. 
Finally we thank the hospitality of 
the Lawrence Berkely National Laboratory, 
where some of this work was done.

\end{document}